\documentclass[useAMS,usenatbib]{mn2e}

\usepackage{graphicx}
\usepackage{subfig}
\usepackage{threeparttable}
\usepackage{txfonts}
\usepackage{natbib}
\usepackage{fixltx2e}
\usepackage{ownmacros}
\usepackage[T1]{fontenc}
\usepackage[compatibility = false]{caption}
\usepackage{aecompl}

\title[Astrometric follow-up observations of sub-stellar companions]{Astrometric follow-up observations of directly imaged sub-stellar companions to young stars and brown dwarfs\thanks{Based on observations made with ESO Telescopes at the La Silla Paranal Observatory under program IDs 081.C-0393(A), 083.C-0283(A), 085.C-0012(A), 087.C-0111(A), 088.C-0013(A), 089.C-0133(A), 089.C-0675(A), 090.C-0448(A).}}

\author[C. Ginski et al.]{C. Ginski$^{1}$\thanks{E-mail:
ginski@mail.strw.leidenuniv.nl}, T. O. B. Schmidt$^{2}$, M. Mugrauer$^{3}$, R. Neuh\"auser$^{3}$, N. Vogt$^{4}$,\newauthor R. Errmann$^{3,5}$, and A. Berndt$^{3}$\\
$^{1}$Sterrewacht Leiden, P.O. Box 9513, Niels Bohrweg 2, 2300RA Leiden, The Netherlands\\
$^{2}$Hamburger Sternwarte, Gojenbergsweg 112, 21029 Hamburg, Germany\\
$^{3}$Astrophysikalisches Institut und Universit\"ats-Sternwarte Jena, Schillerg\"asschen 2, 07745 Jena, Germany\\
$^{4}$Instituto de F\'isica y Astronom\'ia, Universidad de Valpara\'iso, Avenida Gran Breta\~na 1111, Valpara\'iso, Chile\\
$^{5}$Abbe Center of Photonics, Friedrich-Schiller-Universit\"at Jena, Max-Wien-Platz 1, 07743 Jena, Germany}
\begin{document}

\date{Accepted 2014 August 4.  Received 2014 August 1; in original form 2014 July 15}

\pagerange{\pageref{firstpage}--\pageref{lastpage}} \pubyear{2002}

\maketitle

\label{firstpage}

\begin{abstract}
The formation of massive planetary or brown dwarf companions at large projected separations from their host star is not yet well understood. In order to put constraints on formation scenarios we search for signatures in the orbit dynamics of the systems. We are specifically interested in the eccentricities and inclinations since those parameters might tell us about the dynamic history of the systems and where to look for additional low-mass sub-stellar companions. For this purpose we utilized VLT/NACO to take several well calibrated high resolution images of 6 target systems and analyze them together with available literature data points of those systems as well as Hubble Space Telescope archival data. We used a statistical Least-Squares Monte-Carlo approach to constrain the orbit elements of all systems that showed significant differential motion of the primary star and companion.\\
We show for the first time that the GQ\,Lup system shows significant change in both separation and position angle. Our analysis yields best fitting orbits for this system, which are eccentric (e between 0.21 and 0.69), but can not rule out circular orbits at high inclinations. Given our astrometry we discuss formation scenarios of the GQ\,Lup system. In addition, we detected an even fainter new companion candidate to GQ\,Lup, which is most likely a background object. We also updated the orbit constraints of the PZ\,Tel system, confirming that the companion is on a highly eccentric orbit with \textit{e}$>$0.62. Finally we show with a high significance, that there is no orbital motion observed in the cases of the DH\,Tau, HD\,203030 and 1RXS\,J160929.1-210524 systems and give the most precise relative astrometric measurement of the UScoCTIO\,108 system to date.
\end{abstract}


\section{Introduction}

Direct imaging surveys around nearby stars are revealing a growing number of sub-stellar companions. These massive (a few Jupiter masses up to 80\,M$_{\textrm{Jup}}$) objects  at large projected separations from their host stars form a complementary sample to the short period objects discovered by radial velocity and transit measurements. It is particularly interesting to examine if both groups of objects can form in a similar way or if these are really two physically distinctive populations. Determining the eccentricities of wide directly imaged companions could shed some light on this question. If these objects would have formed in-situ by core accretion in a disk, one would expect them to have low eccentricities due to the dampening effect that the disk material has on eccentricity excitations (see e.g. \citealt{1993ARA&A..31..129L}). Similarly, in-situ formation by gravitational instability is expected to form objects with low eccentricity as was found by \cite{2011ApJ...731...74B}. High eccentricities, on the other hand, would point towards dynamical interactions like planet-planet scattering events (see e.g. \citealt{2009ApJ...696.1600V} or \citealt{2011ApJ...742...72N}). In a very recent study, \cite{2013A&A...552A.129V} finds that gravitational instability can not form wide sub-stellar companions around stars with masses less than 0.7\,M$_\odot$ or with semi-major axes smaller than 170\,au. If objects within this parameter range exhibit high eccentricities this would again point towards dynamical interactions and against in-situ formation. In addition to the dynamical knowledge of the system that we gain from determining the eccentricities of wide companions, high eccentricities might also be a tell-tale sign for additional further-in planets, which may be detectable with indirect methods such as the transit or radial velocity method. \\
 There have been a number of recent studies on constraining the orbits of sub-stellar companions, for example the studies of the HR\,7672 system (\citealt{2012ApJ...751...97C}), the $\beta$\,Pic system (\citealt{2012A&A...542A..41C}) and the GJ\,504 system (\citealt{2013ApJ...774...11K}) in addition to some of our own studies of the HR7329 system (\citealt{2011MNRAS.416.1430N}) or the HD\,130948 system (\citealt{ginski-hd130}). All these studies utilized high precision relative astrometry to follow the apparent orbits of the sub-stellar companions around their host stars. In some of the mentioned cases it was already possible to put constraints on the eccentricity and inclination of the studied systems. \\
In this study we present new high precision astrometric measurements of six sub-stellar companions to stars and brown dwarfs. All observations were conducted with VLT/NACO (\citealt{2003SPIE.4839..140R}, \citealt{2003SPIE.4841..944L}) between 2009 and 2012. In the following sections we introduce the target systems in detail and describe our data reduction and astrometric calibration. We then analyze our astrometry of the system together with available literature and archival data and determine if significant orbital motion can be detected. Where this is the case we perform a statistical orbit analysis to constrain the possible orbit elements. Finally we compute detection limits for the systems and discuss our findings.\\


\section{The target systems}
\label{target-section}

\subsection{GQ\,Lup}
\label{sec: target: gqlup}
GQ\,Lup is a classical T Tauri star of spectral type K\,7 (\citealt{2009yCat.1280....0K}). It exhibits a proper motion of -15.1\,$\pm$\,2.8\,mas/yr in right ascension and -23.4\,$\pm$\,2.5\,mas/yr in declination as given in \cite{2010AJ....139.2184Z}\footnote{In \cite{2005AN....326..701M} proper motion measurements of GQ\,Lup of various authors are compared. They arrive at a weighted mean of -19.15\,$\pm$\,1.67\,mas/yr in right ascension and -21.06\,$\pm$\,1.69\,mas/yr in declination. However, each individual measurement is less precise than the one given by \cite{2010AJ....139.2184Z} and five out of six are consistent within 1\,$\sigma$ with this measurement.}. GQ\,Lup is located in the Lupus\,I cloud (\citealt{1996PASJ...48..489T}), a region of ongoing star formation. The distance to GQ\,Lup is generally assumed to be the average distance of objects in the Lupus\,I cloud of 140\,$\pm$\,50\,pc (see e.g. \citealt{1998A&A...330L..29N}). Parallax measurements of \cite{2008AA...484..281N} place GQ\,Lup at 156\,$\pm$\,50\,pc consistent with the average distance of Lupus\,I. Depending on the utilized evolutionary models for pre main sequence stars as well as the (variable) photometry of GQ\,Lup and its spectral type, the mass of GQ\,Lup is located between 0.3 and 0.9\,M$_\odot$ (\citealt{1994AJ....108.1071H}, \citealt{2005AN....326..701M}, \citealt{2008A&A...489..349S}). In this study we adopt a value of 0.7\,M$_\odot$ as was done by \cite{2005AA...435L..13N} or more recently by \cite{2010AJ....139..176F}. The age of the GQ\,Lup system according to \cite{2005AA...435L..13N} is 1\,$\pm$\,1\,Myr. The study by \citealt{2008A&A...489..349S} finds a slightly higher age of 3\,$\pm$\,2\,Myr from photometry in combination with evolutionary models. A similar result of 2\,$\pm$\,3\,Myr was found in the more recent study by \cite{2010A&A...517A..88W} via spectroscopy and template spectrum fitting. GQ\,Lup exhibits a strong mid- and far-infrared excess as was first detected by \cite{1994AJ....108.1071H}. This infrared excess points to the existence of warm dust and thus a disk around GQ\,Lup. Among others this has been confirmed by a recent study of \cite{2012ApJ...757....7M} employing data from the WISE (\citealt{2010AJ....140.1868W}) satellite mission. The circumstellar disk around GQ\,Lup has been marginally resolved by \cite{2010AJ....139..626D} using the SubMillimeter Array at a wavelength of 1.3\,mm. They find that their data fits with an outer disk radius of up to 75\,au. Additionally they state that they find no indication for gaps or holes in the disk. The inclination of the star's rotational axis (and thus the circumstellar disk assuming co-alignment) was first determined by \cite{2007A&A...468.1039B} via photometric rotation period determination in combination with the $v\,sin\,i$ measured by \cite{2005AN....326..958G}. They arrived at a value of \textit{i}\,=\,27$^\circ$\,$\pm$\,5$^\circ$. This result is consistent with a later study by \cite{2009A&A...498..793H} who find an inclination of $\sim$22$^\circ$ as best fit of GQ\,Lup VLT/CRIRES spectra to their disk model. However, \cite{2008A&A...489..349S} derive a much higher inclination of 53$^\circ$\,$\pm$\,18$^\circ$ as best fit to their spectrophotometric data, mainly because they find a longer rotation period than \cite{2007A&A...468.1039B}. This higher inclination is excluded by \cite{2009A&A...498..793H}.   \\
The sub-stellar companion to GQ\,Lup was discovered by \cite{2005AA...435L..13N} with Hubble Space Telescope (HST), Subaru and VLT observations. The authors used differential photometry and comparison with evolutionary models to constrain the companion mass to a range between 1 and 42\,M$_{\textrm{Jup}}$. A later study by \cite{2007ApJ...654L.151M} used the same available HST and Subaru archival data but different evolutionary models and found a mass range between 10 and 20\,M$_{\textrm{Jup}}$. They note that the mass estimate is strongly model dependent. They also found that the companion is significantly overluminous at $\sim$600nm which they attribute to strong ongoing accretion on the companion. A study by \cite{2007A&A...463..309S} used integral field spectroscopy to obtain a spectrum of the companion. They then proceeded to fit this spectrum to theoretical model atmospheres and derive a most likely mass of $\sim$25\,M$_{\textrm{Jup}}$ with a large uncertainty range between 4 and 155\,M$_{\textrm{Jup}}$ due to the large uncertainty of the surface gravity which they measure. However, they compare the spectrum of the companion to spectra of the eclipsing brown dwarf binary 2M0535 for which dynamical masses are known. They find, that the companion to GQ\,Lup is less luminous than either component of the eclipsing binary system and thus they constrain the upper mass of the companion to $\leq$36\,M$_{\textrm{Jup}}$. They also find a strong Pa\,$\beta$ emission line which they also interpret as sign of ongoing accretion on the companion. The ongoing accretion was also confirmed in a most recent photometric study by \cite{2014ApJ...783L..17Z}. It was shown by precise parallax measurements in \cite{2008AA...484..281N} that the companion and its host star are most likely located at the same distance and additionally that the companion shows some small change in relative separation to the host star over time which could be attributed to orbital motion.   

\subsection{PZ\,Tel}
\label{sec: target: pztel}
PZ\,Tel is a young nearby (51.5\,$\pm$\,2.5\,pc, \citealt{2007A&A...474..653V}) solar analog star of spectral type G\,9\,IV. It is located in the constellation of Telescopium and shows a proper motion of 17.64\,$\pm$\,1.13\,mas/yr in right ascension and -83.63\,$\pm$\,0.76\,mas/yr in declination as was determined by \cite{2007A&A...474..653V}. PZ\,Tel is part of the young 12$^{+8}_{-4}$\,Myr $\beta$\,Pic moving group (\citealt{2001ApJ...562L..87Z}). There have been various age determinations of PZ\,Tel putting it in an age range between 5 and 27\,Myr (see \citealt{2012MNRAS.420.3587J} for details), which is in good agreement with the average age range of $\beta$\,Pic moving group members. The mass of the star was recently estimated to lie between 1.1 and 1.3\,M$_\odot$ by \cite{2011MNRAS.410..190T}.\\
The sub-stellar companion to PZ\,Tel was discovered in parallel by \cite{2010A&A...523L...1M} and \cite{2010ApJ...720L..82B} using VLT/NACO and Gemini/NICI imaging respectively. Initial mass estimates relying on the age estimates of the host star as well as on theoretical evolutionary models by \cite{2000ApJ...542..464C}, \cite{2002A&A...382..563B} and \cite{2003A&A...402..701B} and photometry put the companion mass between 24 and 42\,M$_{\textrm{Jup}}$. Recently there was a new study by \cite{2014A&A...566A..85S} which used VLT/SINFONI and atmospheric DRIFT-PHOENIX models (see \citealt{1999JCoAM.109...41H} and \citealt{2008A&A...485..547H}) to determine the mass of the companion candidate independent of the age of the system. They arrived at a mass range between 3.2 and 24.4\,M$_{\textrm{Jup}}$ with a most probable mass of 21\,M$_{\textrm{Jup}}$, thus the companion is most likely a brown dwarf but could in principle also be a planet with a mass below the Deuterium burning limit. The spectral type of the companion is best fitting between M\,6 and L\,0. The authors derive an effective temperature of 2500$^{+138}_{-115}$\,K and a surface gravity of \textit{log\,g} = 3.50$^{+0.51}_{-0.30}$\,dex. We previously examined the orbit dynamics of the system in \cite{2012MNRAS.424.1714M}. The system shows significant orbital motion including orbit curvature. In our analysis we could constrain the possible orbit elements of the system with astrometric data taken between 2007 and 2011. We found that the system is most likely highly eccentric with an eccentricity above 0.6.

\subsection{DH\,Tau}
\label{sec: target: dhtau}

DH\,Tau is a classical T Tauri star of spectral type M\,1 (\citealt{2009ApJ...694..546W}). It is part of the Taurus Molecular Cloud (TMC), which had its spectroscopic distance determined by \cite{1994AJ....108.1872K} to be 140\,$\pm$\,10\,pc. It exhibits an average proper motion of 12\,$\pm$\,4\,mas/yr in right ascension and -25\,$\pm$\,3\,mas/yr in declination, as determined by \cite{2003AJ....125..984M}, \cite{2003yCat.1283....0H} and \cite{2004AJ....127.3043Z}. Mass and age of DH\,Tau have been estimated by \cite{1994ApJ...427..961H}, using spectroscopy as well as optical and infrared photometry. Utilizing the models by \cite{1994ApJS...90..467D} and \cite{1994ApJ...425..286S}, they derive a mass range for DH\,Tau of 0.24 to 0.32\,M$_{\odot}$ and an age range of 0.1 to 0.7\,Myr. A similar study has been conducted by \cite{2001ApJ...556..265W}, who used the BCAH98 models by \cite{1998A&A...337..403B}. They estimate a higher mass for DH\,Tau of 0.53\,M$_{\odot}$ and a much higher age of 4.4\,Myr.\\     
The companion to DH\,Tau was discovered by \cite{2005ApJ...620..984I} using the CIAO instrument (\citealt{2004PASJ...56..509M}) on the Subaru Telescope. It is located 2.3\,arcsec to the southeast of the primary at a position angle (PA) of 139.8$^{\circ}$. They confirm companionship by common proper motion using HST archival images from 1999, in which the companion is resolved as well. To estimate the mass of the companion, near-infrared spectra were taken with the OHS/CISCO instrument (\citealt{2002PASJ...54..315M}), also on the Subaru Telescope. By comparison with model spectra by \cite{2004ApJ...607..511T}, an effective temperature between 2700\,K and 2800\,K, and a surface gravity of $log\,g\,$=\,4.0 to 4.5 are derived for the companion. Using these and the models by \cite{1997MmSAI..68..807D} and \cite{2003A&A...402..701B}, a mass range of 0.03 to 0.05\,M$_{\odot}$ and an age range of 3 up to 10\,Myr are calculated. This places the companion in the brown dwarf regime, however \cite{2012arXiv1201.3537N} calculate a lower minimum mass of 0.006\,M$_{\odot}$ using various evolutionary models and taking a bolometric correction into account (the maximum mass is still up to 0.05\,M$_{\odot}$). \\

\subsection{HD\,203030}

HD\,203030 is a G\,8 dwarf (\citealt{1978BICDS..15..121J}) located in the constellation of Vulpus. The parallax of 24.46\,$\pm$\,0.74\,mas (corresponding to 40.9\,pc) and proper motion of 132.84\,$\pm$\,0.79\,mas/yr in right ascension and 8.44\,$\pm$\,0.65\,mas/yr in declination were measured by the Hipparcos satellite mission. The mass of HD\,203030 was determined independently in several studies and ranges from 0.93 to 1\,$M_{\odot}$ (\citealt{1999A&A...352..555A}, \citealt{2009yCat..21810062M}, \citealt{2011A&A...530A.138C}). The age was first estimated by \cite{2001MNRAS.328...45M}, who claim a likely membership of HD\,203030 in the young supercluster IC\,2391 by kinematics. The age of IC\,2391 members varies between 35 and 55\,Myr (\citealt{1991AJ....102.2028E}). This could not be confirmed by \cite{2006ApJ...651.1166M}, who did a detailed study of age indicators of the star. They find that chromospheric and coronal activity correspond to an age of 130 to 400\,Myr, consistent also with rotational period and Li abundance (\citealt{2000A&AS..142..275S}). Additionally, they find that optical colors and luminosity place HD\,203030 on the main sequence at an age range of 0.1 to 10\,Gyr, i.e. in agreement with the higher age estimate.\\
The companion to HD\,203030 was discovered by \cite{2006ApJ...651.1166M}, using the Hale 200\,inch and KeckII 10\,m telescope at the Palomar Observatory. It is located at an angular separation of $\sim$\,11.9\,arcsec (corresponding to $\sim$\,487\,AU) to the southeast (PA\,$\sim$\, 108.8$^{\circ}$) of the primary. Companionship of the object was confirmed by common proper motion with the primary in the same study. Using near infrared photometry and the models by \cite{1997ApJ...491..856B}, \cite{2000ApJ...542..464C} and \cite{2003A&A...402..701B}, a mass range of 0.012\,$M_{\odot}$ to 0.031\,$M_{\odot}$ is given, provided that the age range of the object is 130\,Myr to 400\,Myr. This places the companion with a high probability in the brown dwarf regime. The spectral type of the companion was determined by near infrared spectroscopy in the \textit{K}-band to be L\,7.5\,$\pm$\,0.5 (\citealt{2006ApJ...651.1166M}).\\

\subsection{1RXS\,J160929.1-210524}
\label{sec: target: rxj1609}

The K\,7 dwarf 1RXS\,J160929.1-210524 (\citealt{2008ApJ...689L.153L}) is located in the constellation of Scorpius. Its proper motion was determined by \cite{2009yCat.1315....0Z} to be $-11.2\,\pm\,1.5$\,mas/yr in right ascension and $-21.9\,\pm\,1.5$\,mas/yr in declination. It is a member of the young upper Scorpius OB association (US, \citealt{1999AJ....117.2381P}).\\
The mean distance of US was inferred by the measurement of Hipparcos parallaxes of member stars to be 145\,$\pm$\,2\,pc (\citealt{1999AJ....117..354D}), with an intrinsic scatter not larger than 20\,pc (\citealt{2002AJ....124..404P}). The age of US has recently become a matter of discussion. Originally it was determined in \cite{1985ASSL..120...95D} and \cite{1989A&A...216...44D} to be about 5 to 6\,Myr, by the H-R main-sequence turn-off point of high mass member stars. This was later confirmed in \cite{1999AJ....117.2381P} and \cite{2002AJ....124..404P}. There is, however, a recent paper by \cite{2012ApJ...746..154P}, stating that they found US members of spectral type F to be underluminous by a factor of $\sim$\,2.5, given the young age. They placed the various US members in H-R diagrams and thereby determined a mean age of 11\,$\pm$\,2\,Myr.\\
The substellar companion to 1RXS\,J160929.1-210524 was discovered by \cite{2008ApJ...689L.153L}, using the Gemini North Telescope with its AO system ALTAIR (\citealt{1998SPIE.3353..600R}) and the NIRI instrument (\citealt{2003PASP..115.1388H}). The companion is located 2.22\,arcsec north-east of the primary ($\sim$\,330\,AU), at a PA of 27.7$^\circ$. From near-infrared spectroscopy \cite{2008ApJ...689L.153L} inferred a spectral type of L4$^{+1}_{-2}$. Given the near-infrared photometry of the companion and using the age of 5\,Myr for US and a distance of 145\,$\pm$\,2\,pc, \cite{2008ApJ...689L.153L} estimate a mass of the companion of 8$^{+4}_{-2}$\,M$_{\textrm{Jup}}$, utilizing the DUSTY models by \cite{2000ApJ...542..464C}. However, they are using the in principle unjustified assumptions that there would be no negligible extinction and that the primary star would be constant. They are also using the 2MASS magnitudes of A, and the models of \cite{1998A&A...337..403B} to estimate the mass of the primary to be 0.85\,$^{+0.2}_{-0.1}$\,M$_{\odot}$. Since these mass limits for the companion are below the mass limit for Deuterium burning of about 13\,M$_{\textrm{Jup}}$, \cite{2008ApJ...689L.153L} state that the companion should be a planetary mass object. Given the recent revision of the age for US, \cite{2012ApJ...746..154P} recalculated the mass, using the luminosities by \cite{2008ApJ...689L.153L} and the DUSTY models by \cite{2000ApJ...542..464C}. They calculate a mass of 14$^{+2}_{-3}$\,M$_{\textrm{Jup}}$, placing the companion just above the Deuterium burning mass limit, and hence state that it seems more likely to be a brown dwarf, rather than a planetary mass object.\\
The common proper motion of the companion with the primary was more recently confirmed in \cite{2010ApJ...719..497L} with a significance of 6\,$\sigma$. 

\subsection{UScoCTIO\,108}

UScoCTIO\,108 is a brown dwarf and also member of US. It was discovered in the survey by \cite{2000AJ....120..479A} carried out at the Cerro Tololo Inter-American Observatory (CTIO). The membership in US was determined by photometry and low resolution spectroscopy, and the mean distance of the association of 145\,$\pm$\,2\,pc was adopted, as well as the age of the association of about 5\,Myr.\\
The companion to UScoCTIO\,108 was discovered by \cite{2008ApJ...673L.185B}, using 2MASS images. They then did follow-up observations using the Wilhelm Herschel Telescope and the Telescopio Nazionale Galileo. Additionally, they used the NIRSPEC instrument on the Keck\,II Telescope to obtain high resolution near-infrared spectra. The companion is located at a distance of 4.6\,arcsec to the south of the primary, at a PA of 177$^\circ$.\\
\cite{2008ApJ...673L.185B} fit the spectrum of UScoCTIO\,108 and its companion by comparison with other young and field dwarfs. They give a spectral type of M\,7\,$\pm$\,0.5 for the primary and M\,9.5 for the companion. Using their photometry and the COND models by \cite{2003A&A...402..701B}, as well as the distance and age of US, they estimate a mass of 60\,$\pm$\,20\,M$_{\textrm{Jup}}$ for the primary and 14$^{+2}_{-8}$\,M$_{\textrm{Jup}}$ for the companion, placing the companion just above the Deuterium burning mass limit. Therefore, the companion is most likely a low mass brown dwarf. This mass estimate changes slightly if the revised age for US is used as given in \cite{2012ApJ...746..154P}. They calculate a companion mass of 16$^{+3}_{-2}$\,M$_{\textrm{Jup}}$ using the DUSTY models by \cite{2000ApJ...542..464C}, making it even more likely that UScoCTIO\,108\,B is indeed a brown dwarf, rather than a planetary mass object.\\
\cite{2008ApJ...673L.185B} perform no proper motion analysis to confirm companionship of UScoCTIO\,108\,B, but rather calculate the probability for another US member to be within 10\,arcsec of UScoCTIO\,108 as only 1.3\,\%, given the density of US. If indeed bound, UScoCTIO\,108\,A and B form one of the widest substellar binaries known to date.\\


\section{Observations and data reduction}
\label{sec: obs}

All observations in this study were conducted using VLT/NACO. Generally the \textit{K}$_\textrm{s}$-band filter was used due to the better Strehl ratio in this band and the better contrast ratio between companion and primary star as compared to bands at shorter wavelengths. In some cases a neutral density filter was used in combination with the K$_s$ filter to prevent saturation of the bright primary stars. In the specific case of the HD\,203030 system the \textit{NB\,2.17} narrow band filter was used instead of the \textit{K}$_\textrm{s}$-band filter, again to prevent saturation of the bright primary star but at the same time to get enough flux from the faint companion in order to take a precise astrometric measurement.\\
Since most companions are located at very small angular separations from their host stars ($\leq$5\,arcsec) we used the S13 objective with a field of view of 14$\times$14\,arcsec for all but one of the target systems. The companion to HD\,203030 is approximately 12\,arcsec separated from its primary star and thus we utilized the S27 objective with a larger field of view of 28$\times$28\,arcsec in this case. \\
We used the jitter observation technique to subtract the bright infrared sky background. Due to the different target characteristics and observation program requirements, the detector integration times vary, but were generally chosen to provide a high signal-to-noise on the faint companions while avoiding saturation of the bright primary stars. Total integration times thus varied between 47.25\,min and 0.6\,min. Details of all observations are listed in Tab.~\ref{pm: tab: observation-summary}.\\
For data reduction we used the ESO-Eclipse software package (\citealt{2001ASPC..238..525D}). All individual images were flatfielded and dark-subtracted, then consecutive images taken at different dither positions were subtracted from each other to remove the infrared sky background. The final reduced images were shifted and co-added. To gain astrometric accuracy and improve detection limits the primary stars PSF (Point Spread Function) was subtracted using a roll subtraction technique. For this purpose the reduced images were artificially rotated in steps of 2$^\circ$ and at each step the rotated images was subtracted from the original images. This was done for rotation angles up to 360$^\circ$. The resulting difference images were then median combined to create a final image in which the radial symmetric part of the stellar PSF is removed. 

\begin{table*}
\centering
  \caption{Observation summary}
  \label{pm: tab: observation-summary}
  \begin{threeparttable}
  \begin{tabular}{@{}llllccc@{}}
  \hline   
Date							& Target					& Instrument			& Filter			& Exposure Time\,[s] \tnote{1}	& Pixel Scale\,[mas/pixel]					&  DPA\,$[^\circ]$			\\
 \hline
2008-06-14 				& GQ\,Lup					& VLT/NACO S13		&	\textit{K}$_\textrm{s}$				&	40$\times$24$\times$2.5				& 13.243\,$\pm$\,0.086							& +0.73\,$\pm$\,0.40		\\
2009-06-29 				&									& VLT/NACO S13		&	\textit{K}$_\textrm{s}$				&	23$\times$175$\times$0.3454		& 13.234\,$\pm$\,0.018							& +0.42\,$\pm$\,0.10		\\
2010-05-05 				&									& VLT/NACO S13		&	\textit{K}$_\textrm{s}$				&	5$\times$126$\times$0.347			& 13.231\,$\pm$\,0.020							& +0.67\,$\pm$\,0.13		\\
2011-06-05  			&									& VLT/NACO S13		&	\textit{K}$_\textrm{s}$				&	45$\times$126$\times$0.5			& 13.234\,$\pm$\,0.021							& +0.65\,$\pm$\,0.14		\\
2012-03-03				&									& VLT/NACO S13		&	\textit{ND\_short} / \textit{K}$_\textrm{s}$				&	36$\times$3$\times$20					& 13.234\,$\pm$\,0.022							& +0.68\,$\pm$\,0.15		\\
\hline
2012-06-08 				& PZ\,Tel					& VLT/NACO S13		&	\textit{K}$_s$				&	9$\times$51$\times$1					& 13.234\,$\pm$\,0.022							& +0.68\,$\pm$\,0.15		\\
2012-06-08 				& PZ\,Tel					& VLT/NACO S13		&	\textit{ND\_short} / \textit{K}$_\textrm{s}$				&	5$\times$10$\times$5					& 13.234\,$\pm$\,0.022							& +0.68\,$\pm$\,0.15		\\
\hline
1999-01-17				& DH\,Tau					&	HST/WFPC2				&	\textit{814w}				&	20 / 200 \tnote{2}						& 45.539 \tnote{3}									& 130.37 \tnote{3}								\\ 
2009-10-01 				&									& VLT/NACO S13		&	\textit{K}$_\textrm{s}$				&	22$\times$60$\times$1					& 13.234\,$\pm$\,0.018							& +0.42\,$\pm$\,0.10		\\
2012-01-22				& 								&	HST/WFC3				&	\textit{336w}				&	1400													& 39.617 \tnote{3}									& 40.24 \tnote{3}									\\ 
2012-01-22				& 								&	HST/WFC3				&	\textit{475w}				&	280														& 39.617 \tnote{3}									& 40.24 \tnote{3}									\\ 
2012-01-22				& 								&	HST/WFC3				&	\textit{625w}				&	100														& 39.617 \tnote{3}									& 40.24 \tnote{3}									\\ 
2012-01-22				& 								&	HST/WFC3				&	\textit{673n}				&	500														& 39.617 \tnote{3}									& 40.24 \tnote{3}									\\
2012-01-22				& 								&	HST/WFC3				&	\textit{775w}				&	80														& 39.617 \tnote{3}									& 40.24 \tnote{3}									\\ 
2012-01-22				& 								&	HST/WFC3				&	\textit{850lp}				&	40														& 39.617 \tnote{3}									& 40.24 \tnote{3}									\\  
2012-12-05				& 								&	VLT/NACO S13		&	\textit{K}$_\textrm{s}$				&	6$\times$12$\times$5					& 13.235\,$\pm$\,0.030							& +0.65\,$\pm$\,0.11		\\
\hline
2009-08-20				& HD\,203030			&	VLT/NACO S27		&	\textit{NB\,2.17}		&	23$\times$120$\times$0.5			& 27.15	\tnote{3}										& +0.42\,$\pm$\,0.10		\\
\hline
2009-08-15 				&	1RXS\,J1609			& VLT/NACO S13		&	\textit{K}$_\textrm{s}$				&	23$\times$60$\times$1					& 13.234\,$\pm$\,0.018							& +0.42\,$\pm$\,0.10		\\
\hline
2009-08-16 				&	UScoCTIO\,108		& VLT/NACO S13		&	\textit{K}$_\textrm{s}$				&	31$\times$1$\times$60					& 13.234\,$\pm$\,0.018							& +0.42\,$\pm$\,0.10		\\									
									
 \hline\end{tabular}
\begin{tablenotes}\scriptsize
\item[1] given in NEXP $\times$ NDIT $\times$ DIT if applicable, wherein DIT is the single integration time, NDIT the number of co-adds at one dither position and NEXP the number of dither positions
\item[2] both images were taken at the same pointing, the long exposure was used to measure the companion position and the short exposure to measure the primary position
\item[3] as provided in the respective image headers, no calibrators were available in these cases thus no uncertainties were calculated
\end{tablenotes}
\end{threeparttable}
\end{table*}


\section{Astrometric calibration and measurements}
\label{sec: astrometry}

\begin{figure}

\includegraphics[scale=0.5]{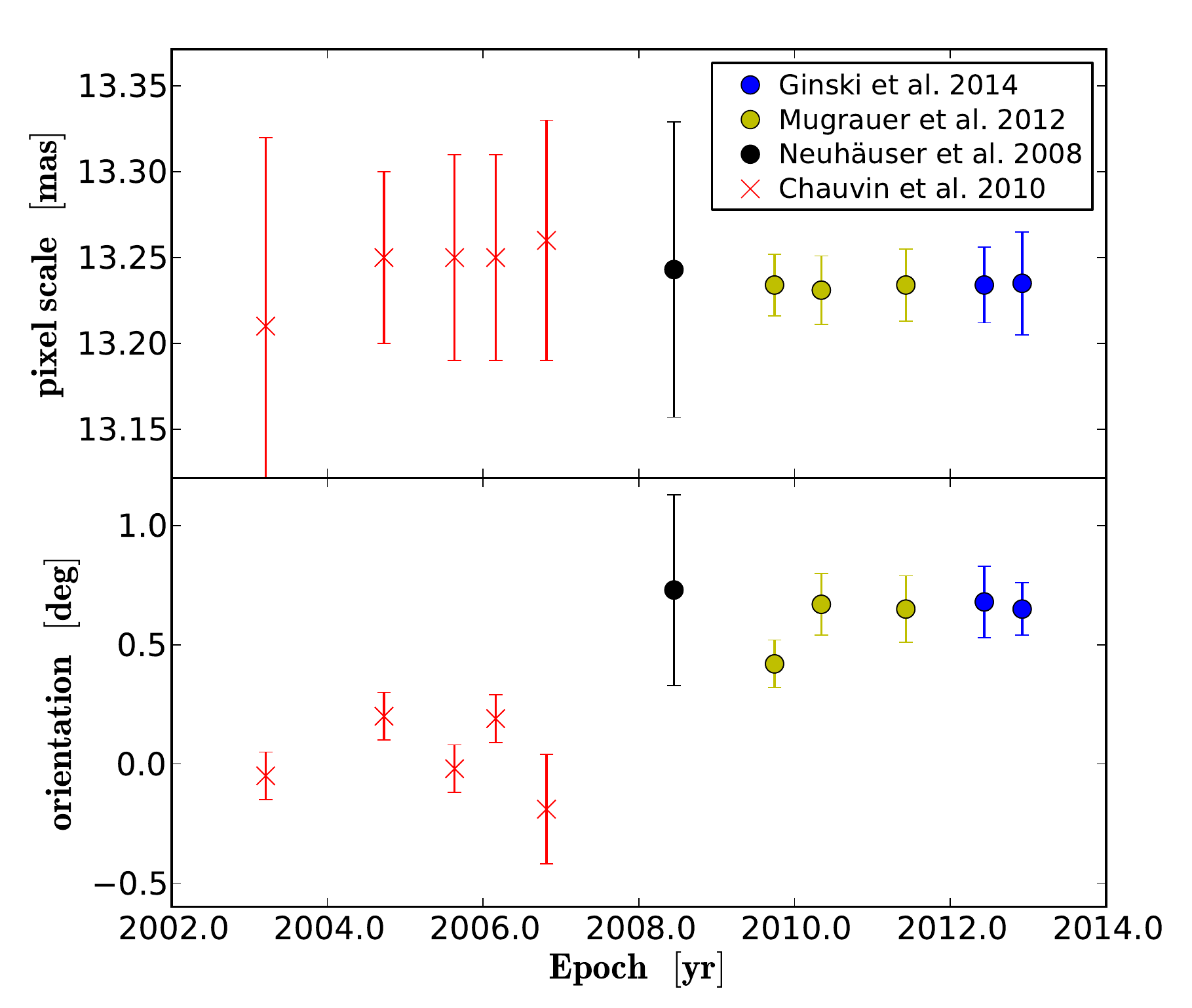}

\caption[]{Astrometric calibration of VLT/NACO in different observing epochs. The circle markers show astrometric solutions used in this work, all of them derived with the core of 47\,Tuc as astrometric calibrator. The crosses show earlier astrometric solutions by \cite{2010A&A...509A..52C}, which use Trapezium as astrometric calibrator. } 
\label{astrocal}
\end{figure}

In order to do high precision astrometric monitoring of our targets we need to do frequent astrometric calibrations of the used detector. For this purpose we imaged the center of the globular cluster 47\,Tuc, for which precise astrometry from HST observations is available. \textsc{GAIA} (Graphical Astronomy and Image Analysis tool, \citealt{2000ASPC..216..615D}) and the included \textsc{SExtractor} (Source Extractor, \citealt{1996A&AS..117..393B}) were used to extract the star positions from the HST image, and to create an astrometric reference catalog. The same was done for the NACO images of 47\,Tuc. The NACO catalogs were then matched with the reference, and pixel scales and detector orientations for each pair of stars were computed. Sigma clipping was then applied to exclude all stars from the NACO catalogs, which produced significantly different pixel scales and orientations. These discrepancies are most likely caused by a higher proper motion of such stars and hence larger deviations from the measured HST positions. The standard deviations of the pixel scales and detector orientations were adopted as uncertainties for both values.\\  
Images were usually taken in the same night (or within a few nights) as the science targets. However, the observations in 2009 as well as in early 2012 were conducted in service mode and thus it was not possible to obtain astrometric calibrators at the same time as the science targets. We thus relied on astrometric calibrations that have been taken with a time difference of up to 3 months. For NACO's S27 objective, which was employed on 2009-08-20 to image HD\,203030, there was no calibrator available to compute the pixel scale and its uncertainty (the detector orientation could still be calibrated with the 2009 calibrator). Thus in this case we relied on the information provided in the image header. All of the astrometric calibrations used in this work have been previously published in either \cite{2008AA...484..281N}, \cite{2012MNRAS.424.1714M} or \cite{2014MNRAS.438.1102G}. 
In Fig.~\ref{astrocal} we show the utilized astrometric solutions versus time. In addition we show astrometric solutions by \cite{2010A&A...509A..52C} which utilized the Trapezium cluster as astrometric calibrator. As can clearly be seen the pixel scale and orientation of NACO were stable between 2010 and 2012 with variations much smaller than the expected uncertainties. Thus our approach is justified. The final astrometric solutions used in each observation epoch are listed in Tab.~\ref{pm: tab: observation-summary}.\\
To measure the relative positions of primary star and companion we fitted a two dimensional gaussian to both components. To ensure that the bright halo of the primary stars is not influencing the position measurement of the faint companions, we removed the primary stars' PSF before measuring the companions' positions as described in the previous section. Each individual measurement was repeated several times with varying start parameters in order to ensure the stability of the fitted position. The measured image positions were then translated into angular separation and relative position angle on the sky using the astrometric solutions listed in Tab.~\ref{pm: tab: observation-summary}. The final results are listed in Tab.~\ref{tab: measurements}. The given uncertainties were calculated in each case by taking into account the average uncertainty of all individual image position measurements and the uncertainties of the astrometric calculations.

\section{Photometry}

For all our new VLT/NACO observation epochs we performed relative aperture photometry between primary star and companion. For this purpose we utilized the Aperture Photometry Tool (\cite{2012PASP..124..737L}). The aperture size was in each case adjusted to two times the full width at half maximum of the companion PSF. Companion and primary star instrumental magnitudes were then always measured with the same aperture size. Care was taken to ensure that the region used for background estimation was not polluted by light from the primary star or the companion. To prevent contamination of the companion flux from the bright primary halo, the primary PSF was always subtracted as described in section \ref{sec: obs} before the companion's instrumental magnitude was measured. The final results are listed in Tab.~\ref{tab: measurements}.\\ 
The apparent magnitudes of the companions were calculated using the obtained differential magnitudes as well as the 2MASS measurements of the primary stars. In the case of UScoCTIO\,108 no 2MASS measurement was available and we thus relied on the primary star's magnitude as given in \cite{2008ApJ...673L.185B}. In the case of the GQ\,Lup system the companion is in all observation epochs located close to a diffraction spike of the primary star. Thus the described PSF subtraction technique could not completely remove the flux contamination of the companion. The given magnitude differences should therefore be regarded as lower limits. Furthermore, in all cases where only lower limits are listed, the primary star was saturated or out of the linear regime of the detector and thus a precise determination of a differential magnitude was not possible. \\
The given uncertainties for the differential magnitudes include the uncertainties of the instrumental magnitudes of primary and companion. In addition, we conservatively included the small deviations in our error budget that were caused by a change of the aperture size of one pixel. The uncertainties of the given apparent magnitudes then also include the uncertainty of the respective 2MASS (or \citealt{2008ApJ...673L.185B}) measurement. \\
In the case of the GQ\,Lup system we measure a maximum variation of the magnitude difference of 0.115\,$\pm$\,0.032\,mag. This variation could be introduced by the primary star and would then be consistent with the brightness variations of GQ\,Lup\,A found by \cite{2007A&A...468.1039B}, which showed an amplitude of up to 0.44\,mag in K. A similar variation of 0.154\,$\pm$\,0.032\,mag is observed in our measurements of the DH\,Tau system. Due to the nature of our measurements it is not possible to decide if the variation is due to variations of the primary star or the companion. However, T\,Tauri stars are known to exhibit variations much stronger than this (see e.g. \citealt{2007A&A...461..183G}).

\begin{table*}
 \centering
  \caption{Astrometric and photometric measurements}
	\label{tab: measurements}
  \begin{threeparttable}
  \begin{tabular}{@{}llcclcc@{}}
  \hline
      
 Date									& Primary 				& Separation\,[arcsec]	 			& Position Angle\,$[^\circ]$	& Filter 								& $\Delta$\,mag						&  mag of comp.\,[mag] 										\\
\hline
2008-06-14 						& GQ\,Lup					&	0.7255\,$\pm$\,0.0050				& 276.66\,$\pm$\,0.50					&	\textit{K}$_\textrm{s}$									&	 $\geq$6.1 							& $\geq$13.2				\\
2009-06-29						& 								&	0.7264\,$\pm$\,0.0016				& 276.54\,$\pm$\,0.17					&	\textit{K}$_\textrm{s}$									&	 6.378\,$\pm$\,0.024 			& 13.474\,$\pm$\,0.031					\\
2010-05-05						& 								&	0.7256\,$\pm$\,0.0014				& 276.86\,$\pm$\,0.18					&	\textit{K}$_\textrm{s}$									&	 6.290\,$\pm$\,0.025 			& 13.386\,$\pm$\,0.032					\\
2011-06-05						& 								&	0.7240\,$\pm$\,0.0020				& 276.94\,$\pm$\,0.23					&	\textit{K}$_\textrm{s}$									&	 6.400\,$\pm$\,0.046 			& 13.496\,$\pm$\,0.050					\\
2012-03-03						& 								&	0.7240\,$\pm$\,0.0020				& 277.04\,$\pm$\,0.24					&	\textit{ND\_short} / \textit{K}$_\textrm{s}$			&	 6.405\,$\pm$\,0.020 			& 13.501\,$\pm$\,0.028					\\
\hline
2012-06-08 						& PZ\,Tel					&	0.4201\,$\pm$\,0.0013				& 59.55\,$\pm$\,0.19					&	\textit{K}$_\textrm{s}$									&	$\geq$4.6								& $\geq$10.99					\\
2012-06-08 						& PZ\,Tel					&	0.4188\,$\pm$\,0.0014				& 59.61\,$\pm$\,0.24					&	\textit{ND\_short} / \textit{K}$_\textrm{s}$			&	5.160\,$\pm$\,0.060			& 11.530\,$\pm$\,0.060				\\
\hline
1999-01-17						& DH\,Tau					&	2.3320\,$\pm$\,0.0099				& 138.68\,$\pm$\,0.19					&	\textit{814W}									&	-												& -					\\
2009-10-01						& 								&	2.3393\,$\pm$\,0.0041				& 138.63\,$\pm$\,0.14					&	\textit{K}$_s$									&	5.995\,$\pm$\,0.018			& 14.173\,$\pm$\,0.032					\\
2012-01-22						& 								&	2.3323\,$\pm$\,0.0061			 	&	138.76\,$\pm$\,0.16					&	averaged							&	-												& -					\\
2012-12-05						& 								&	2.3427\,$\pm$\,0.0057				& 138.61\,$\pm$\,0.15					& \textit{K}$_\textrm{s}$									&	6.149\,$\pm$\,0.026			& 14.327\,$\pm$\,0.037					\\
\hline	
2009-08-20						& HD\,203030			&	11.9764\,$\pm$\,0.0290			& 108.92\,$\pm$\,0.14					&	\textit{NB\,2.17}							&	9.33\,$\pm$\,0.13				& -					\\
\hline
2009-08-15						& 1RXS\,J1609			&	2.1989\,$\pm$\,0.0035				& 27.08\,$\pm$\,0.13					&	\textit{K}$_\textrm{s}$									&	7.391\,$\pm$\,0.012			& 16.307\,$\pm$\,0.024					\\
\hline
2009-08-16						& UScoCTIO\,108		&	4.5641\,$\pm$\,0.0112				& 176.79\,$\pm$\,0.19					&	\textit{K}$_\textrm{s}$									&	2.789\,$\pm$\,0.031			& 15.299\,$\pm$\,0.131					\\
\hline\end{tabular}

\end{threeparttable}
\end{table*}


\section{Archival observation epochs and literature data points}

\subsection{GQ\,Lup}
\label{sec: archive: gqlup}

In addition to our observations of the GQ\,Lup system, we utilized several literature data points. There were a total of 7 observation epochs available taken between 2004-06-24 and 2007-02-16 with VLT/NACO and published by \cite{2008AA...484..281N}. For all those data points the astrometric calibration was done by the authors with images of the binary star HIP\,73357 imaged in the same night as the science targets. The uncertainties include the uncertainties of the gaussian centering as well as the uncertainties of the astrometric solution. The derived astrometric solutions are given in \cite{2008AA...484..281N} and show no strong offsets to our own astrometric calibrations for our later observation epochs. \\
Furthermore we include the original Subaru/CIAO data point of 2002-06-17 and the HST/WFPC2 measurement of 1999-04-10 in our analysis that were published in the discovery paper of this system by \cite{2005AA...435L..13N}. For the Subaru data point the authors used the astrometric calibration provided in \cite{2003ApJ...590L..49F} while the HST data point was calibrated with the astrometric solution given in \cite{1995PASP..107..156H}. In both cases the uncertainties of the data points include the uncertainties of the astrometric solutions as well as the uncertainties of the gaussian centering. For the Subaru observation we only take the separation measurement into account since no position angle measurement was provided in the original study due to missing astrometric calibrators.\\
Finally we also included the tentative pre-discovery observations of GQ\,Lup\,b by \cite{2006AA...453..609J} in ComeOn+/Sharp2 data taken at the ESO LaSilla 3.6\,m telescope on 1994-04-02 in our study. The data was astrometrically calibrated by the authors of that study by observations of the binary system IDS\,17430S6022. The uncertainties of this data point are significantly larger than any of our other measurements due to the spurious nature of the detection. 

\subsection{PZ\,Tel}

In this study we utilized six additional literature data points for the PZ\,Tel system, all taken with VLT/NACO and published by us in \cite{2012MNRAS.424.1714M}. The measurements were taken between 2007-06-13 and 2011-06-06. All of the images were taken in \textit{K}$_\textrm{s}$-band, sometimes with an additional neutral density filter in place to prevent the bright primary star from saturating. In all but the 2007 epoch NACO's S13 objective was utilized. In the 2007 epoch the S27 objective was used. Astrometric calibration was performed as described in section \ref{sec: astrometry}. The only exception is again the 2007 epoch for which the astrometric solution was taken from \cite{2010A&A...509A..52C}. All utilized data points are listed in Tab.~\ref{tab: astrometric-datapoints-literature}.

\subsection{DH\,Tau}
\label{sec: dhtau: literature}

In addition to our own VLT/NACO observations we used available HST archival data of the DH\,Tau system. The observations were carried out on 1999-01-17 and 2012-01-22 using the Planetary Camera detector of WFPC2 (\citealt{1994ApJ...435L...3T}) and WFC3 (\citealt{2008SPIE.7010E..43K}) respectively. In the 2012 observation epoch, data in multiple bands was available, while in 1999 data was taken in the \textit{F814w} wide band filter at integration times of 20\,s and 200\,s. Details of these observations are listed in Tab.~\ref{pm: tab: observation-summary}. In all cases we used the standard reduced data as provided by the HST science archive for our analysis. Since no astrometric calibrators were directly available we rely in both cases on the astrometric solution given in the image headers. In the case of the 1999 measurements we determined the primary star's position in the short exposure image and the companion position in the long exposure image. This was necessary because the star is heavily saturated in the long exposure, but the companion is not visible in the short exposure image. We checked the pointing accuracy for both images with the data given in the image headers and found that the two pointing positions differ only by $\sim$0.5\,mas, i.e. differential pointing effects between the two exposures are negligible. We also checked for differential image orientation and found that the discrepancy is of the order of $10^{-6}$$^\circ$, i.e. also negligible. In the case of the 2012 observations, we measured the separation and relative position angle of the companion and the primary star in each filter. We then computed the average values for both quantities. The results for both epochs are listed in Tab.~\ref{tab: measurements}. The uncertainties that are listed are the uncertainties resulting from the image position measurements of primary and companion, i.e. no uncertainty of the astrometric solution is included since it is unknown. In the case of the 2012 epoch the average value of the measurements in different filters is given. The standard deviation between those measurements is slightly higher in separation (0.010\,arcsec) but lower in position angle (0.08$^\circ$).\\
The HST data of 1999-01-17 was already used by \cite{2005ApJ...620..984I} to perform an astrometric measurement. The original result is listed in Tab.~\ref{tab: astrometric-datapoints-literature}. They stated that they used the astrometric solution provided in \cite{1995PASP..107..156H} to calibrate the HST measurement.
\cite{2005ApJ...620..984I} also present two additional observations carried out with the Subaru telescope and the CIAO instrument. The observations were executed on 2002 November 23 and 2004 January 08. However, in 2003 the detector of the CIAO instrument was replaced with a new infrared array and hence the astrometric solution of the instrument changed slightly. Both CIAO data points should therefore be regarded as systematically uncorrelated measurements.\\
For the astrometric calibration of the 2002 data point, the authors used the astrometric solution presented in \cite{2002PASJ...54..963I}, wherein observations of the Trapezium cluster with the CIAO instrument are compared with reference observations of \cite{1999AJ....117.1375S}. They calculated a pixel scale of 21.250$\pm$0.025\,mas/pixel and provided a general uncertainty of the detector orientation of 0.073$^\circ$. It should be noted that these astrometric calibrations were done in January of 2001, whereas the science observation was executed in November 2002, almost two years later. There is no information about the astrometric stability of the instrument provided in \cite{2005ApJ...620..984I}.\\
For the 2004 data point, a changed pixel scale of 21.33$\pm$0.02\,mas/pixel after the instrument refurbishment is provided. There is, however, no information given as to how this astrometric solution was computed, especially if the Trapezium cluster was used again for calibration and if the astrometric calibration was done in the same night as the science observation. It is therefore possible that there are systematic offsets between these two data points.\\
For both CIAO data points, the total astrometric uncertainties include the uncertainty of the astrometric solution as well as the standard deviation of multiple measurements of the object positions. They do not include the measurement uncertainty of each single measurement. Hence it could be that the uncertainties of these astrometric measurements are underestimated.\\
All astrometric measurements are listed in Tab.~\ref{tab: astrometric-datapoints-literature}

\subsection{HD\,203030}

There are four additional literature data points available for the HD\,203030 system by \cite{2006ApJ...651.1166M}. They observed the system three times between 2002 and 2005 with the Hale 200\,inch (5\,m) telescope at the Palomar observatory, utilizing the PALAO AO system (\citealt{2000SPIE.4007...31T}) and the PHARO camera (\citealt{2001PASP..113..105H}). They took combinations of deep (long time exposed) coronagraphic images and short exposure non-coronagraphic images in J-, H- and \textit{K}$_\textrm{s}$-band to measure the positions of the bright primary and the faint companion. The pixel scale of the instrument is about 25.09\,mas/pixel.\\ 
They also observed the system on 2005-07-12 with the Keck 10\,m telescope, using the KeckII AO system (\citealt{2000PASP..112..315W}) and the NIRC2 instrument  (see e.g. \citealt{2003SPIE.4841....1M}). They employed the 20\,mas/pixel pixel scale of the instrument for coronagraphic observations in the \textit{K}$_\textrm{s}$-band.\\
All literature epochs were calibrated astrometrically by the authors with images of the visual binary WDS\,18055+0230 taken in the same nights as the science observations. For this binary, a high grade astrometric orbit solution is available in the Sixth\footnote{See http://ad.usno.navy.mil /wds/orb6.html} Catalog of Orbits of Visual Binary Stars. Furthermore, there were distortion solutions available for both instruments by \cite{2006PhDT.........1M}, which the authors used for geometric distortion correction. They considered the uncertainties of the astrometric calibration and of the individual measurements for calculation of the total astrometric uncertainty of each data point.\\
Of the mentioned data points, only the first one taken on 2002-08-28 was explicitly given in \cite{2006ApJ...651.1166M}, but the others could be extracted from a figure therein with a precision of $\sim$\,1.5\,mas in separation and $\sim$\,0.01$^\circ$ in PA. All data points are listed in Tab.~\ref{tab: astrometric-datapoints-literature}. Given the listed astrometric uncertainties, the additional uncertainties from the extraction of the data points is negligible.\\

\subsection{1RXS\,J160929.1-210524}

In addition to the VLT/NACO observation presented in section \ref{sec: obs}, there were astrometric data points of the 1RXS\,J160929.1-210524 system available in publications by \cite{2010ApJ...719..497L} and \cite{2011ApJ...726..113I}. \\
\cite{2010ApJ...719..497L} present astrometric measurements obtained with the Gemini-North Telescope and the NIRI instrument in combination with the ALTAIR AO system. The observations were executed between early 2008 and mid-2009. All data points are listed in Tab.~\ref{tab: astrometric-datapoints-literature}. They did not observe any astrometric calibrators and could hence only use the stored header information for pixel scale and detector orientation. They also note that the pixel scale of 21.4\,mas/pixel that they used is not well calibrated in their chosen observation mode with the ALTAIR field lens in place (see also the ALTAIR instrument webpage\footnote{http://www.gemini.edu/sciops/instruments/altair/field-lens-option}). However, in all of the Gemini-North observations, six background stars were present in the field of view. Separation and PA of these background stars with respect to 1RXS\,J160929.1-210524\,A were monitored throughout the different observation epochs and compared with the predicted values, given the proper motion of 1RXS\,J160929.1-210524\,A. While the pixel scale and orientation can not be absolutely calibrated this way, its is possible to monitor changes between the observing epochs. The maximum changes that were measured by \cite{2010ApJ...719..497L} correspond to 6\,mas in separation and 0.1$^\circ$ in PA, hence these values were adopted by the authors as the maximum uncertainties of their astrometric measurements.\\
\cite{2011ApJ...726..113I} observed the 1RXS\,J160929.1-210524 system with the Keck 10\,m telescope and the NIRC2 instrument. Their two data points taken mid-2008 and mid-2009 are listed in Tab.~\ref{tab: astrometric-datapoints-literature}. They do not report any astrometric calibrators imaged, but rather use the astrometric solution of the instrument as provided by \cite{2008ApJ...689.1044G}. They therein report a pixel scale of 9.963\,$\pm$\,0.005\,mas/pixel and a detector orientation of 0.13$^\circ$\,$\pm$\,0.02$^\circ$ as calculated by their high precision observations of the galactic center. These calibration observations were conducted between mid-2005 and late 2007. The astrometric solution was, in this timeframe, stable within the given uncertainties. The total uncertainties of the astrometric measurements by \cite{2011ApJ...726..113I} include the uncertainties of the astrometric solution and the standard deviation of multiple position measurements, both added in quadrature. However, they neglected the actual uncertainty of each individual position measurement, which was most likely significantly larger than the standard deviation of multiple measurements.\\ 

\subsection{UScoCTIO\,108}

There is only one astrometric data point of UScoCTIO\,108\,A and B available in the literature. \cite{2008ApJ...673L.185B} measured the separation and PA in their observations with the IAC (Instituto de Astrof\'{i}sica de Canarias) 80\,cm telescope in the \textit{I}-band on 2007 July 05. They used a pixel scale of 305\,mas/pixel, but did not provide any information on the astrometric calibration of their observations. Their result is listed in Tab.~\ref{tab: astrometric-datapoints-literature}.

\begin{table*}
 \centering
  \caption{Astrometric literature data points}
	\label{tab: astrometric-datapoints-literature}
  \begin{threeparttable}
  \begin{tabular}{@{}llccl@{}}
  \hline
      
 Date									& Primary 					& Separation\,[arcsec]	 			& Position Angle\,$[^\circ]$	& Reference\\
\hline
	1994-04-02										& GQ\,Lup					&	0.7138\,$\pm$\,0.0355				&	275.5\,$\pm$\,1.1						&	\cite{2006AA...453..609J}		\\
	1999-04-10										& 								&	0.739\,$\pm$\,0.011					&	275.62\,$\pm$\,0.86					&	\cite{2005AA...435L..13N}		\\
	2002-06-17										& 								&	0.7365\,$\pm$\,0.0057				&		-													&	\cite{2005AA...435L..13N}		\\
	2004-06-24										& 								&	0.7347\,$\pm$\,0.0031				&	275.48\,$\pm$\,0.25					&	\cite{2008AA...484..281N}		\\
	2005-05-25										& 								&	0.7351\,$\pm$\,0.0033				&	276.00\,$\pm$\,0.34					&	\cite{2008AA...484..281N}		\\
	2005-08-06										& 								&	0.7333\,$\pm$\,0.0039				&	275.87\,$\pm$\,0.37					&	\cite{2008AA...484..281N}		\\
	2006-02-20										& 								&	0.7298\,$\pm$\,0.0033				&	276.14\,$\pm$\,0.35					&	\cite{2008AA...484..281N}		\\
	2006-05-18										& 								&	0.7314\,$\pm$\,0.0035				&	276.06\,$\pm$\,0.38					&	\cite{2008AA...484..281N}		\\
	2006-07-15										& 								&	0.7332\,$\pm$\,0.0050				&	276.26\,$\pm$\,0.68					&	\cite{2008AA...484..281N}		\\
	2007-02-16										& 								&	0.7300\,$\pm$\,0.0064				&	276.04\,$\pm$\,0.63					&	\cite{2008AA...484..281N}		\\
\hline
	2007-06-13										& PZ\,Tel					&	0.2556\,$\pm$\,0.0025											&	61.68\,$\pm$\,0.60								&	\cite{2012MNRAS.424.1714M}		\\
	2009-09-27										& 								&	0.3366\,$\pm$\,0.0012											&	60.52\,$\pm$\,0.22								&	\cite{2012MNRAS.424.1714M}		\\
	2010-05-07										& 								&	0.3547\,$\pm$\,0.0012											&	60.34\,$\pm$\,0.21								&	\cite{2012MNRAS.424.1714M}		\\
	2010-10-27										& 								&	0.3693\,$\pm$\,0.0011											&	59.91\,$\pm$\,0.18								&	\cite{2012MNRAS.424.1714M}		\\
	2011-03-25										& 								&	0.3822\,$\pm$\,0.0010											&	59.84\,$\pm$\,0.19								&	\cite{2012MNRAS.424.1714M}		\\
	2011-06-06										& 								&	0.3883\,$\pm$\,0.0005											&	59.69\,$\pm$\,0.10								&	\cite{2012MNRAS.424.1714M}		\\
\hline
1999-01-17						& DH\,Tau					& 2.351\,$\pm$\,0.001			& 139.36\,$\pm$\,0.10		& \cite{2005ApJ...620..984I}\\
2002-11-23						& 								& 2.340\,$\pm$\,0.006			& 139.56\,$\pm$\,0.17		& \cite{2005ApJ...620..984I}\\ 
2004-01-08						& 								&	2.344\,$\pm$\,0.003			& 139.83\,$\pm$\,0.06		& \cite{2005ApJ...620..984I}\\
\hline
2002-08-28						& HD\,203030			& 11.923\,$\pm$\,0.021		& 108.76\,$\pm$\,0.12 	& \cite{2006ApJ...651.1166M}\\
2003-07-15						& 								&	11.918\,$\pm$\,0.056		& 108.67\,$\pm$\,0.19 	& \cite{2006ApJ...651.1166M}\\
2004-06-26						& 								&	11.880\,$\pm$\,0.056		& 108.59\,$\pm$\,0.21 	& \cite{2006ApJ...651.1166M}\\
2005-07-12						& 								&	11.926\,$\pm$\,0.056		& 108.82\,$\pm$\,0.34 	& \cite{2006ApJ...651.1166M}\\
\hline
2008-04-27						& 1RXS\,J1609			&	2.215\,$\pm$\,0.006			&	27.75\,$\pm$\,0.10		&	\cite{2010ApJ...719..497L}\\
2008-06-17						& 								&	2.221\,$\pm$\,0.006			& 27.76\,$\pm$\,0.10		& \cite{2010ApJ...719..497L}\\
2008-06-18						& 								&	2.2101\,$\pm$\,0.001		& 27.62\,$\pm$\,0.04		& \cite{2011ApJ...726..113I}\\
2009-04-05						& 								&	2.222\,$\pm$\,0.006			&	27.65\,$\pm$\,0.10		& \cite{2010ApJ...719..497L}\\
2009-05-30						& 								&	2.2113\,$\pm$\,0.0009		& 27.61\,$\pm$\,0.05		& \cite{2011ApJ...726..113I}\\
2009-06-29						& 								&	2.219\,$\pm$\,0.006			& 27.74\,$\pm$\,0.10		& \cite{2010ApJ...719..497L}\\
\hline
2007-07-05						& UScoCTIO\,108		&	4.6\,$\pm$\,0.1					& 177\,$\pm$\,1					& \cite{2008ApJ...673L.185B}\\

\hline\end{tabular}

\end{threeparttable}
\end{table*}


\section{Proper motion analysis}

\subsection{GQ\,Lup}

\begin{figure}

\includegraphics[scale=0.3]{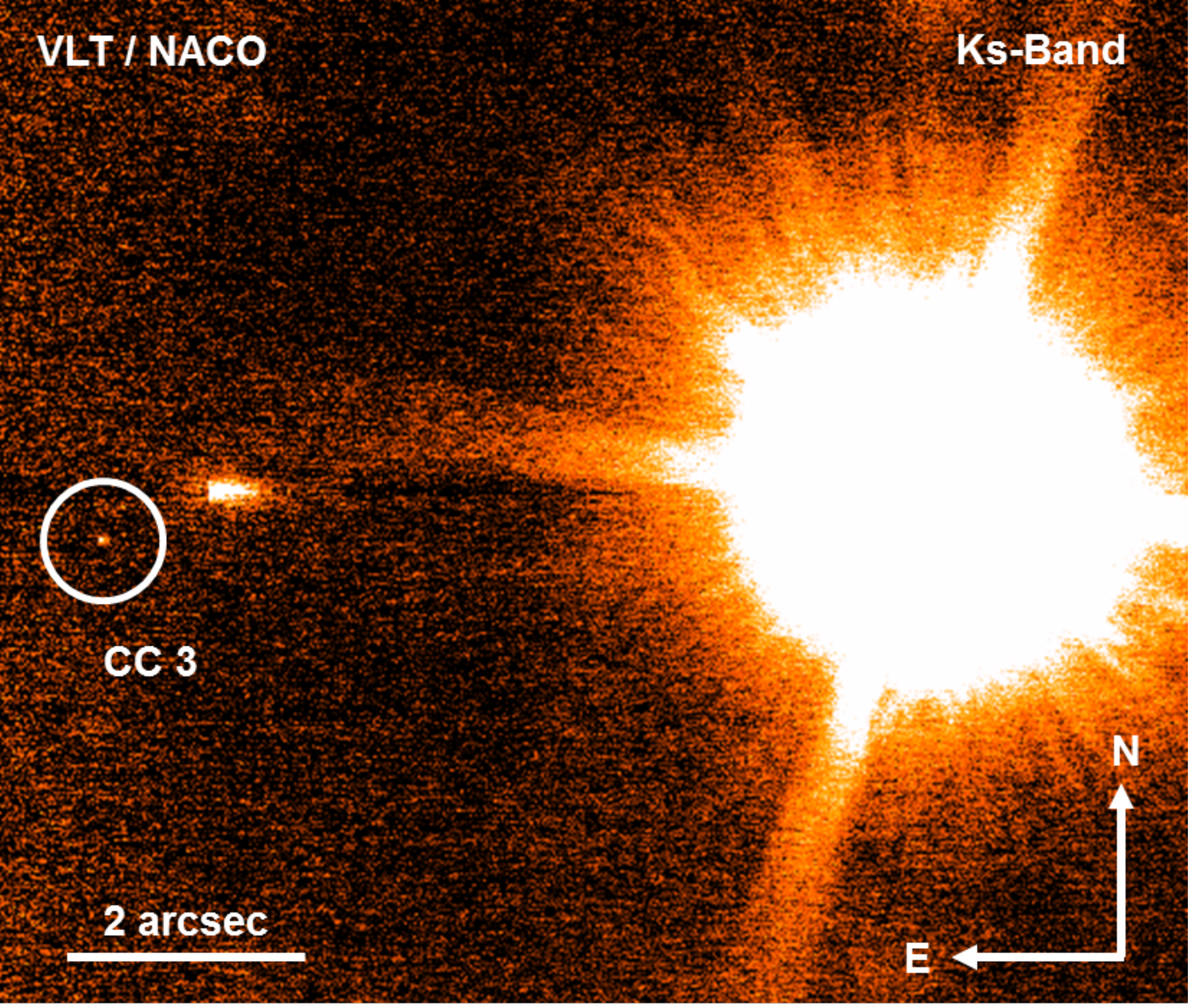}

\caption[]{New companion candidate 3 to GQ\,Lup\,A in the 2004 observation epoch with VLT/NACO.} 
\label{gqlup-cc3}
\end{figure}

The proper motion diagrams of the GQ\,Lup system including all data points from Tab.~\ref{tab: measurements} and Tab.~\ref{tab: astrometric-datapoints-literature} are shown in Fig.~\ref{gqlup-sep-pm} and Fig.~\ref{gqlup-pa-pm}. As was expected for the GQ\,Lup system, the background hypothesis, i.e. the hypothesis that the companion is a non-moving background object, can be rejected with a very high significance of 12.3\,$\sigma$ in separation and $>$\,13.1\,$\sigma$ in PA. At the same time a clear decline of separation and an increase of the PA are visible, especially if the measurements of \cite{2008AA...484..281N} and the new high precision measurements of this study are considered. As already mentioned in section \ref{sec: archive: gqlup} it is unlikely that the observed differential motion is caused by a systematic offset since it is detectable in both independent datasets, which were taken with the same instrument in similar configurations. \\
To determine the differential velocity in separation and PA we fitted a linear function to the astrometric measurements, which is indicated as (red) line in Fig.~\ref{gqlup-sep-pm} and Fig.~\ref{gqlup-pa-pm}. The fit yielded a decline in separation of 1.4\,$\pm$\,0.3\,mas/yr and an increase of PA of 0.16$^\circ$\,$\pm$\,0.03$^\circ$\,/yr. This is highly significant with 4.7\,$\sigma$ and 5.3\,$\sigma$, respectively. The early astrometric measurement by \cite{2006AA...453..609J} does not fit well with these values, however, this measurement also has very large uncertainties and is thus consistent with the solution within 1.3\,sigma.\\
The strong detection of differential motion encouraged us to fit relative Keplerian orbits to the astrometric data points. This assumes that GQ\,Lup\,A and B form indeed a gravitationally bound system, which is likely, but can only be finally confirmed once significant curvature in the orbit can be detected. The resulting orbit analysis is discussed in the following section.\\
In addition to the well known co-moving companion to GQ\,Lup\,A, we found a previously not recognized faint companion candidate in the 2004-06-24 and 2008-06-14 observation epochs. It is not detected in other observation epochs because they are either not deep enough or an unfavorable jitter pattern was employed that left the companion candidate outside the field of view. The object is located at a separation of 6.931\,$\pm$\,0.027\,arcsec and a PA of 93.52\,$\pm$\,0.27 in the 2004 observation epoch and is indicated in Fig.~\ref{gqlup-cc3}. We refer to it as "cc3", since it is the third companion candidate discovered in the close vicinity of GQ\,Lup,A (cc2 in \citealt{2008AA...484..281N}). The corresponding proper motion diagrams are shown in Fig.~\ref{gqlup-cc3-pm}. The astrometric measurements are slightly more consistent with a non-moving background object than with an additional co-moving companion. The background hypothesis cannot be rejected with any reasonable significance (0.6\,$\sigma$ and 0.02\,$\sigma$, for separation and PA respectively). However, an additional astrometric measurement should be performed to exclude without doubt that the object is co-moving with GQ\,Lup\,A. Since it seems more likely that the object is a background object, it was treated as such and thus ignored in the subsequent orbit analysis of GQ\,Lup\,B.   

\subsection{PZ\,Tel}

In the case of the PZ\,Tel system we added one new astrometric data point to our astrometric monitoring campaign of this object published in \cite{2012MNRAS.424.1714M}. The corresponding proper motion diagrams can be found in Fig.~\ref{pztel-sep-pm} and Fig.~\ref{pztel-pa-pm}. The new data point of 2012-06-08  follows very well (within 1\,$\sigma$) the expected increase of separation of 30.5\,$\pm$\,0.3\,mas/yr that was calculated in \cite{2012MNRAS.424.1714M} and is indicated by the (red) line in the diagram (calculations excluded the 2007 data point which clearly deviates from the linear fit). In position angle the new data point deviates by about 2\,$\sigma$ from the linear decrease of 0.5$^\circ$\,$\pm$\,0.2$^\circ$\,/yr fitted also by \cite{2012MNRAS.424.1714M}. Since the measured position angle in this epoch is larger than predicted by a simple linear decrease fit, this could be an indication for further deceleration, as was already confirmed with high significance for the timeframe between the first epoch in 2007 and the successive observation on 2009-09-27 in \cite{2012MNRAS.424.1714M}. Further deceleration would be expected if the companion moves towards its apastron. However, the deviation from a constant decrease of position angle is only $\sim$2\,$\sigma$ and is thus not yet significant. In section \ref{sec: orbit: pztel} we use the literature data points along with our new measurement to update the orbit analysis of the PZ\,Tel system.

\subsection{DH\,Tau}

In Fig.~\ref{dhtau-sep-pm} and Fig.~\ref{dhtau-pa-pm} proper motion diagrams containing all discussed astrometric data points of the DH\,Tau system are shown. If we consider only our own measurements, then the background hypothesis can be rejected with 4.0\,$\sigma$ in separation and 7.6\,$\sigma$ in position angle. However, if the Subaru data points by \cite{2005ApJ...620..984I} are considered and compared with our own VLT/NACO measurement, then we can only reject the background hypothesis in separation. The change of position angle between the NACO and Subaru measurements are actually much more consistent with a non-moving background object. This could be related to problems with the astrometric calibration of this data set, as was discussed in section \ref{sec: dhtau: literature}. It seems highly probable that the uncertainties of these measurements are underestimated. Most specifically the uncertainty given for the 2004 Subaru position angle measurement is only 0.06$^\circ$, while the average uncertainty of the detector orientation that was given by the same authors is 0.073$^\circ$.\\ 
Even more puzzling than the Subaru measurements is the 1999 HST measurement. Using the same data set, our own measurement deviates from the measurement by \cite{2005ApJ...620..984I} by 0.68$^\circ$ in position angle, i.e. more than 3\,$\sigma$. This could be partially due to the fact that different geometric distortion solutions were applied in both cases. However, the discrepancy seems too large to be explained by this. In any case, since we used the latest geometric distortion solution available we are quite confident in our own result. This confidence seems justified given that all our measurements show a consistent behavior of the system, i.e. all our measurements are consistent within 1\,$\sigma$ for position angle and within 2\,$\sigma$ for separation. We thus conclude that no orbital motion can be detected in the DH\,Tau system with a time baseline of $\sim$13\,yr. This is not entirely surprising given the low mass of primary and companion as well as the large projected separation of $\sim$330\,au. 

\subsection{HD\,203030}

Considering the first astrometric data point of epoch 2002.658 and the latest VLT/NACO measurement of epoch 2009.637, the background hypothesis can be rejected with 23.5\,$\sigma$ in separation and with 4.5\,$\sigma$ in PA. If all data points are taken into account, it is possible to fit a linear increase in separation of 7.9\,$\pm$\,2.9\,mas/yr and a linear increase in PA of 0.045$^\circ$\,$\pm$\,0.025$^\circ$\,/yr. However, the significance of the detected differential motion is low, with only 1.5\,$\sigma$ in separation and 1.2\,$\sigma$ in PA. The Keck data point of epoch 2005.533 and the VLT/NACO data point of epoch 2009.637 would also be consistent with no differential motion. There is also no intrinsic differential motion detected in the data points by \cite{2006ApJ...651.1166M}. This could be due to the larger uncertainties and the shorter time difference between these measurements, but could also indicate that the fitted differential motion is caused by a systematic offset between the literature dataset and the VLT/NACO measurement. The later alternative would be supported by the fact that the pixel scale of NACO with the S27 objective was not calibrated as explained in section \ref{sec: astrometry}, but was only taken from the image header. Furthermore, it was not corrected for geometric distortions of the NACO S27 setup, as there is no geometric distortion correction available. Given the large separation of $\sim$\,11.9\,arcsec between primary and companion, a change in the pixel scale of $\sim$\,1\,\% could already lead to a change of separation in the order of $\sim$\,0.1\,arcsec.\\
Given the above considerations, it remains doubtful if the detected differential motion between 2002 and 2009 is a real effect. An additional measurement with VLT/NACO could shed some light on this question. If the fitted differential motion is taken into account as well as the precision of the VLT/NACO measurement, a significant change in separation should be detectable after a time difference of $\sim$\,3.6\,yr. For PA this takes about $\sim$\,5.5\,yr. The system should hence be observed again in the near future.\\
If the detected differential motion is indeed a real effect, then it would be consistent with an inclined and eccentric orbit, since changes in separation and PA are observed which are both smaller than predicted for a circular orbit.

\subsection{1RXS\,J160929.1-210524}

The corresponding proper motion diagrams to the data points listed in Tab.~\ref{tab: measurements} and Tab.~\ref{tab: astrometric-datapoints-literature} are shown in Fig.~\ref{rxj1609-sep-pm} and Fig.~\ref{rxj1609-pa-pm}. In both separation and PA there is an offset between the datasets of \cite{2010ApJ...719..497L} and \cite{2011ApJ...726..113I}. Given the observations of 2008.460 and 2008.462, which were conducted on consecutive nights where one can assume that separation and PA of the companion with respect to the primary have not changed, the systematic offset in separation is in the order of $\sim$11\,mas and the offset in PA is in the order of $\sim$0.14$^\circ$. Furthermore, both datasets seem to show a systematic offset towards the 2009.623 VLT/NACO measurement presented in this work. The significant differences, especially in PA, when comparing this measurement to the 2009.497 measurement of \cite{2010ApJ...719..497L} and the 2009.415 measurement of \cite{2011ApJ...726..113I} (0.49$^\circ$ and 0.36$^\circ$ respectively), are most likely caused by systematic offsets rather than differential proper motion, given the very short time difference of only a few months.\\
Systematic offsets between the two literature datasets could be caused by the essentially uncalibrated astrometric solution used by \cite{2010ApJ...719..497L} as already discussed. In addition, the astrometric solution used by \cite{2011ApJ...726..113I} was computed more than one year before the science epochs. Although this solution was stable over a timeframe of several years beforehand, there is always the possibility of a glitch in the system causing the solution to change, even more so if there was some maintenance work done on the detector or the AO system between 2007 and mid-2008. Furthermore, given our own experiences the individual measurement errors in one image frame can be as much as ten times as large as the standard deviation of measurements in multiple image frames as considered by \cite{2011ApJ...726..113I} for the total astrometric uncertainty of their measurements. Hence it seems likely that these uncertainties are underestimated. Larger uncertainties of these data points would put them in better agreement with the data points by \cite{2010ApJ...719..497L} and the VLT/NACO measurement.\\
Given the discussed difficulties, it is not appropriate to compare the two literature datasets, or each of the literature datasets individually with the VLT/NACO measurement to derive a potential differential motion between primary and companion. It is, however, possible to compare the data points within each literature dataset independently since there should be negligible systematic effects between measurements done with the same instrument settings and astrometric calibrations. The result shows for neither of the two datasets a significant ($>$1\,$\sigma$) differential motion in the covered timeframe. The largest change can be observed between the two measurements of \cite{2011ApJ...726..113I} in separation, with an increase of 0.012\,arcsec and a corresponding significance of 0.92\,$\sigma$. This significance decreases drastically if the measurement uncertainties are underestimated as suspected. Hence it is very questionable whether or not this is a real effect.\\
Given the proper motion of the primary, which is generally in the direction of the companion, the separation between the two objects should decrease while the PA should stay approximately the same (save parallactic changes), if the companion would be a background object (as indicated by the grey areas in Fig.~\ref{rxj1609-sep-pm} and Fig.~\ref{rxj1609-pa-pm}). Considering the discussed systematic offsets between the different datasets, a prediction of the significance level on which the background hypothesis can be rejected would be very unreliable in PA, with only minimal differences between real companions and background objects. In separation, the effect of the primary's proper motion on a background object is, however, much stronger. If the VLT/NACO measurement and the first measurement of 2008.321 are taken into account, the background hypothesis can be rejected with 4.28\,$\sigma$. Even given the discussed systematic offsets, this should place the significance level with which the background hypothesis can be rejected well above 3\,$\sigma$. Hence, we can independently confirm that primary and companion share a common proper motion and are thus most likely orbiting each other, although no orbital motion can be detected yet.  

\subsection{UScoCTIO\,108}

In Fig.~\ref{usco108-sep-pm} and Fig.~\ref{usco108-pa-pm} both available data points are plotted in proper motion diagrams. The VLT/NACO measurement is approximately ten times more precise in separation and 4 times more precise in PA, and is hence the most precise astrometric measurement of these two objects to date. Due to the large uncertainty of the first astrometric measurement, and the proper motion of the primary (-7.4$\pm$4.6\,mas/yr and -20.4$\pm$4.6 in R.A. and Dec respectively, \citealt{2010AJ....139.2440R}), the background hypothesis can not yet be rejected with any reasonable significance. Given the precision level of our VLT/NACO measurement, a similar measurement in the immediate future would allow for rejection of the background hypothesis in separation with $\sim$3\,$\sigma$ if both objects are indeed co-moving. \\
It is not possible to detect any orbital motion of UScoCTIO\,108\,B around A, for the same reason that the background hypothesis can not yet be rejected. Furthermore, \cite{2008ApJ...673L.185B} calculated that the escape velocity of this alleged wide low-mass binary would only be 0.4\,km/s. Given the projected separation of $\sim$670\,AU, and assuming a face-on orbit, this means that the orbital motion in PA should be smaller than 10$^{-5}$\,$^\circ$/yr. Considering the uncertainties of the VLT/NACO measurement, it would take $\sim$\,47000\,yr to detect orbital motion on the 1\,$\sigma$ level. Similar considerations for an edge-on orbit yield a time baseline of $\sim$\,35000\,yr for an analog detection in separation. However, the two objects could in principle show a significant higher differential motion if they are not bound or in the process of ejection.\\

\begin{figure*}[!h]
\subfloat[GQ\,Lup separation]{
\includegraphics[scale=0.43]{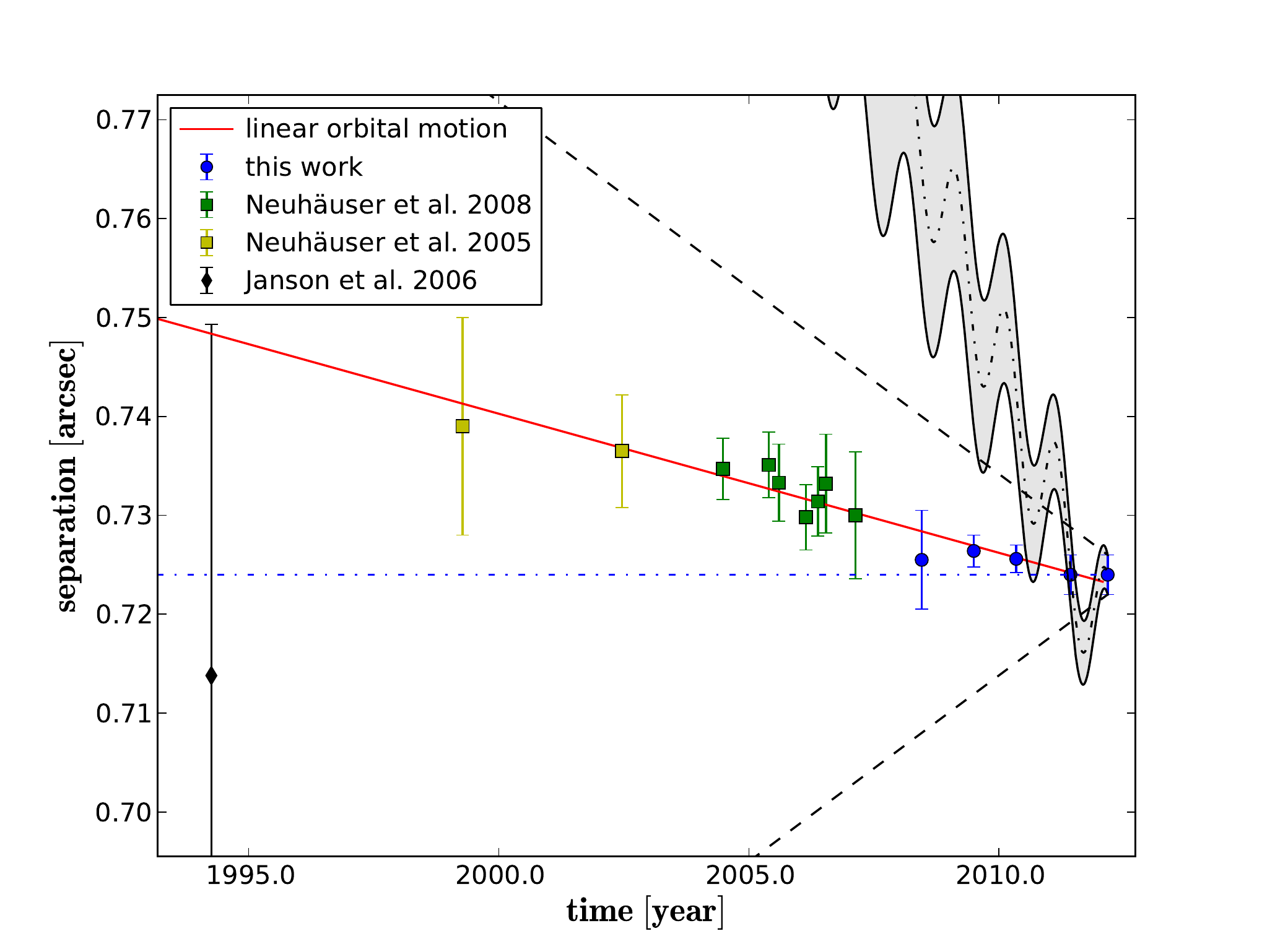}
\label{gqlup-sep-pm}
}
\subfloat[GQ\,Lup position angle]{
\includegraphics[scale=0.43]{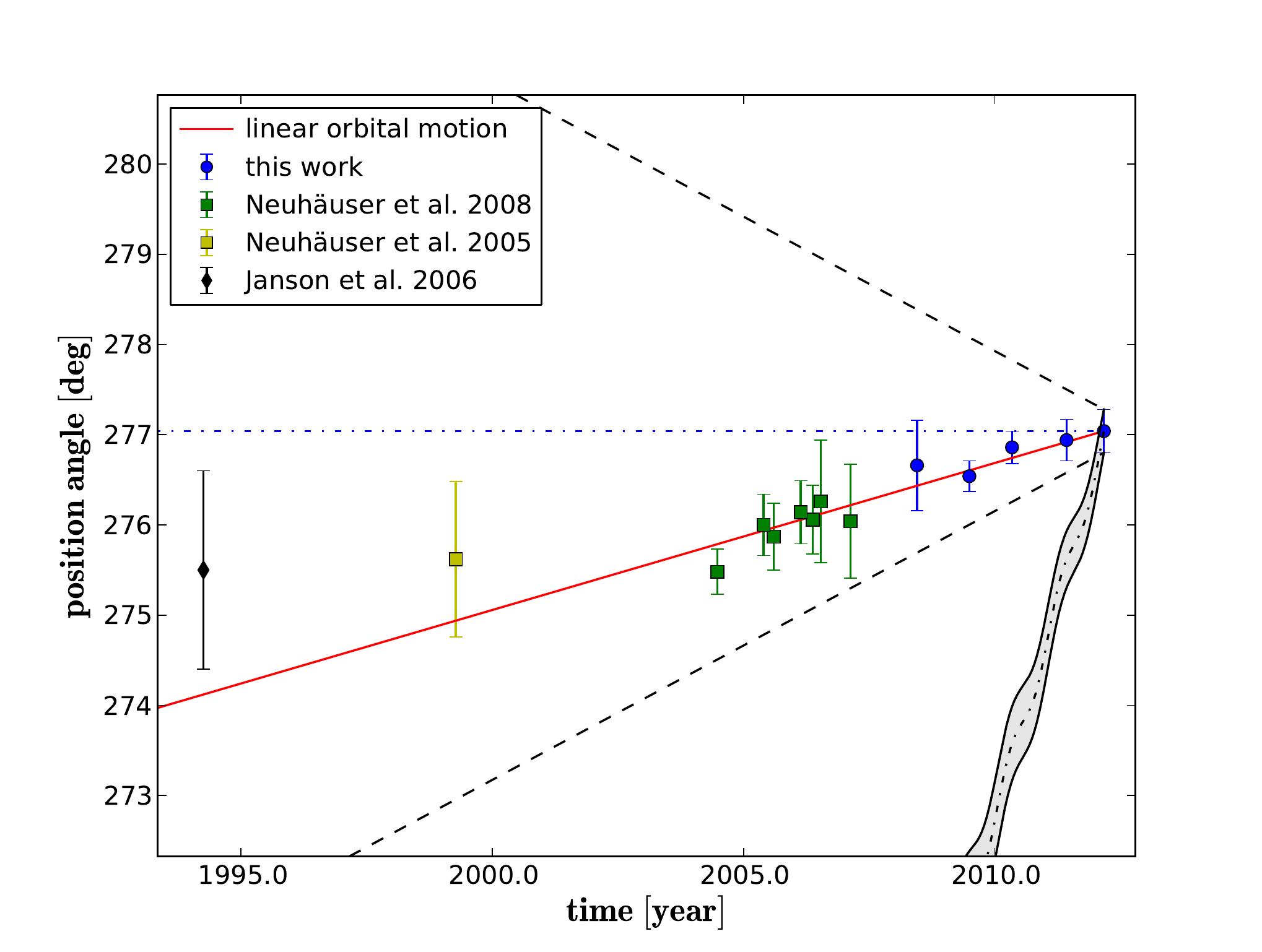}
\label{gqlup-pa-pm}
}

\subfloat[PZ\,Tel separation]{
\includegraphics[scale=0.43]{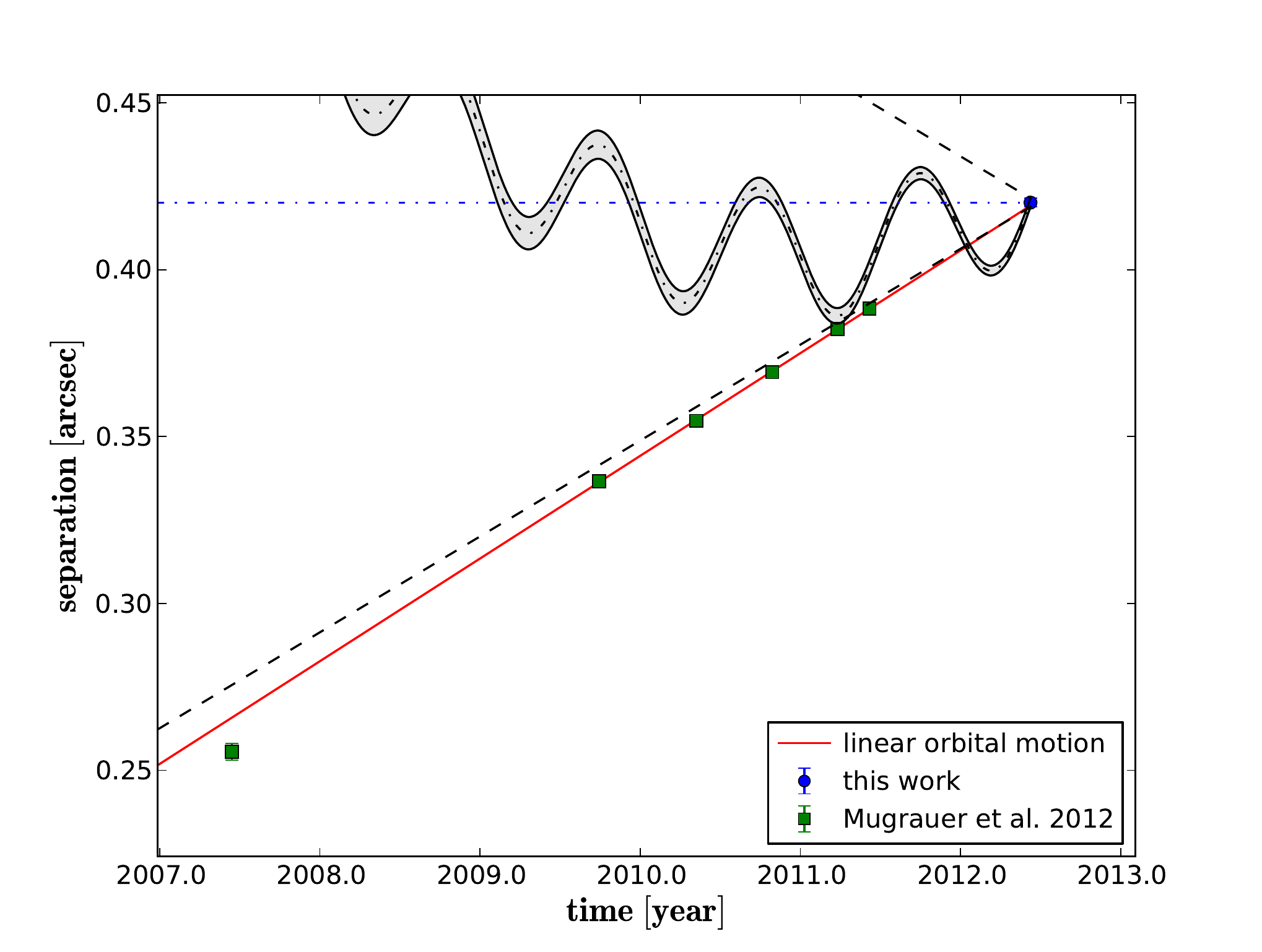}
\label{pztel-sep-pm}
}
\subfloat[PZ\,Tel position angle]{
\includegraphics[scale=0.43]{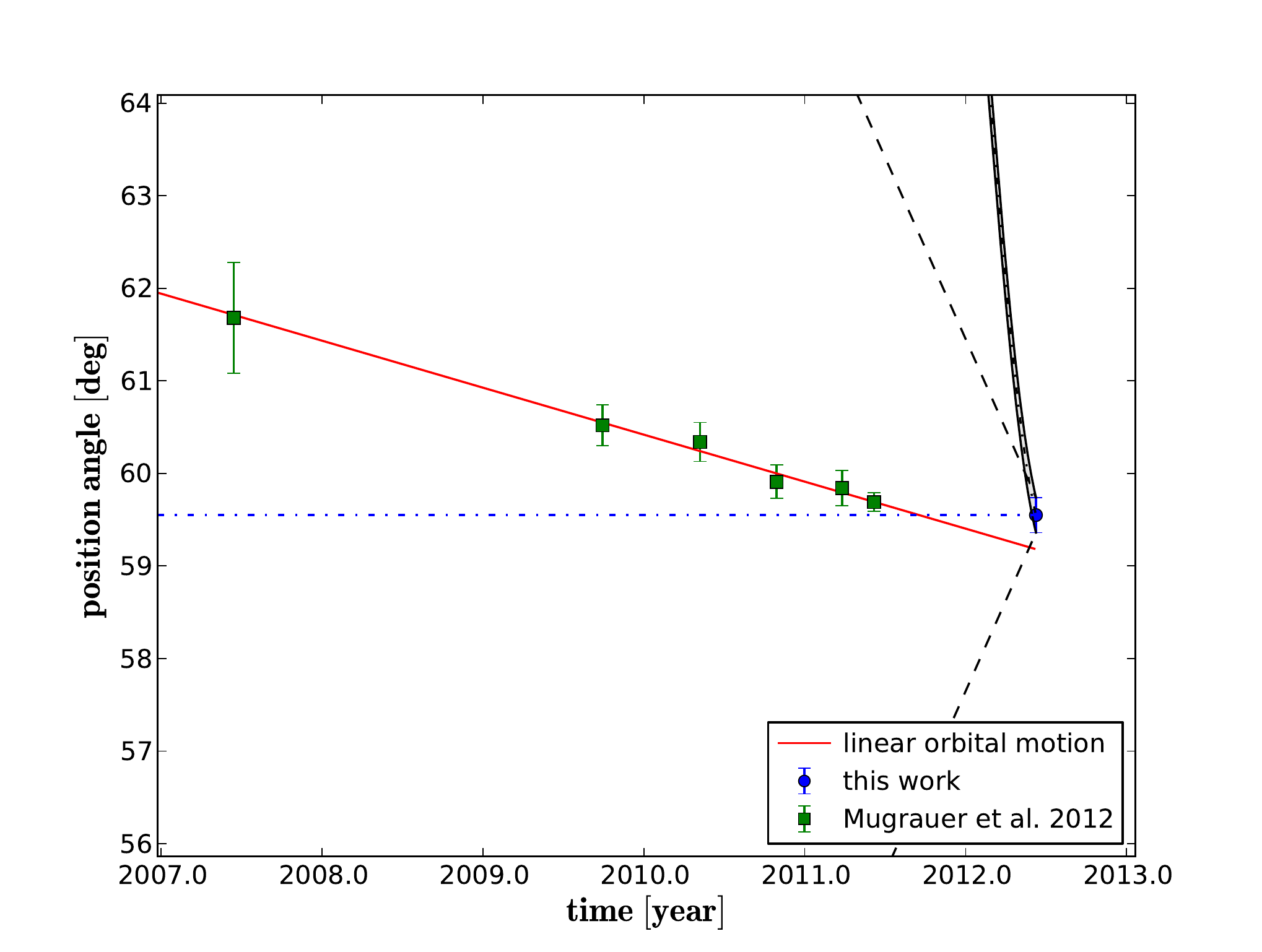}
\label{pztel-pa-pm}
}

\subfloat[DH\,Tau separation]{
\includegraphics[scale=0.43]{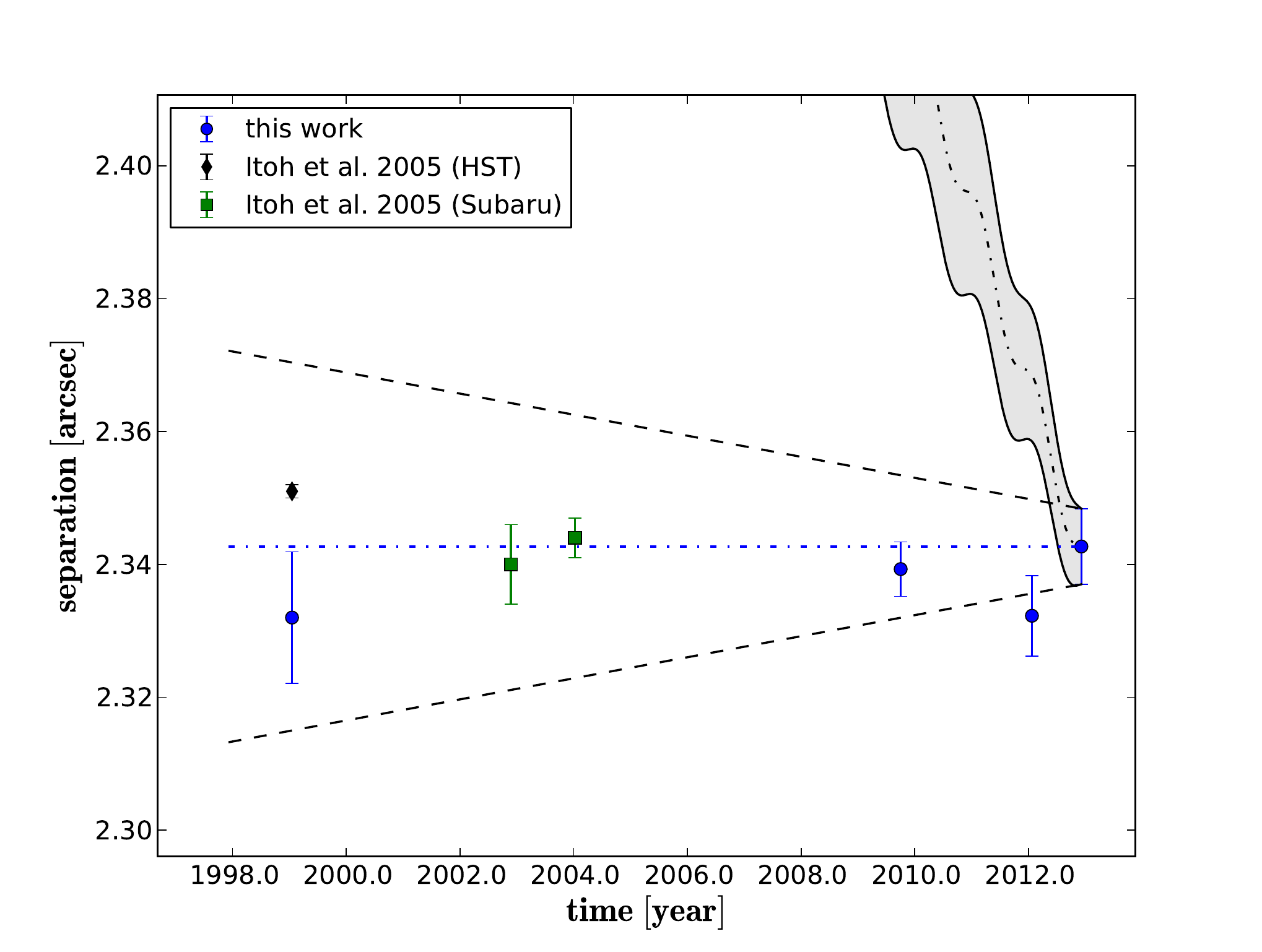}
\label{dhtau-sep-pm}
}
\subfloat[DH\,Tau position angle]{
\includegraphics[scale=0.43]{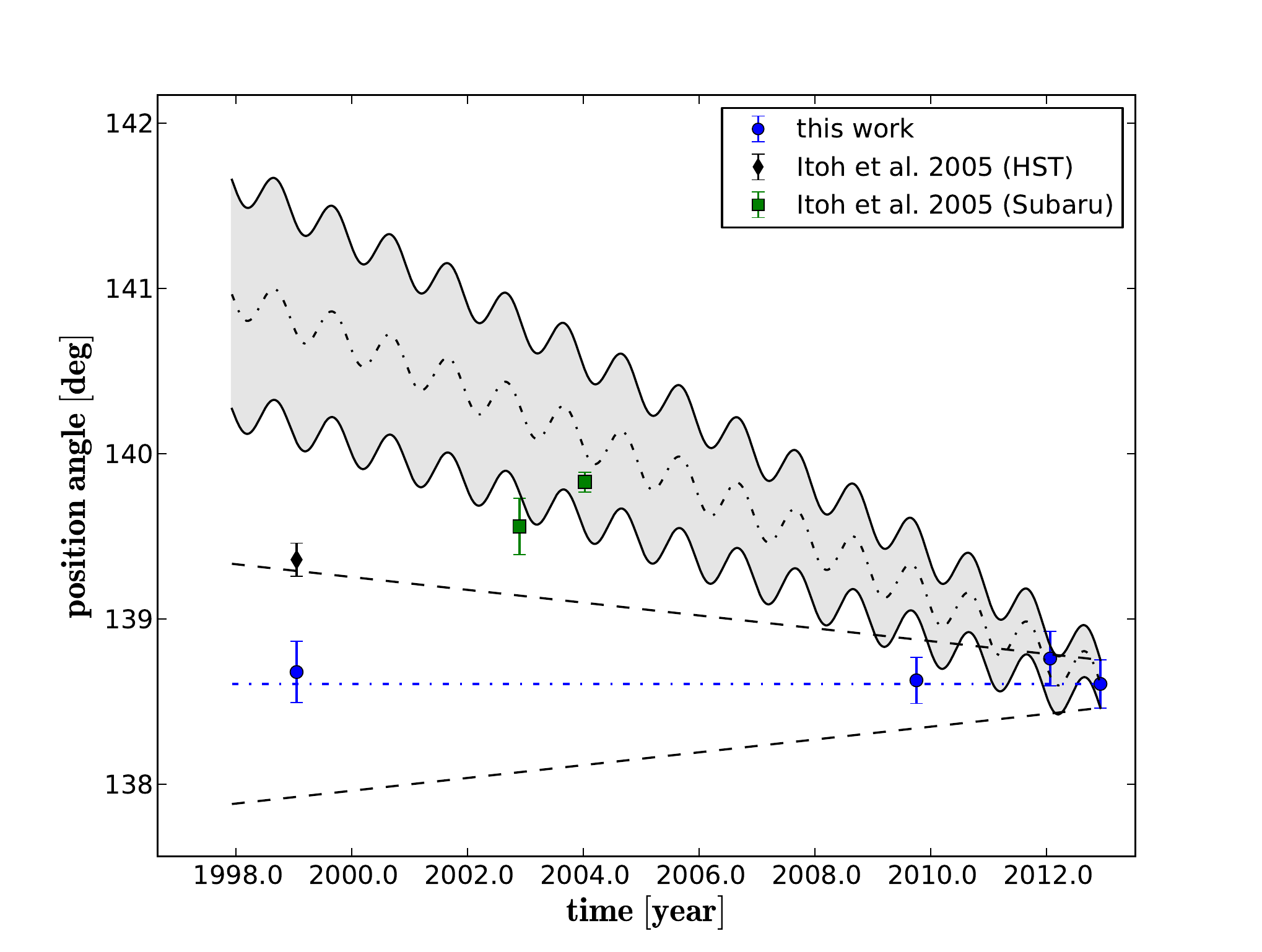}
\label{dhtau-pa-pm}
}

\phantomcaption

\end{figure*}

\begin{figure*}[!h]

\ContinuedFloat

\subfloat[HD\,203030 separation]{
\includegraphics[scale=0.43]{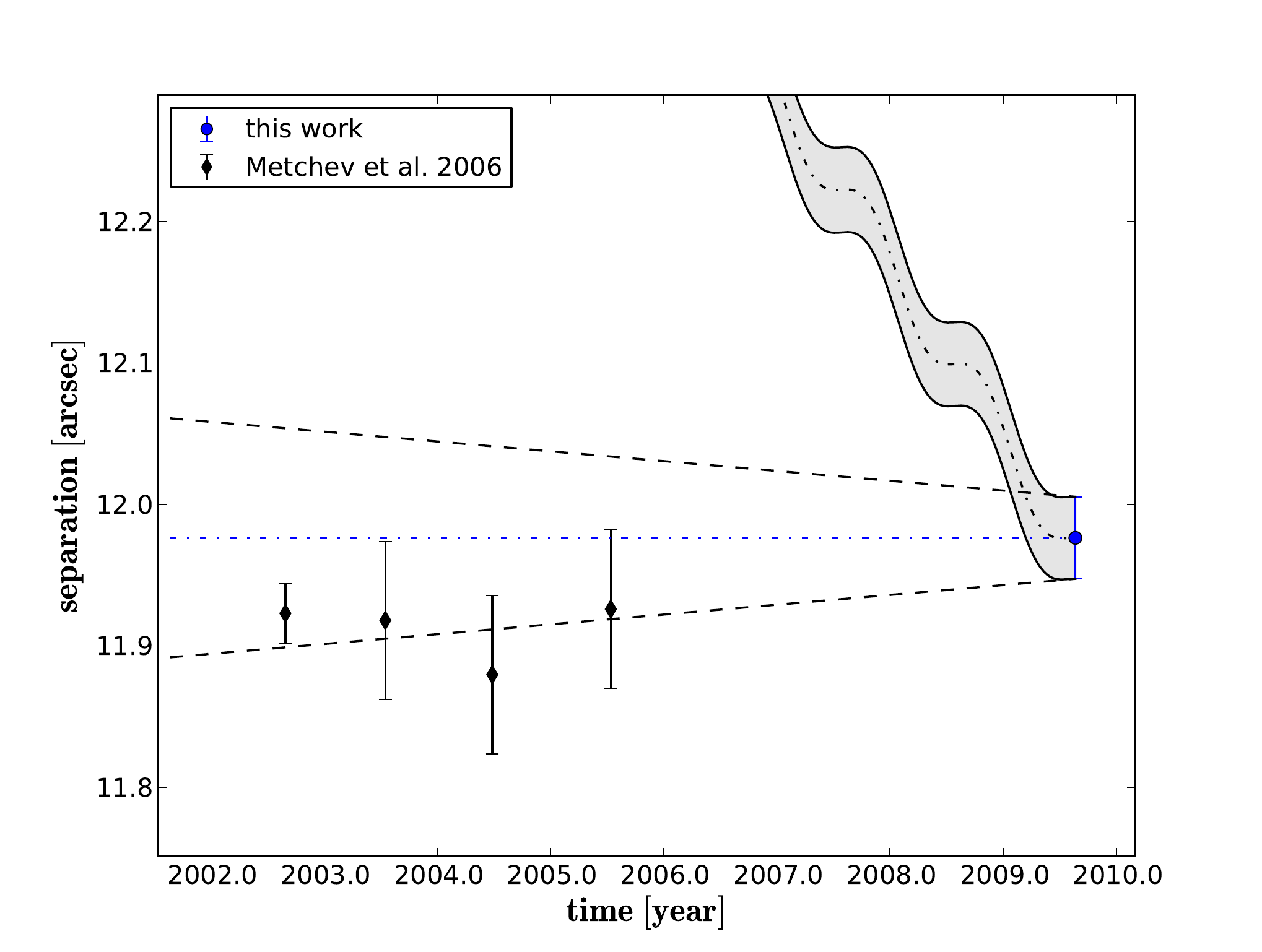}
\label{hd203030-sep-pm}
}
\subfloat[HD\,203030 position angle]{
\includegraphics[scale=0.43]{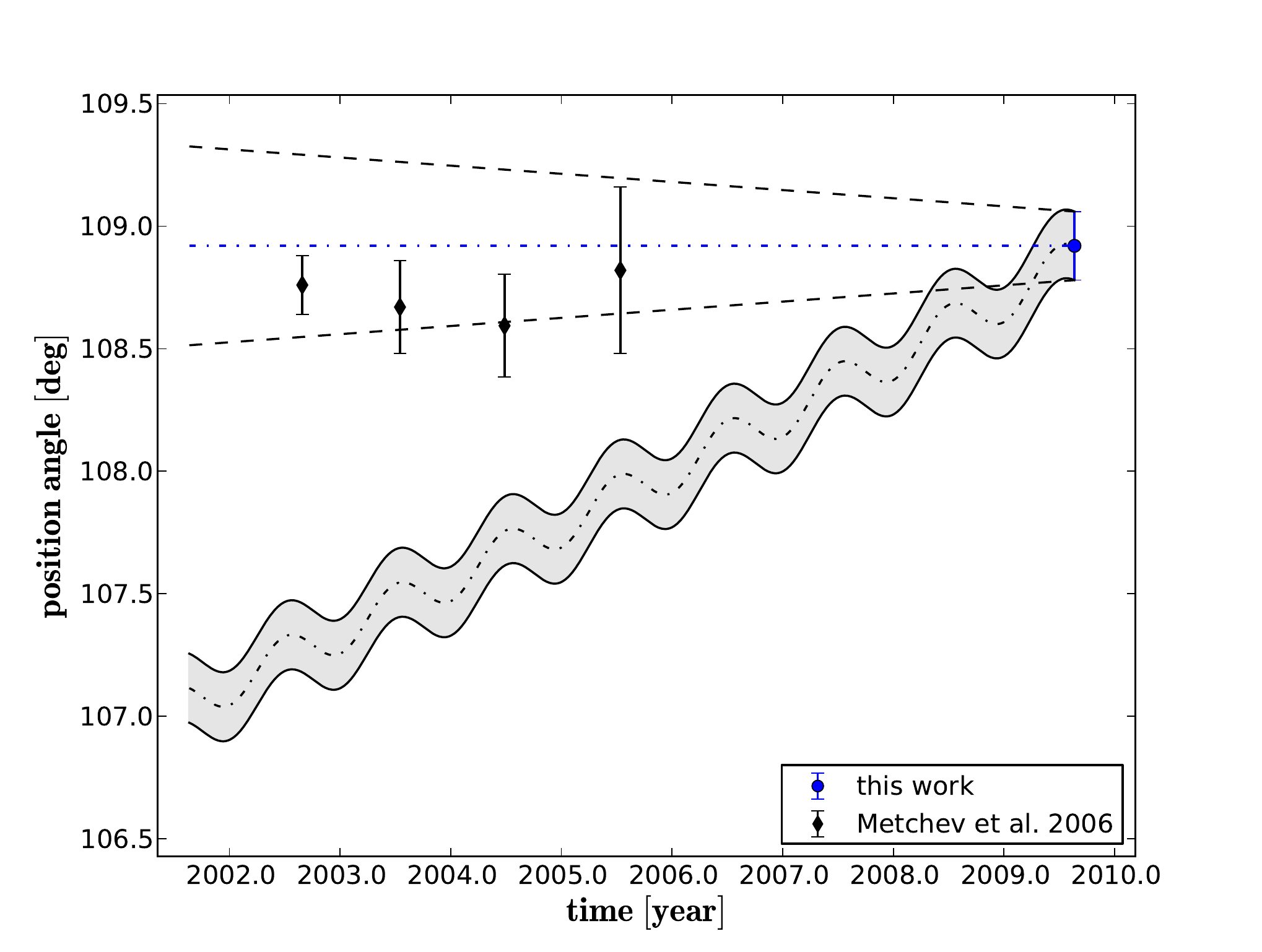}
\label{hd203030-pa-pm}
}

\subfloat[RXJ\,1609 separation]{
\includegraphics[scale=0.43]{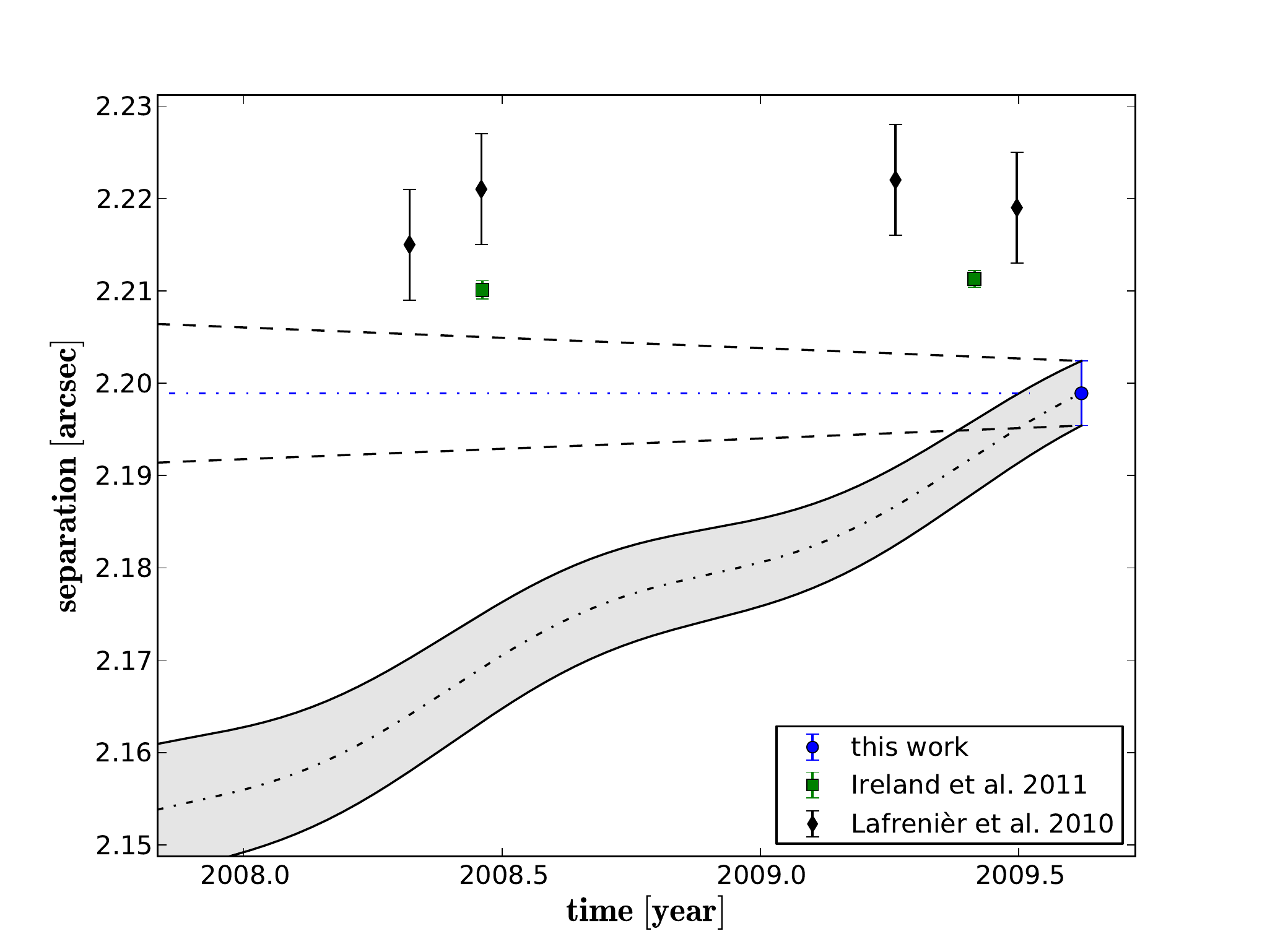}
\label{rxj1609-sep-pm}
}
\subfloat[RXJ\,1609 position angle]{
\includegraphics[scale=0.43]{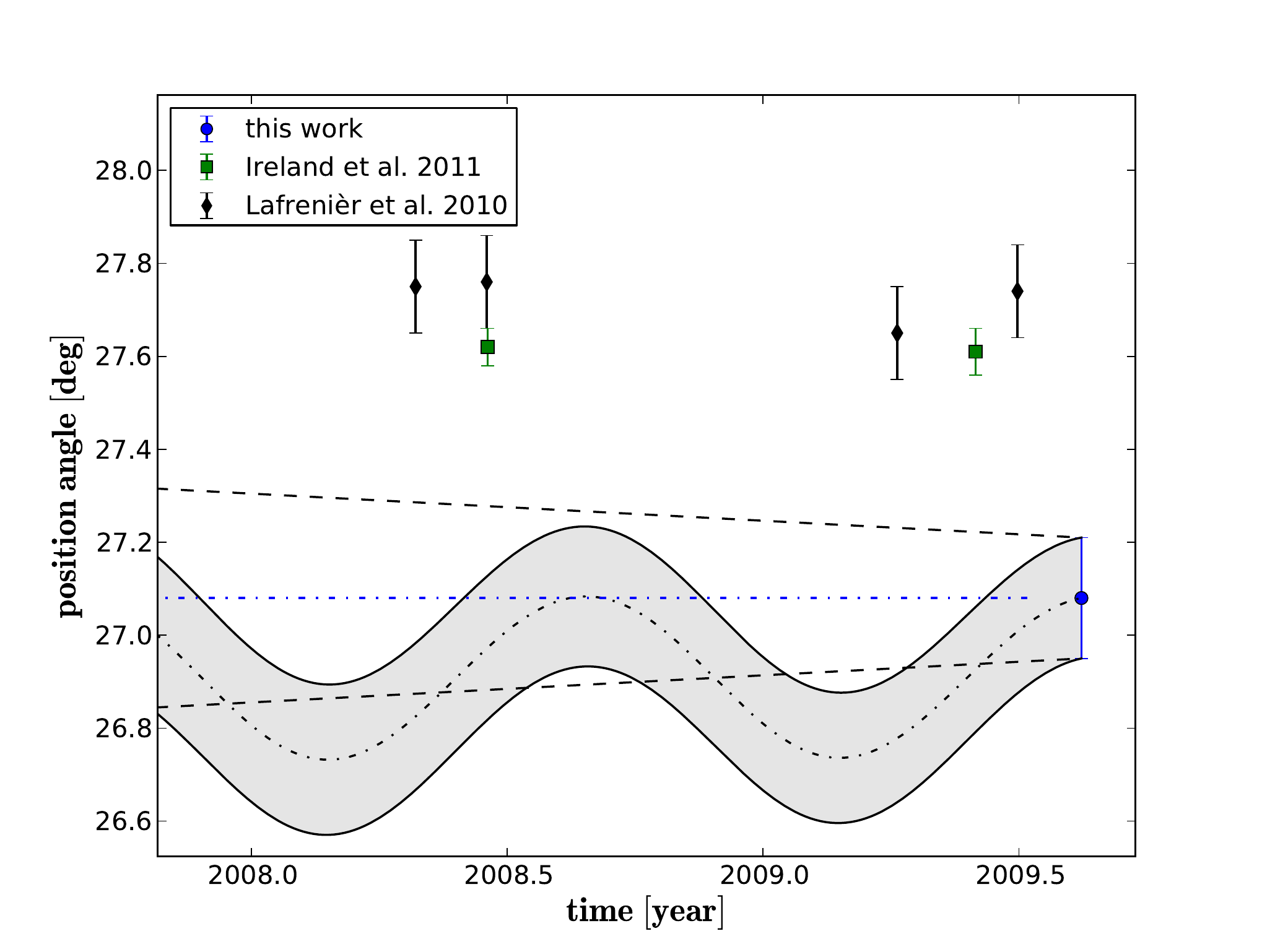}
\label{rxj1609-pa-pm}
}

\subfloat[UScoCTIO\,108 separation]{
\includegraphics[scale=0.43]{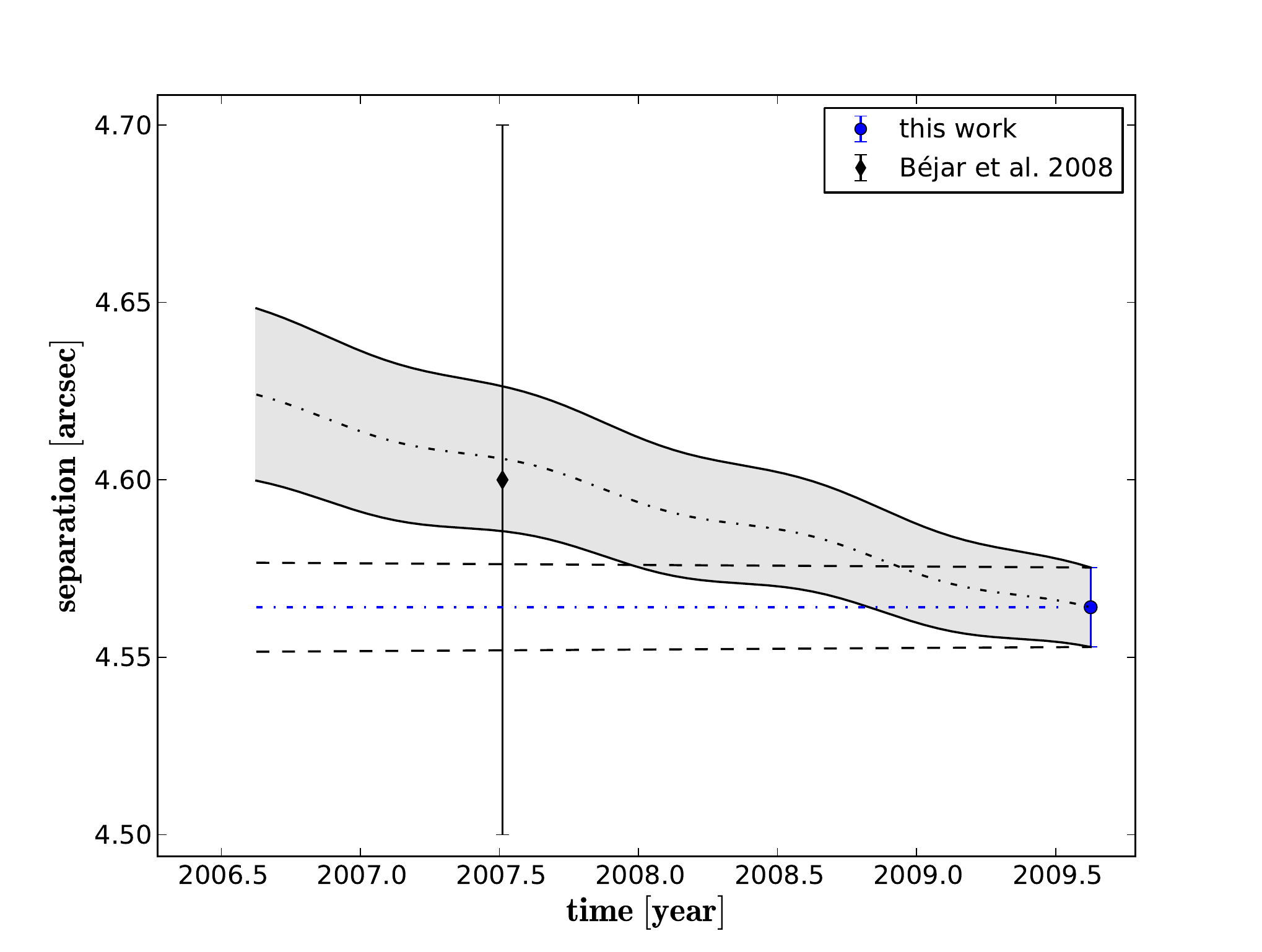}
\label{usco108-sep-pm}
}
\subfloat[UScoCTIO\,108 position angle]{
\includegraphics[scale=0.43]{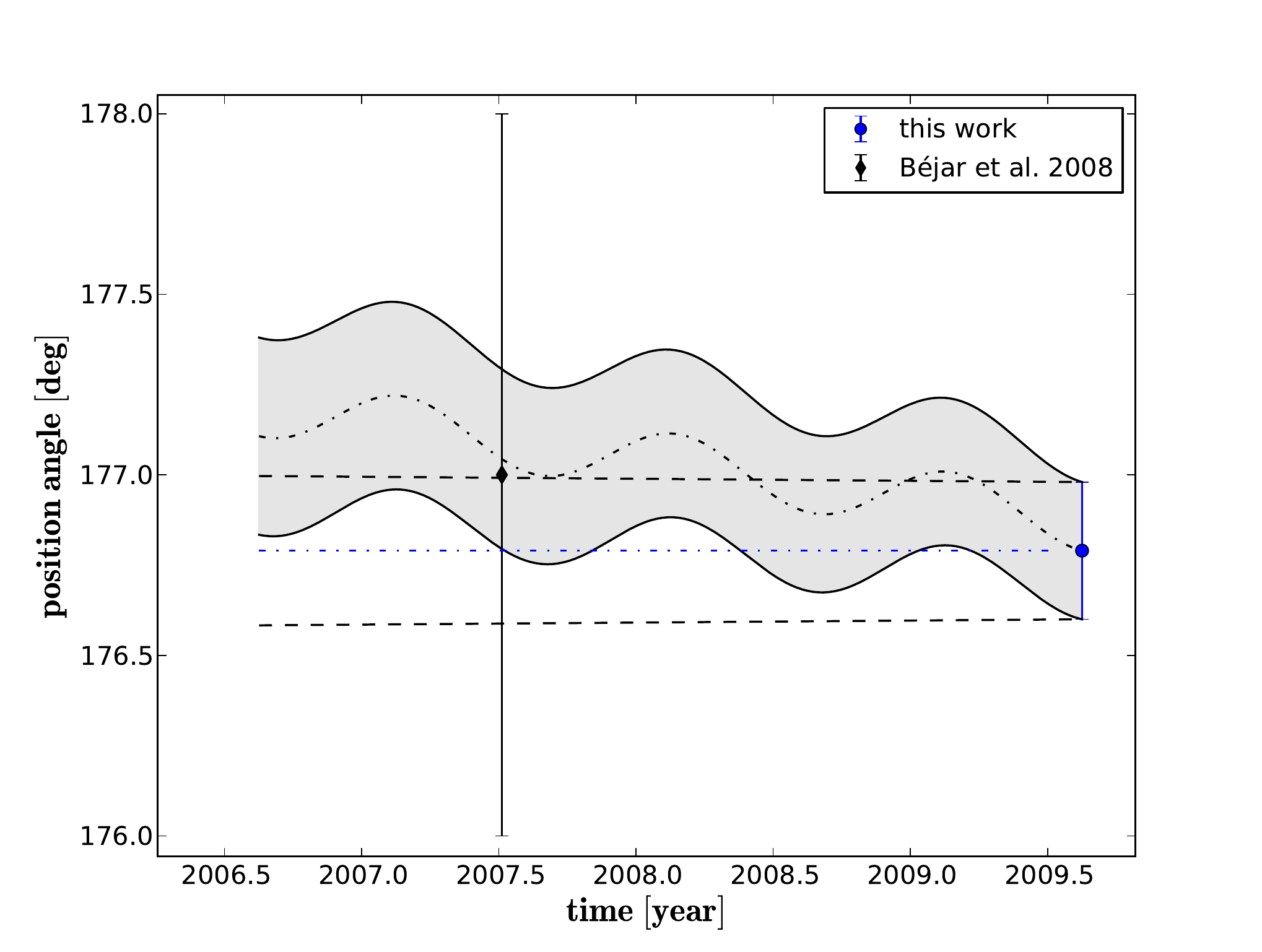}
\label{usco108-pa-pm}
}

\caption[]{Separation (a, left) and PA (b, right) plotted over time. The grey area enclosed by the wobbled lines represents the background hypothesis, i.e. the position that a non-moving background object would have at the given time (given the proper motion of the primary star). The wobble is introduced by the parallactic motion of the star due to earth's revolution around the sun. The dashed lines represent the area for maximum orbital motion in case of a circular orbit (assuming total system masses as discussed in section~\ref{target-section}). For details concerning this method see \cite{2012A&A...546A..63V}.} 
\label{pm-diagram}
\end{figure*}


\begin{figure}

\includegraphics[scale=0.6]{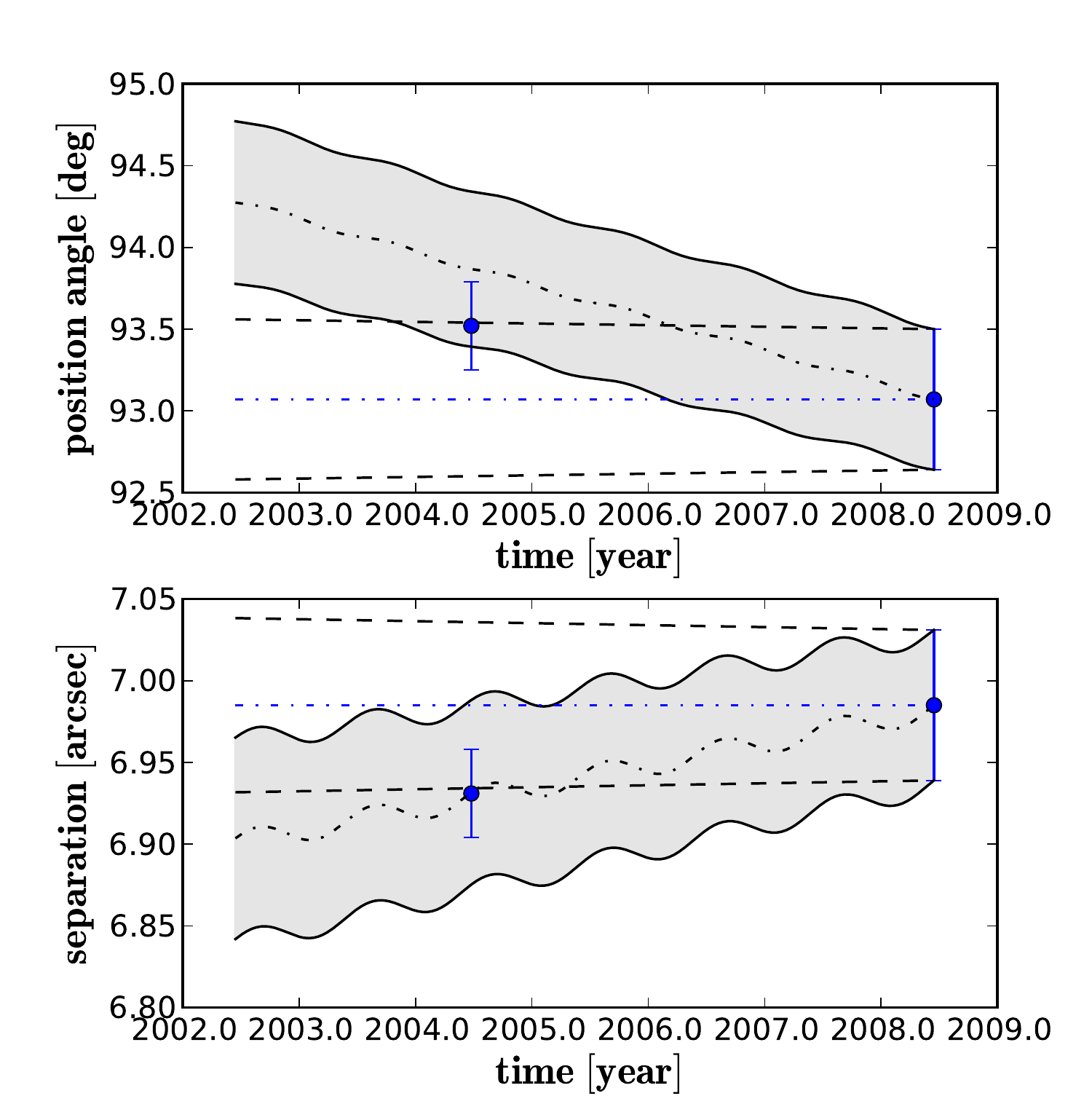}

\caption[]{Separation and PA of the third companion candidate to GQ\,Lup plotted over time. The astrometric results are more compatible with a non-moving background source than with a co-moving object. However, given the precision of our measurements it can not be completely ruled out that the companion candidate is co-moving with GQ\,Lup\,A. Lines as in Fig.~\ref{pm-diagram}} 
\label{gqlup-cc3-pm}
\end{figure}


\section{Orbit constraints}

\subsection{Least-Squares Monte-Carlo approach}
\label{sec: LSMC}

In order to constrain the possible orbits of systems that show significant differential motion we used a Least-Squares Monte-Carlo (LSMC) approach. For each system we randomly generated 5$\times$10$^6$ sets of orbit elements from a uniform distribution. We then used these orbit elements as starting points for least-squares optimizations utilizing the Levenberg-Marquardt algorithm (\citealt{levenberg}). To restrict the possible parameter space, we generally assumed that the systems are long-term stable, i.e. that they are stable against disruptions in the galactic disk. For this purpose we utilized the criterion for the maximum semi-major axis a$_{max}$[au]= 1000 M$_{tot}$/M$_\odot$ of \cite{2003ApJ...587..407C}, where M$_{tot}$ is the combined system mass in solar masses. In addition, we fixed the system mass for both of the discussed systems thereby reducing the number of free parameters by one to a total of six. A more detailed discussion of this approach is given in \cite{ginski-hd130}.

\subsection{GQ\,Lup system}
\label{sec: orbit: gqlup}

\begin{figure*}
\subfloat[][]{
\includegraphics[scale=0.45]{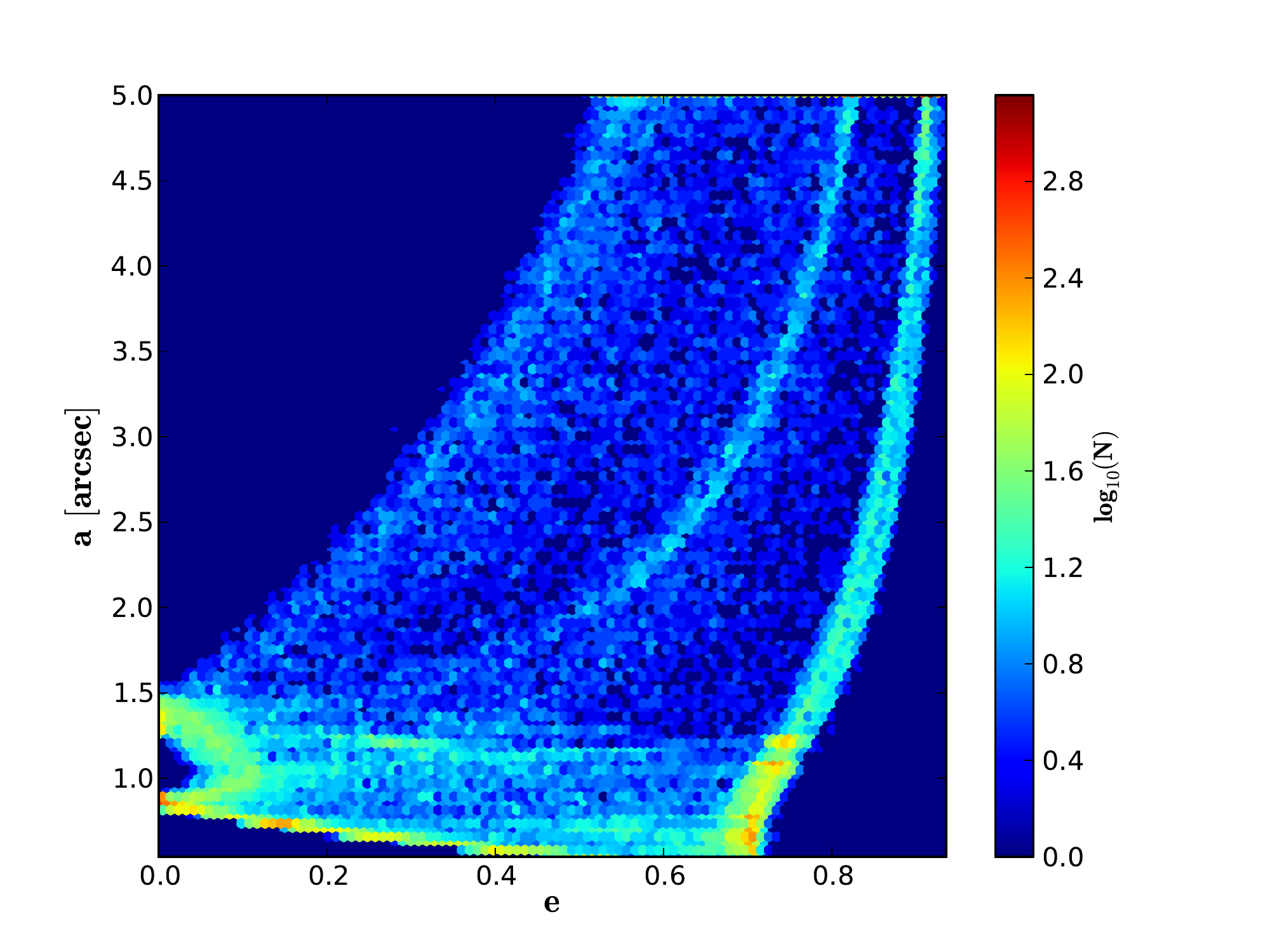}
\label{a-vs-e-gqlup}
}
\subfloat[][]{
\includegraphics[scale=0.45]{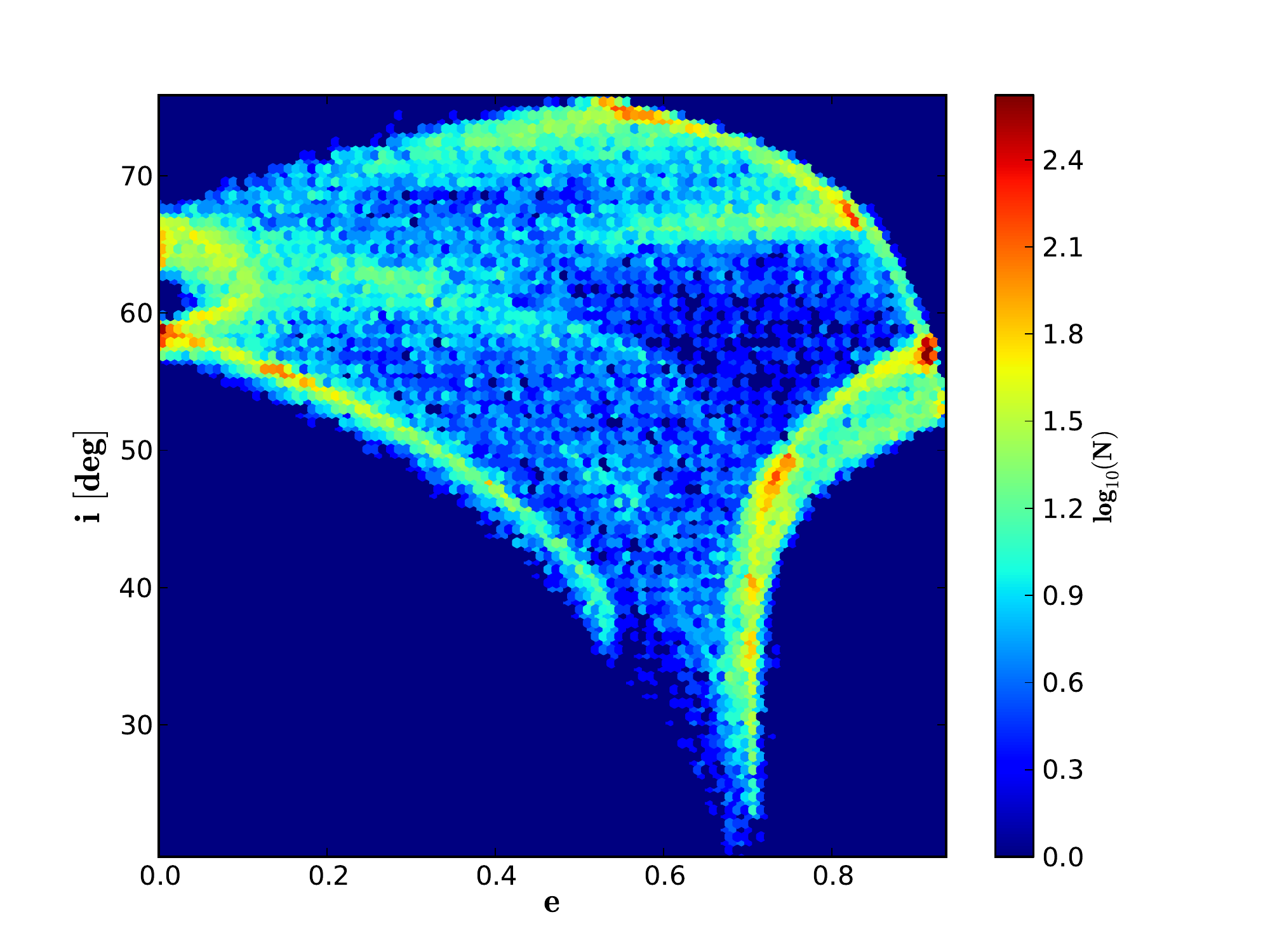}
\label{i-vs-e-gqlup}
}\\
\subfloat[][]{
\includegraphics[scale=0.45]{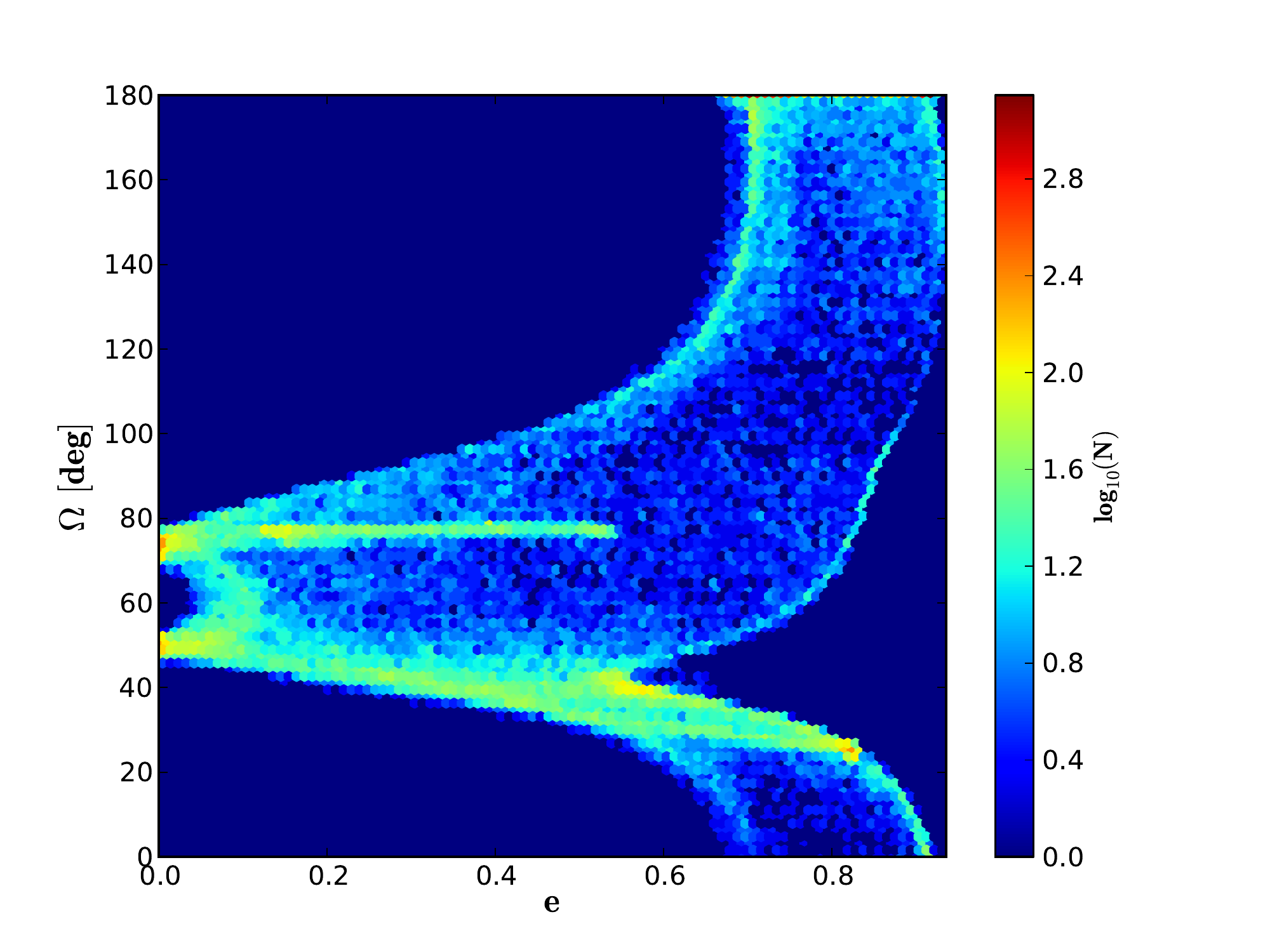}
\label{node-vs-e-gqlup}
}
\subfloat[][]{
\includegraphics[scale=0.45]{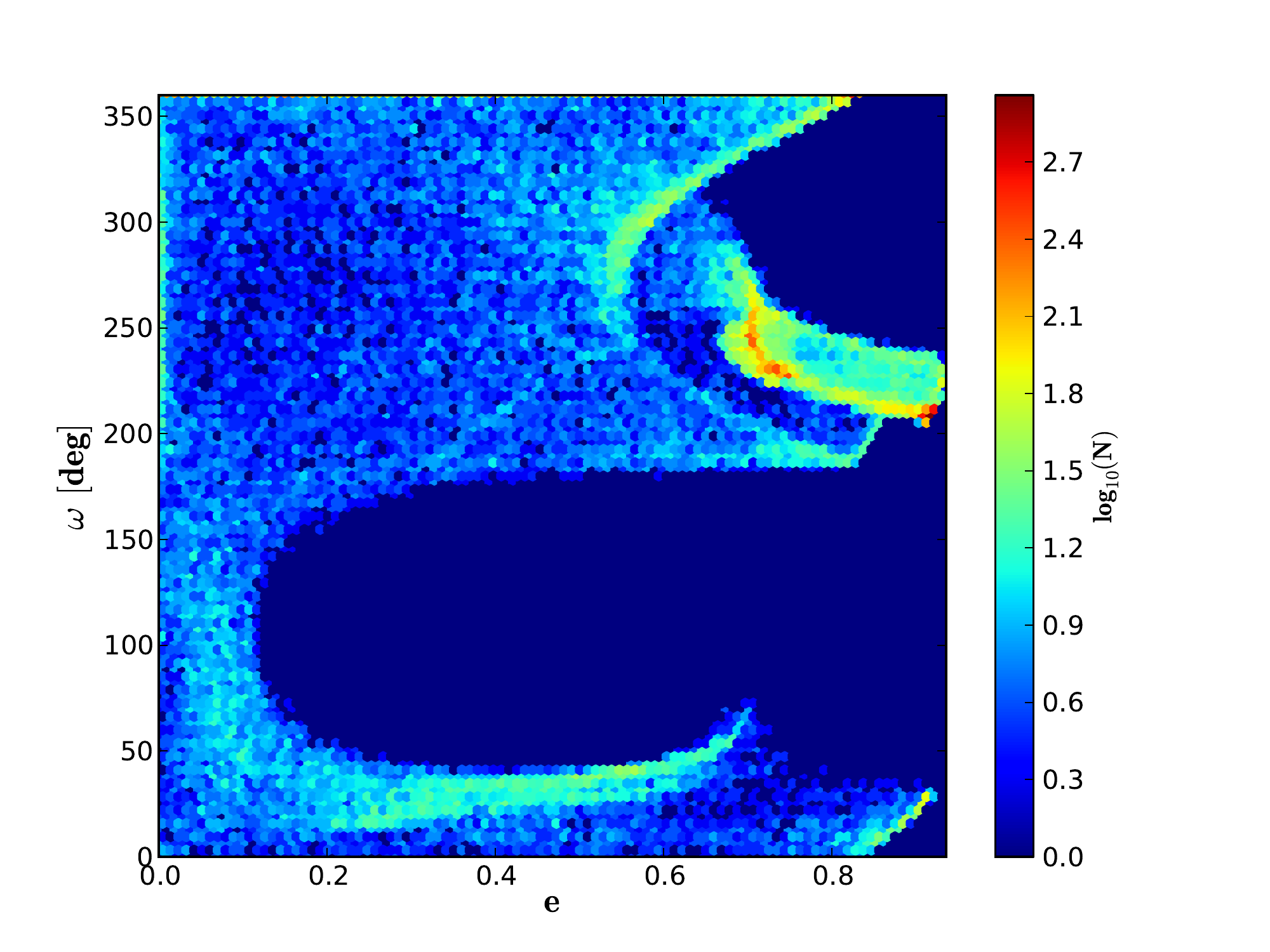}
\label{peri-vs-e-gqlup}
}\\
\subfloat[][]{
\includegraphics[scale=0.45]{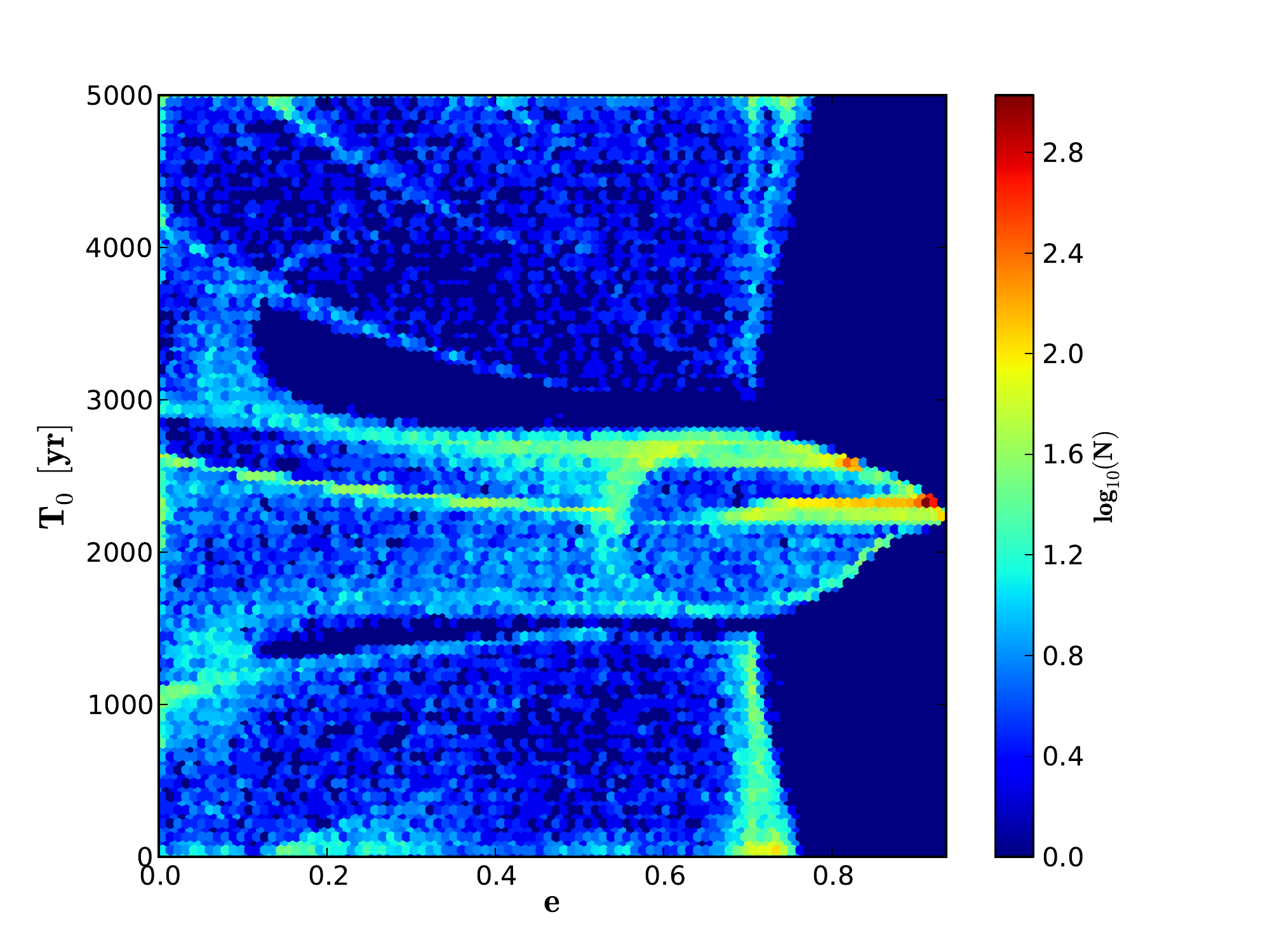}
\label{T0-vs-e-gqlup}
}

\caption[]{Orbital elements as function of eccentricity for the substellar companion of GQ\,Lup around its host star for the 1\,\% best fitting solutions out of 5,000,000 runs of our LSMC fit. Logarithmic density of solutions is indicated by color (a color version of this figure is available in the online version of the journal). } 
\label{fig:orbit-corr-gqlup}
\end{figure*}

\begin{figure*}
\subfloat[][]{
\includegraphics[scale=0.45]{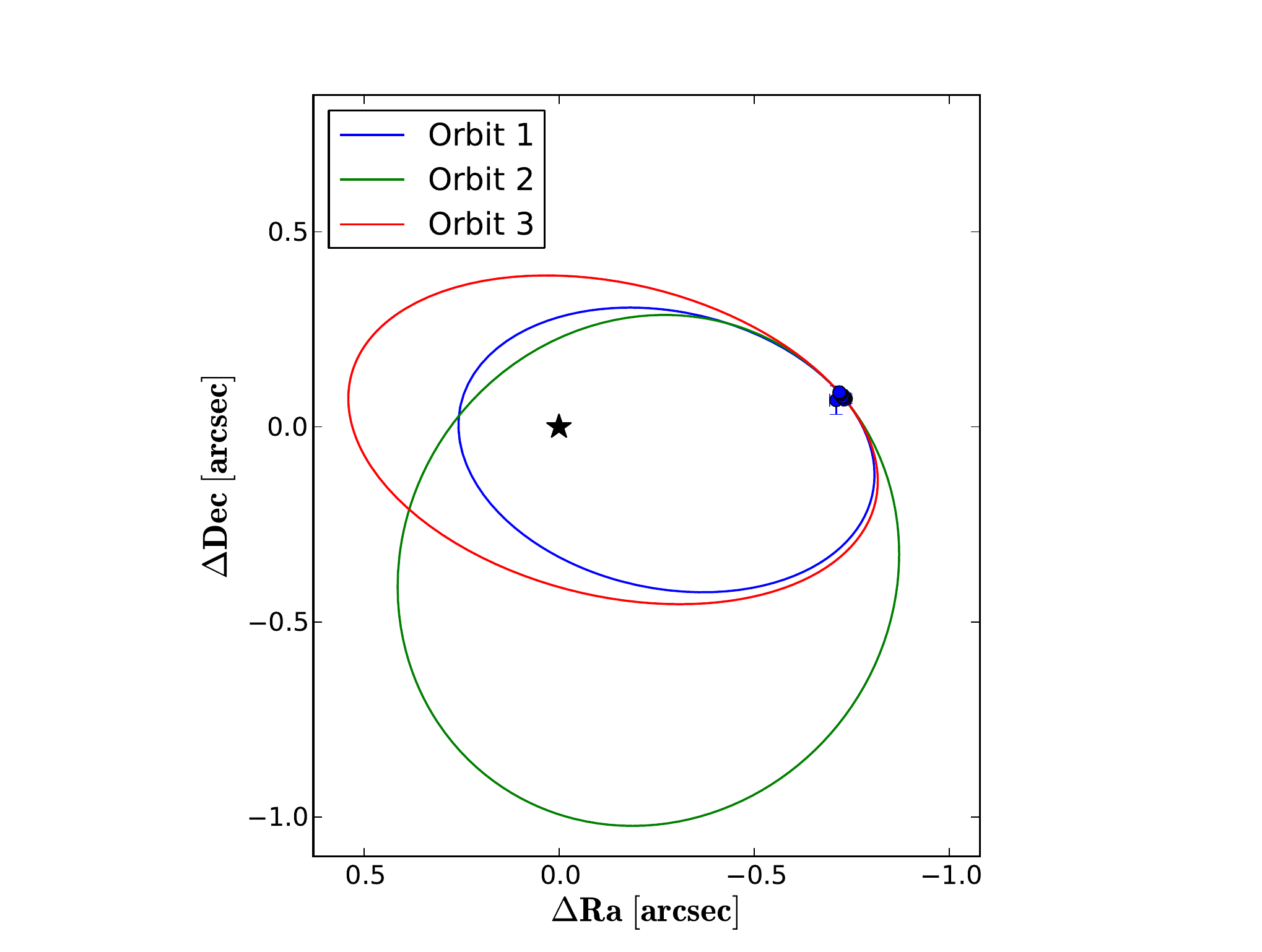}
\label{best-orbit-gqlup}
}
\subfloat[][]{
\includegraphics[scale=0.45]{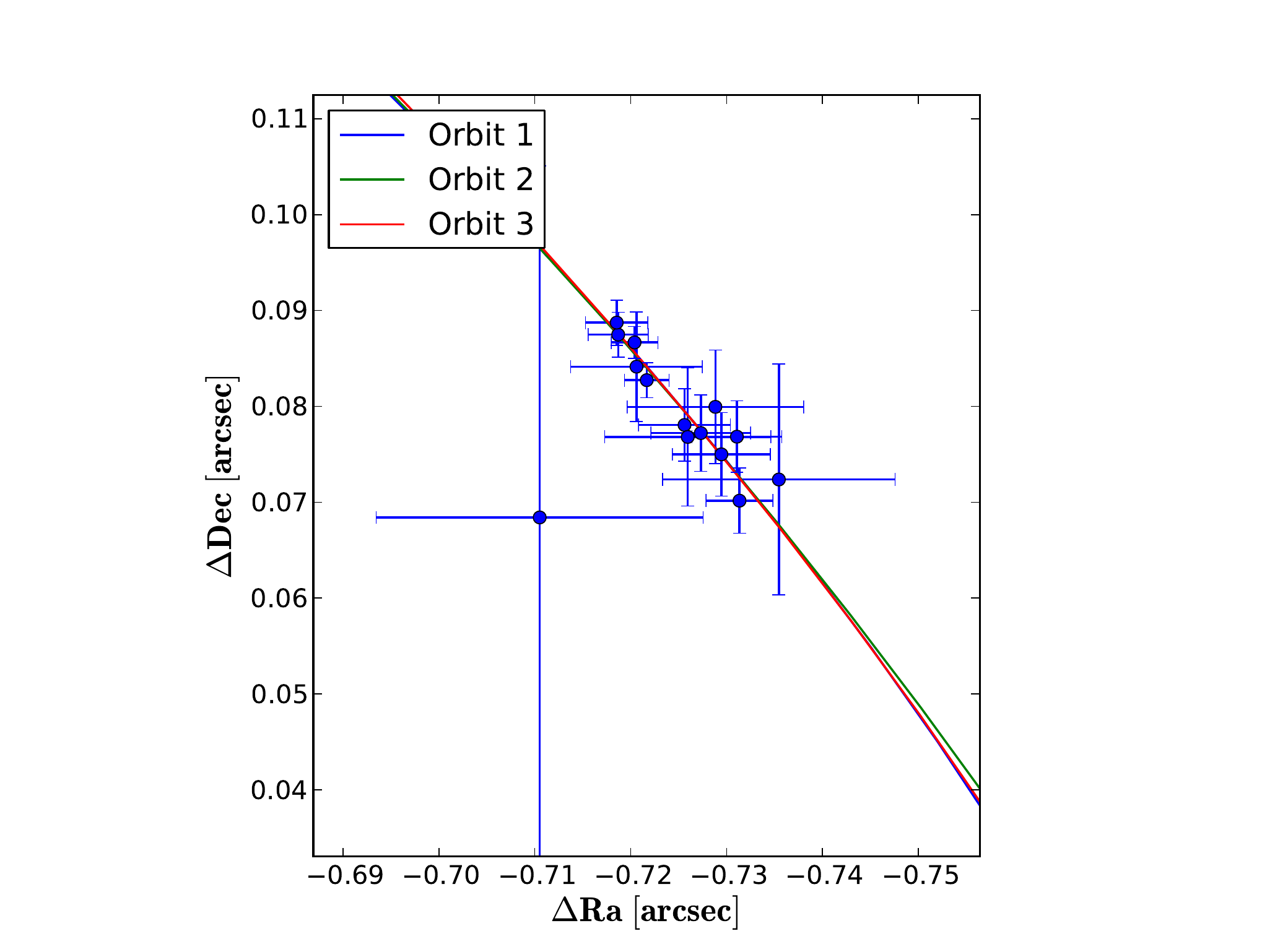}
\label{best-orbit-zoomed-gqlup}
}

\caption[]{Top 3 best-fitting orbits for the substellar companion of GQ\,Lup around its host star out of 5,000,000 unconstrained runs of our LSMC fit. Solid lines represent the apparent orbits. \ref{best-orbit-zoomed-gqlup} is zoomed in on the data points. The corresponding orbit elements are listed in Tab.~\ref{tab: orbit-elements-gqlup}} 
\label{orbits-gqlup}
\end{figure*}

\begin{figure*}
\subfloat[][]{
\includegraphics[scale=0.42]{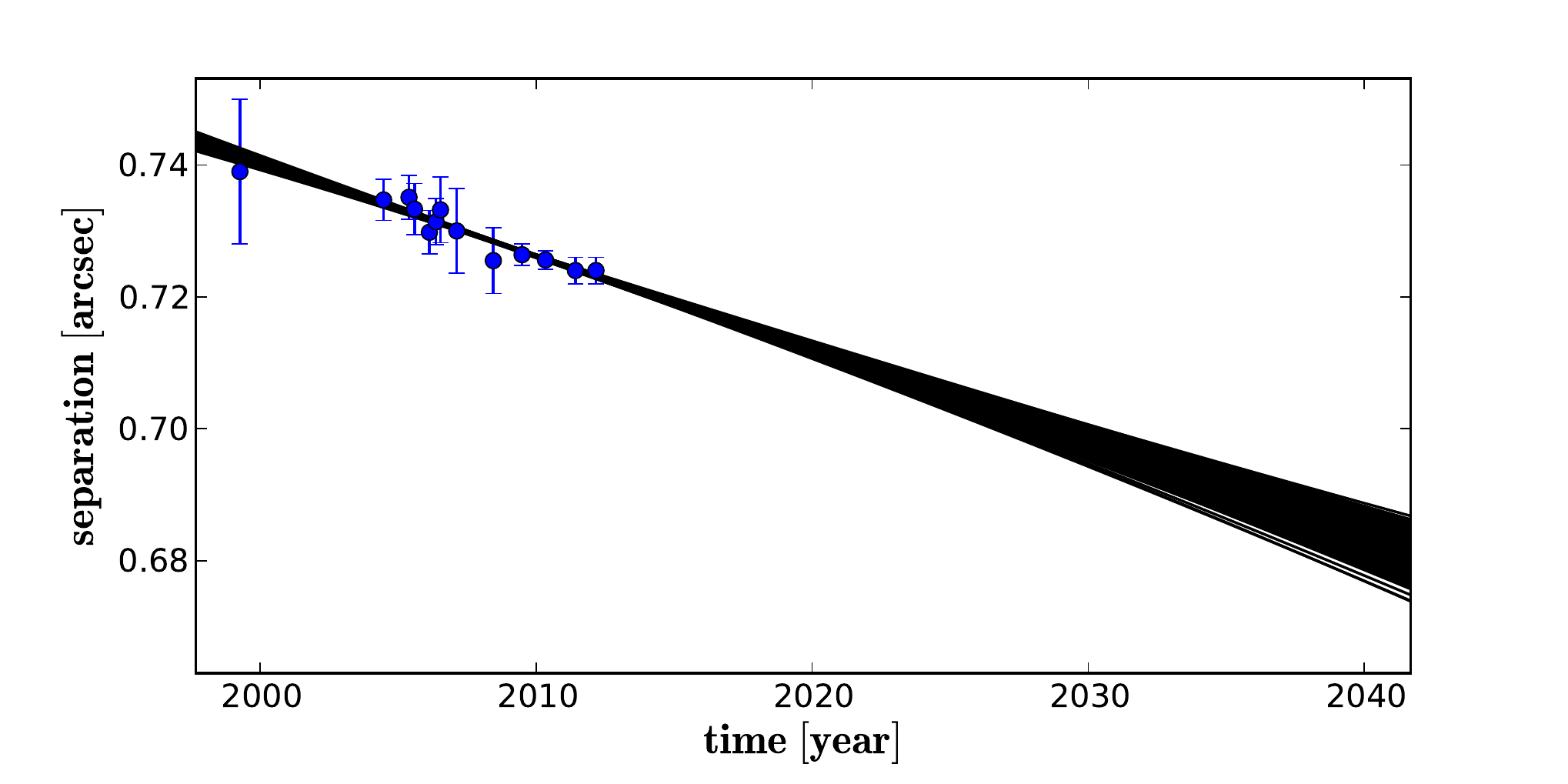}
\label{gqlup-worst-best-orbits-sep}
}
\subfloat[][]{
\includegraphics[scale=0.42]{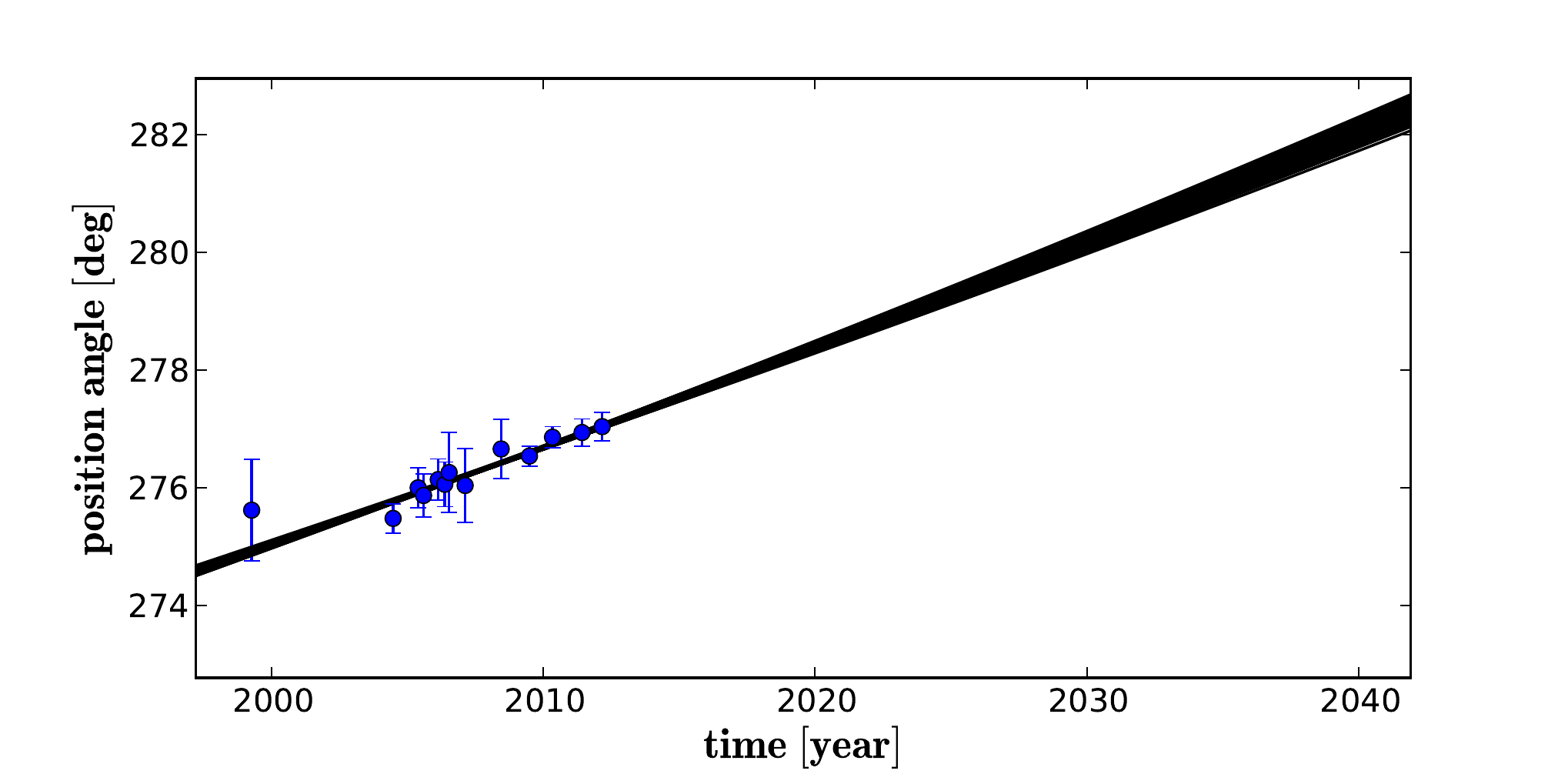}
\label{gqlup-worst-best-orbits-pa}
}

\caption[]{Separation and position angle development of the 300 orbits with the largest reduced $\chi^2$ out of the sample of the 1\,\% best fitting orbit solutions of the GQ\,Lup system shown in Fig.~\ref{fig:orbit-corr-gqlup}} 
\label{gqlup-worst-best-orbits}
\end{figure*}

For the GQ\,Lup system we fixed the total system mass to 0.7\,M$_\odot$ and we restricted the semi-major axis to values smaller than 5\,arcsec (700\,au) by the criterion given in the previous section. Furthermore we also restricted the semi-major axis to values larger than 0.54\,arcsec. This was done because \cite{2010AJ....139..626D} found a circumstellar disk around GQ\,Lup\,A with an outer radius as large as 75\,au which corresponds to 0.54\,arcsec at 140\,pc. While the disk was only marginally resolved by \cite{2010AJ....139..626D} with the SubMillimeter Array at 1.3\,mm they state that their SED modelling shows no indications for gaps or holes in the disk. Thus we find it likely that the orbit of the sub-stellar companion to GQ\,Lup is not intersecting with the disk.\\
In Fig.~\ref{fig:orbit-corr-gqlup} we show the 1\,\% best fitting solutions out of 5,000,000 runs of our LSMC fit, i.e. the solutions with the smallest reduced $\chi^2$. Shown are all orbit elements as a function of eccentricity. As can be seen in Fig.~\ref{a-vs-e-gqlup} we can already exclude some combinations of semi-major axis and eccentricity. Most notably for eccentricities close or equal to 0 we only find orbit solutions with strongly localized peaks in semi-major axis around $\sim$0.9\,arcsec and $\sim$1.4\,arcsec. For larger eccentricities the range of possible semi-major axes increases up to the point where we find orbit solutions that satisfy the astrometric measurements that span the full range of allowed semi-major axes between 0.54\,arcsec and 5\,arcsec. This corresponds to orbit periods between 786.1\,yr and 22149.7\,yr. In addition to wide orbits with low eccentricities we can also exclude short orbits with very high eccentricities. For a semi-major axis smaller than 1.26\,arcsec we only find solutions with eccentricities smaller than 0.8. In general we recover more orbit solutions with small (<1.5\,arcsec) semi-major axes than larger ones, with a total of 57\,\% of our solutions falling in this category. We also observe some notable peaks in the distribution of eccentricities at values of 0, 0.7 and 0.92.\\
In Fig.~\ref{i-vs-e-gqlup} we show the inclination as a function of eccentricity. We find possible inclinations between 20.4$^\circ$ and 75.9$^\circ$ with strong peaks in the distribution at 48.0$^\circ$, 58.8$^\circ$, 66.6$^\circ$ and 74.2$^\circ$. For eccentricities between 0 and 0.4 we can, however, exclude inclinations smaller than $\sim$48$^\circ$. Similarly for large eccentricities above 0.8 we only find inclinations between 47.8$^\circ$ and 69.2$^\circ$. In principle, all solutions with inclinations smaller than $\sim$45$^\circ$ correspond to semi-major axes smaller than 1.25\,arcsec, while for larger inclinations the full range of semi-major axes is possible.\\  
In Fig.~\ref{node-vs-e-gqlup} we show the longitude of the ascending node $\Omega$ versus eccentricity. Since there are no precise radial velocity measurements of the system available, we show only solutions with 0$\leq\Omega\leq$180$^\circ$. Depending on the eccentricity the range of possible longitudes varies significantly. Below an eccentricity of 0.6 we find longitudes between 22.9$^\circ$ and 119.4$^\circ$, while above that threshold longitudes up to 180$^\circ$ are possible. For even larger eccentricities \textit{e}$\sim$0.7 longitudes down to 0$^\circ$ are found.\\
The argument of the periastron and the time of the periastron passage are shown in Fig.~\ref{peri-vs-e-gqlup} and Fig.~\ref{T0-vs-e-gqlup}, respectively. For orbits with very small eccentricities up to $\sim$0.1 we can not put any constraints on either of the two parameters. For large eccentricities above 0.8 we can narrow down the time of the periastron passage to the timeframe between the years 1769 and 2657. In general the distribution of the time of periastron passage shows strong peaks around the years 2347 and 2604 as well as the year 5000. Movement towards periastron is consistent with the decline in separation which we are observing.\\
The three orbit solutions with the lowest reduced $\chi^2$ are shown in Fig.~\ref{orbits-gqlup} and the corresponding set of orbit elements is given in Tab.~\ref{tab: orbit-elements-gqlup}. It should be noted that these orbits are not necessarily the most probable solutions, but that in principle all recovered orbit solutions with a reasonable good fit to the astrometric data points should be regarded as equally likely. To determine when we can put stronger constraints on the GQ\,Lup system orbit, we show in Fig.~\ref{gqlup-worst-best-orbits} the 300 orbits with the largest reduced $\chi^2$ out of the 1\,\% best fitting sample. Orbits are plotted in separation and position angle together with the currently available data points. Given the current level of accuracy of our astrometric measurements, additional data points taken in $\sim$2030 would enable us to further constrain the system's orbit parameters by excluding some of the currently possible orbit solutions.

\begin{table}
  \caption{Orbit elements and $\chi^2_{red}$ of the best-fitting orbits of the substellar companion of GQ\,Lup around its host star shown in Fig.~\ref{orbits-gqlup}}
  \begin{tabular}{@{}lccc@{}}
  \hline   
 	Nr. 							& 1						&  2 						& 3 				\\
 	\hline
	a\,[arcsec] 			& 0.54				&	0.92 					& 0.69			\\
	e									& 0.52				& 0.69					& 0.21			\\
	P\,[yr]						& 786.1				&	1734.1				&	1138.3		\\	
	i\,[$^\circ$]					& 39.6				&	46.4					&	53.7			\\
	$\Omega$\,[$^\circ$]		& 77.9				&	126.6					&	76.5			\\
	$\omega$\,[$^\circ$]		& 360.0				&	266.7					& 360.0			\\
	T$_0$\,[JD]				& 2550088.0		& 2529861.8			& 2609815.2	\\
	$\chi^2_{red}$		&	0.34				&	0.34					&	0.34			\\
 \hline\end{tabular}

\label{tab: orbit-elements-gqlup}
\end{table}

\subsection{PZ\,Tel system}
\label{sec: orbit: pztel}

\begin{figure*}
\subfloat[][]{
\includegraphics[scale=0.45]{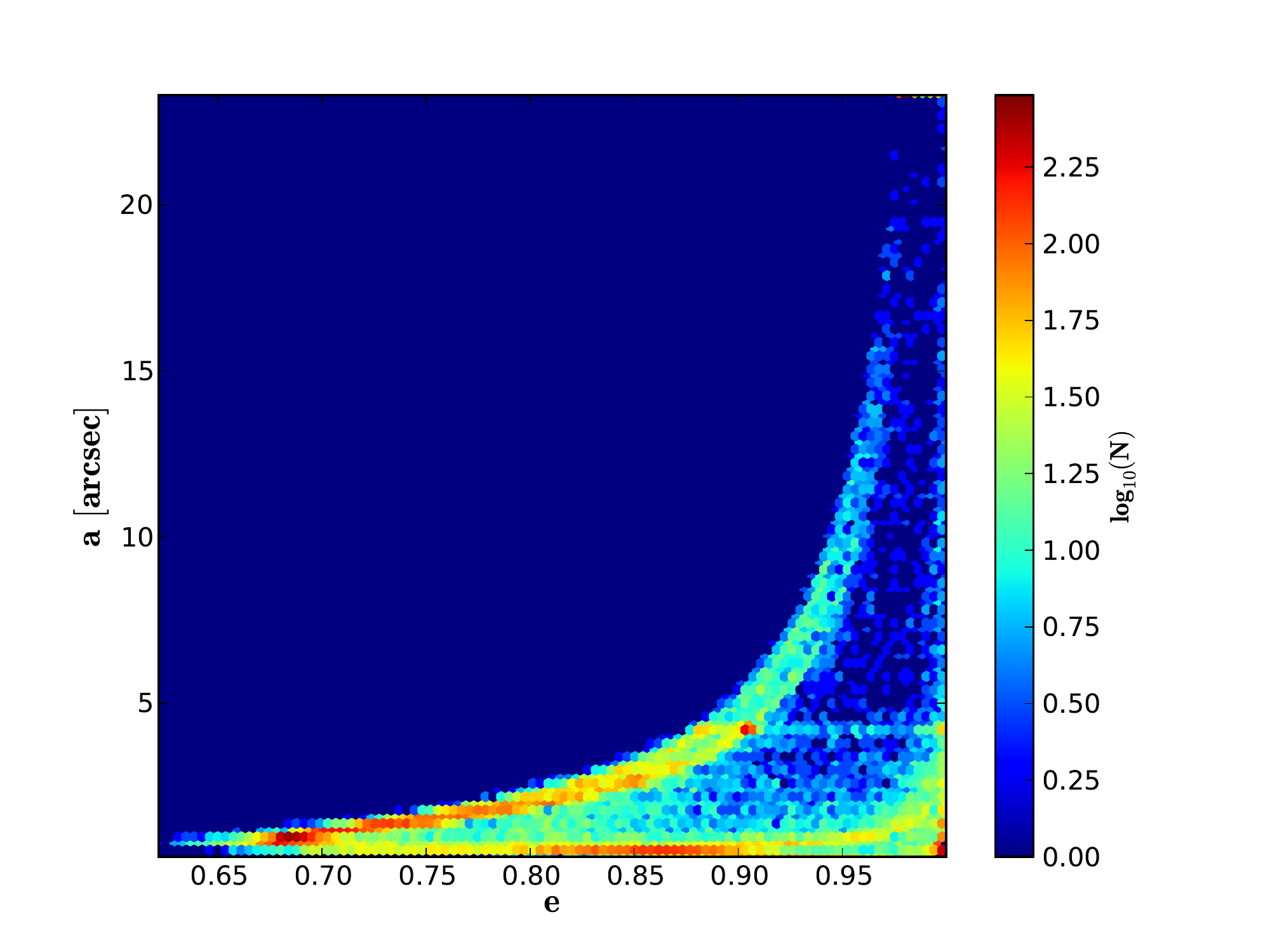}
\label{a-vs-e-pztel}
}
\subfloat[][]{
\includegraphics[scale=0.45]{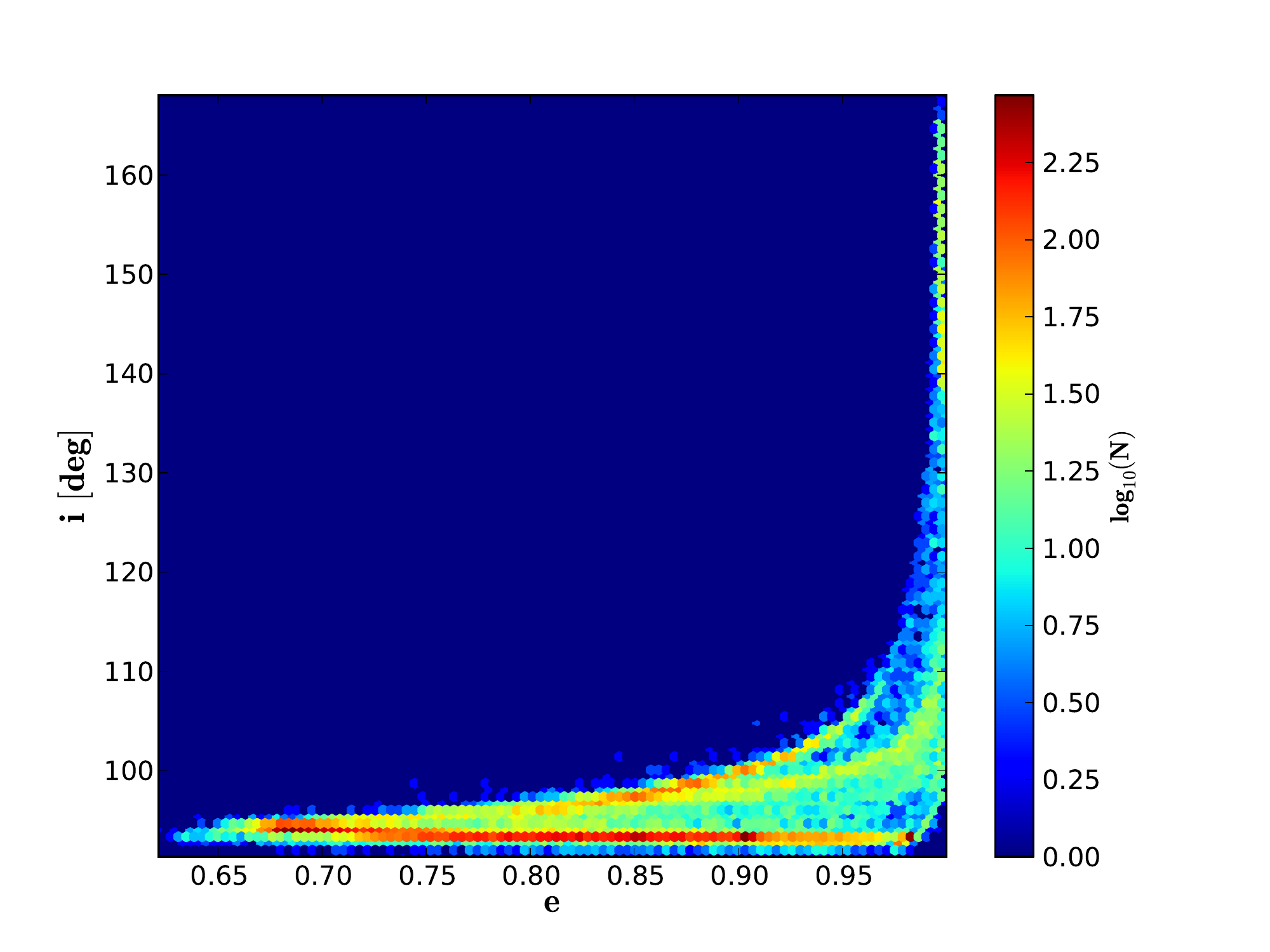}
\label{i-vs-e-pztel}
}\\
\subfloat[][]{
\includegraphics[scale=0.45]{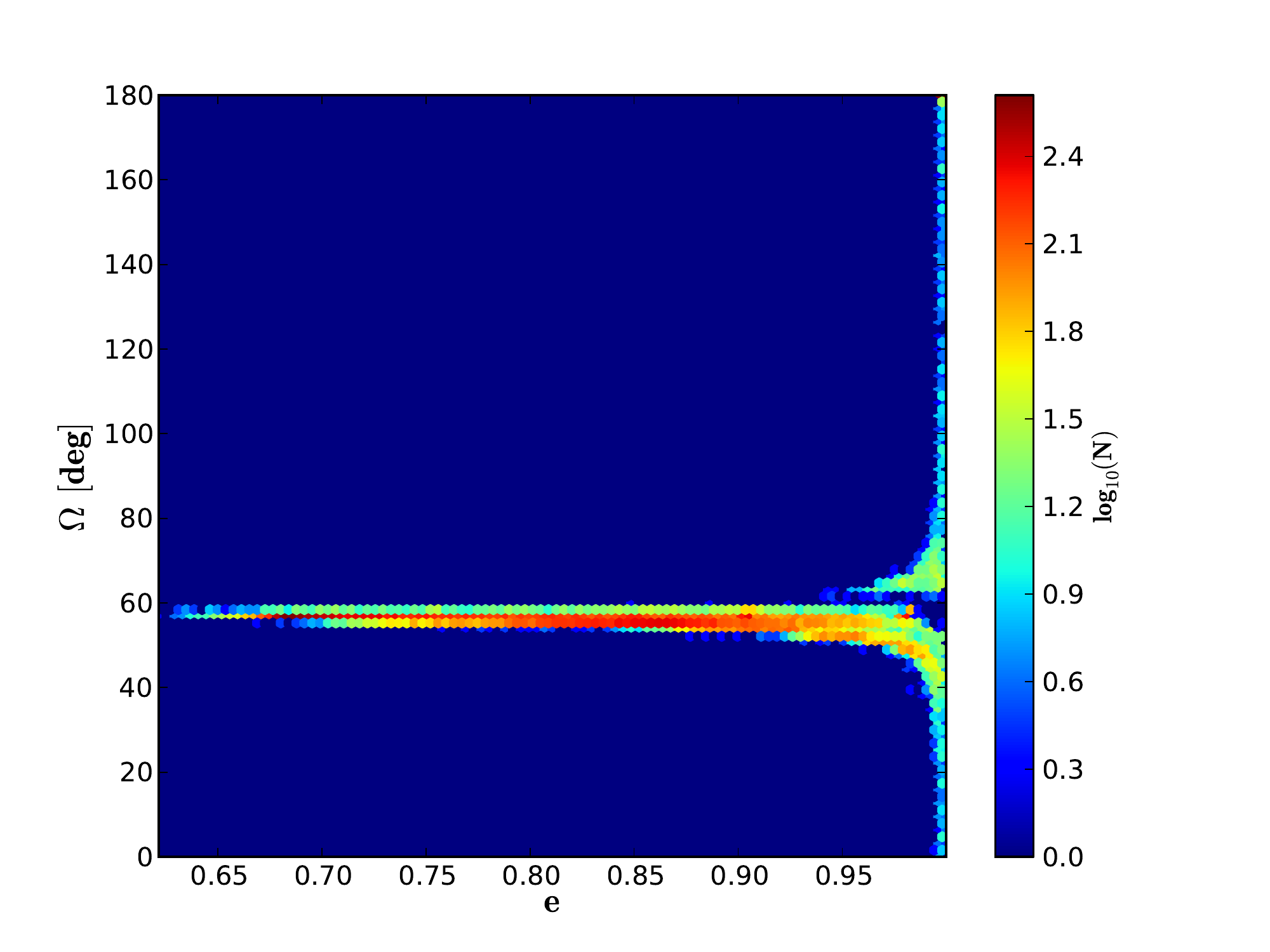}
\label{node-vs-e-pztel}
}
\subfloat[][]{
\includegraphics[scale=0.45]{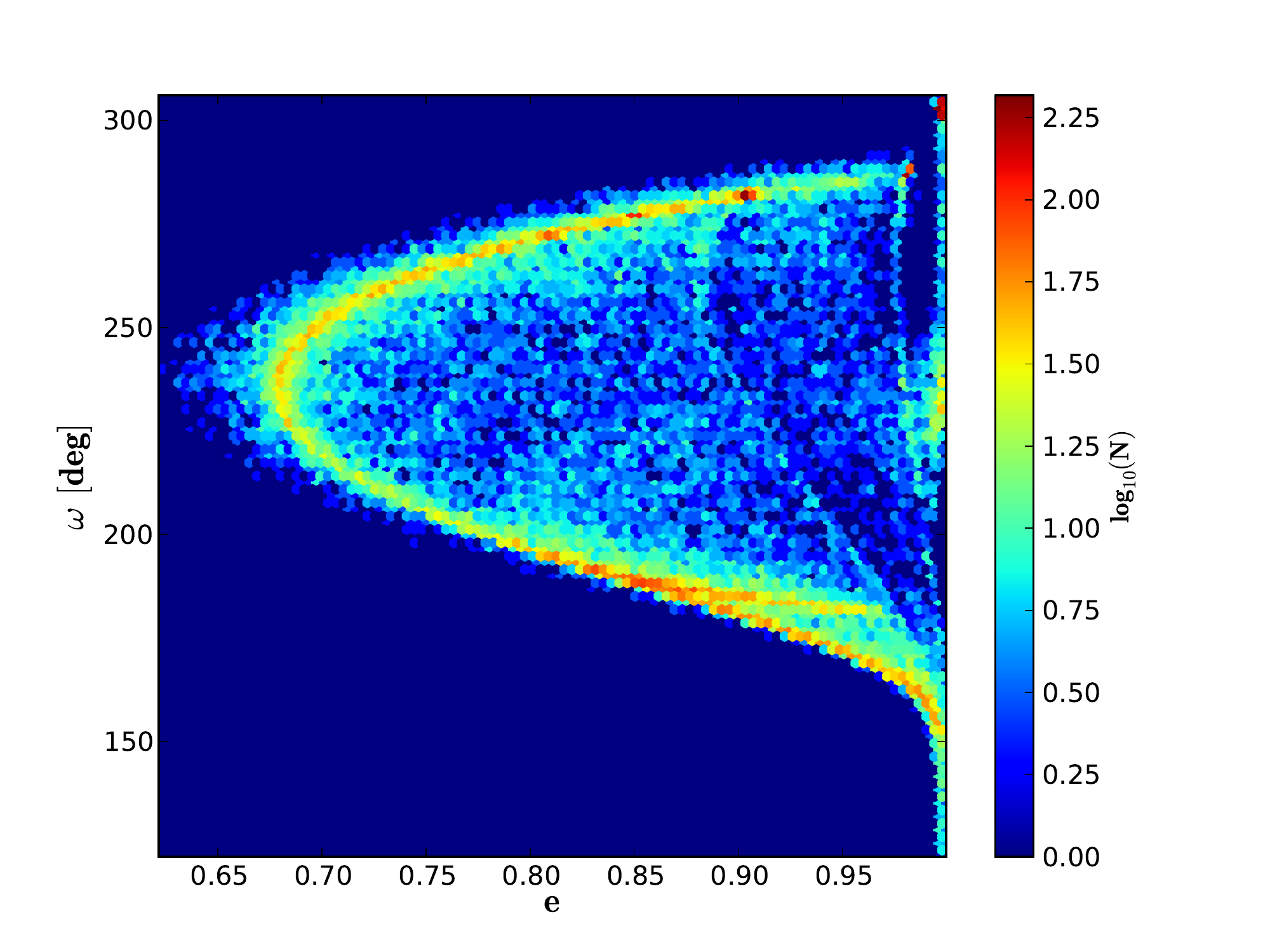}
\label{peri-vs-e-pztel}
}\\
\subfloat[][]{
\includegraphics[scale=0.45]{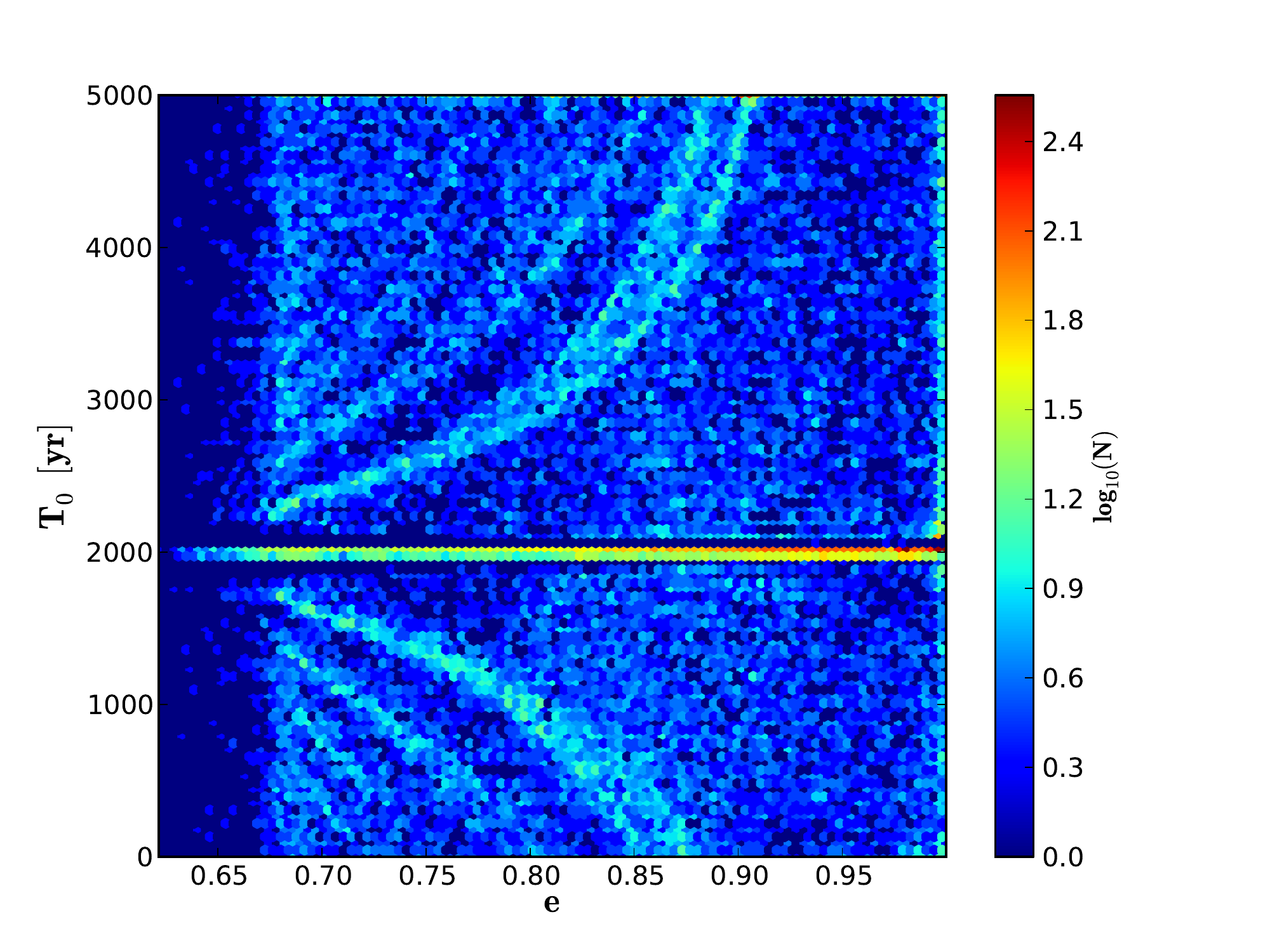}
\label{T0-vs-e-pztel}
}

\caption[]{Orbital elements as function of eccentricity for the substellar companion of PZ\,Tel around its host star for all solutions with $\chi^2_{red}\leq2$ solutions out of 5,000,000 runs of our LSMC fit. Logarithmic density of solutions is indicated by color (a color version of this figure is available in the online version of the journal). } 
\label{fig:orbit-corr-pztel}
\end{figure*}

\begin{figure*}
\subfloat[][]{
\includegraphics[scale=0.45]{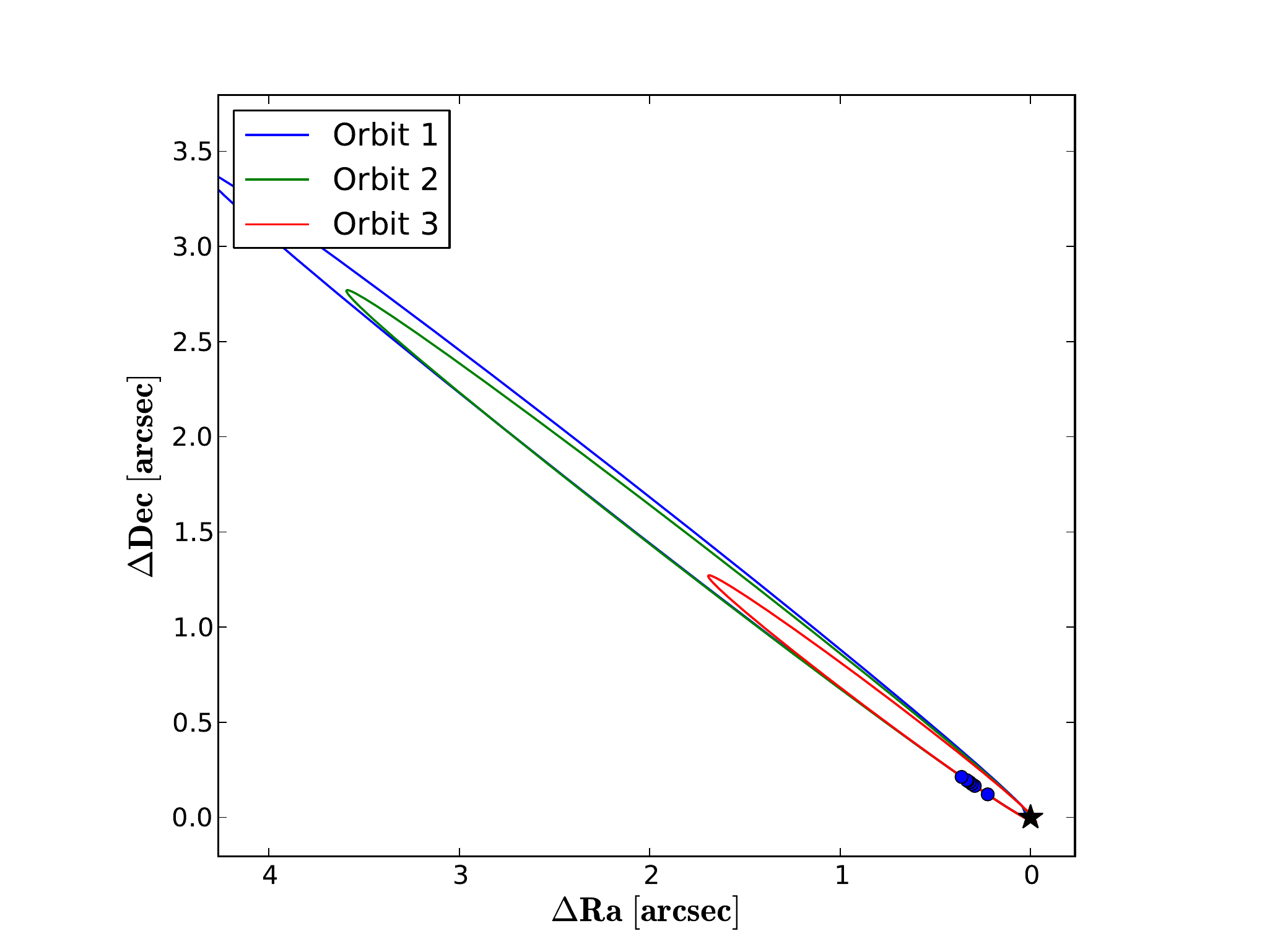}
\label{best-orbit-pztel}
}
\subfloat[][]{
\includegraphics[scale=0.45]{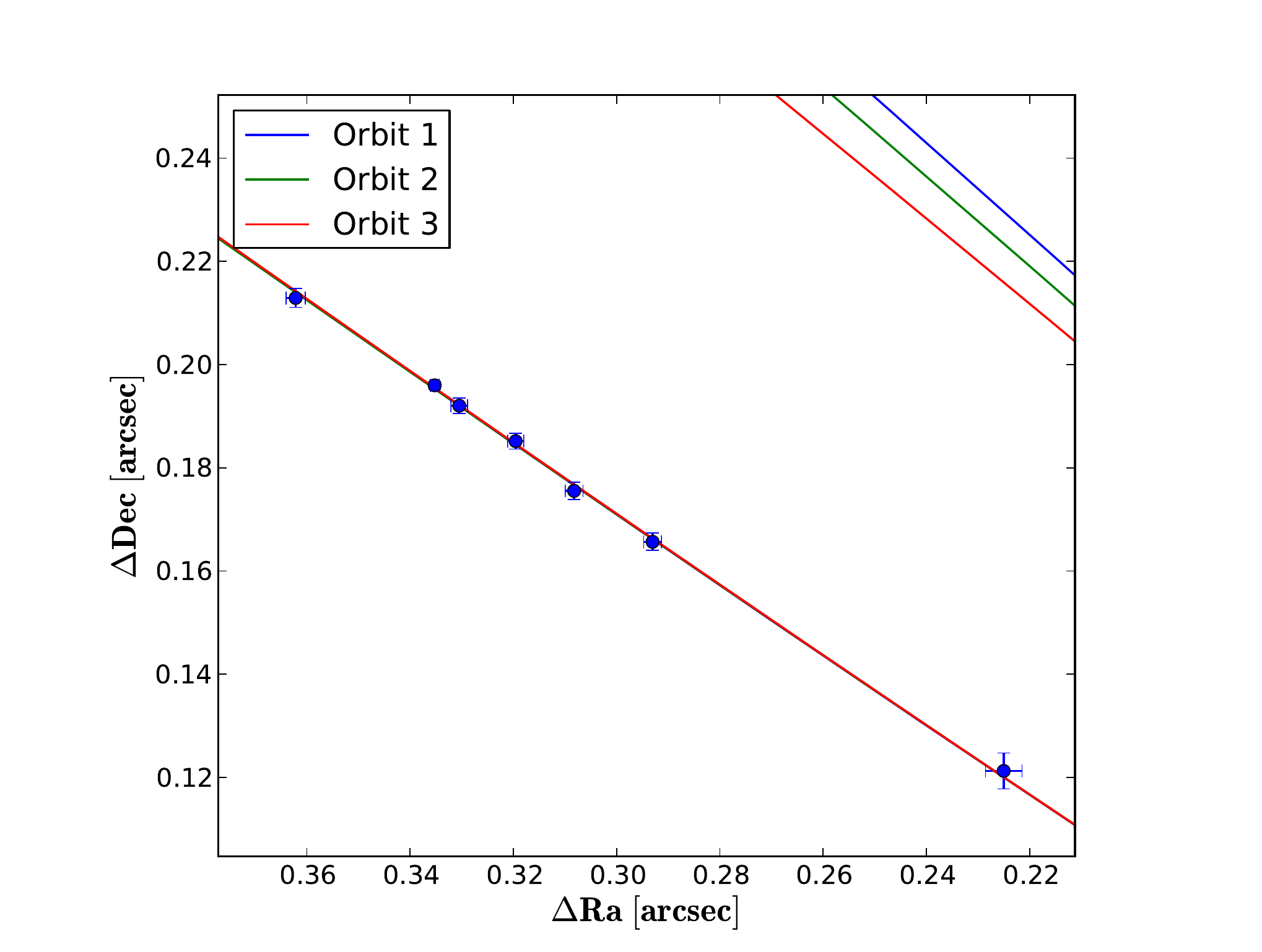}
\label{best-orbit-zoomed-pztel}
}

\caption[]{Top 3 best-fitting orbits for the substellar companion of PZ\,Tel around its host star out of 5,000,000 unconstrained runs of our LSMC fit. Solid lines represent the apparent orbits. \ref{best-orbit-zoomed-pztel} is zoomed in on the data points. The corresponding orbit elements are listed in Tab.~\ref{tab: orbit-elements-pztel}} 
\label{orbits-pztel}
\end{figure*}

\begin{figure*}
\subfloat[][]{
\includegraphics[scale=0.42]{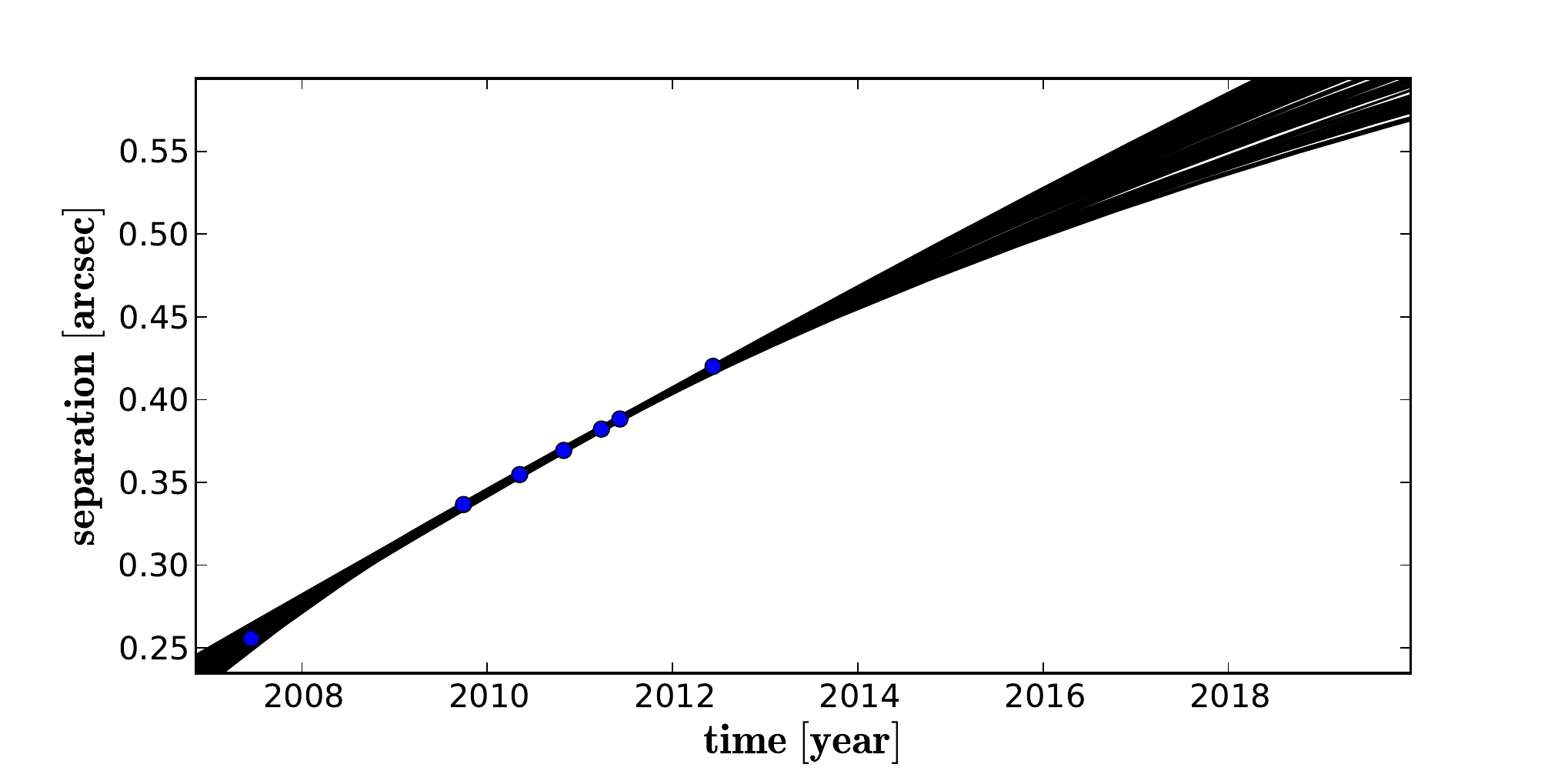}
\label{pztel-worst-best-orbits-sep}
}
\subfloat[][]{
\includegraphics[scale=0.42]{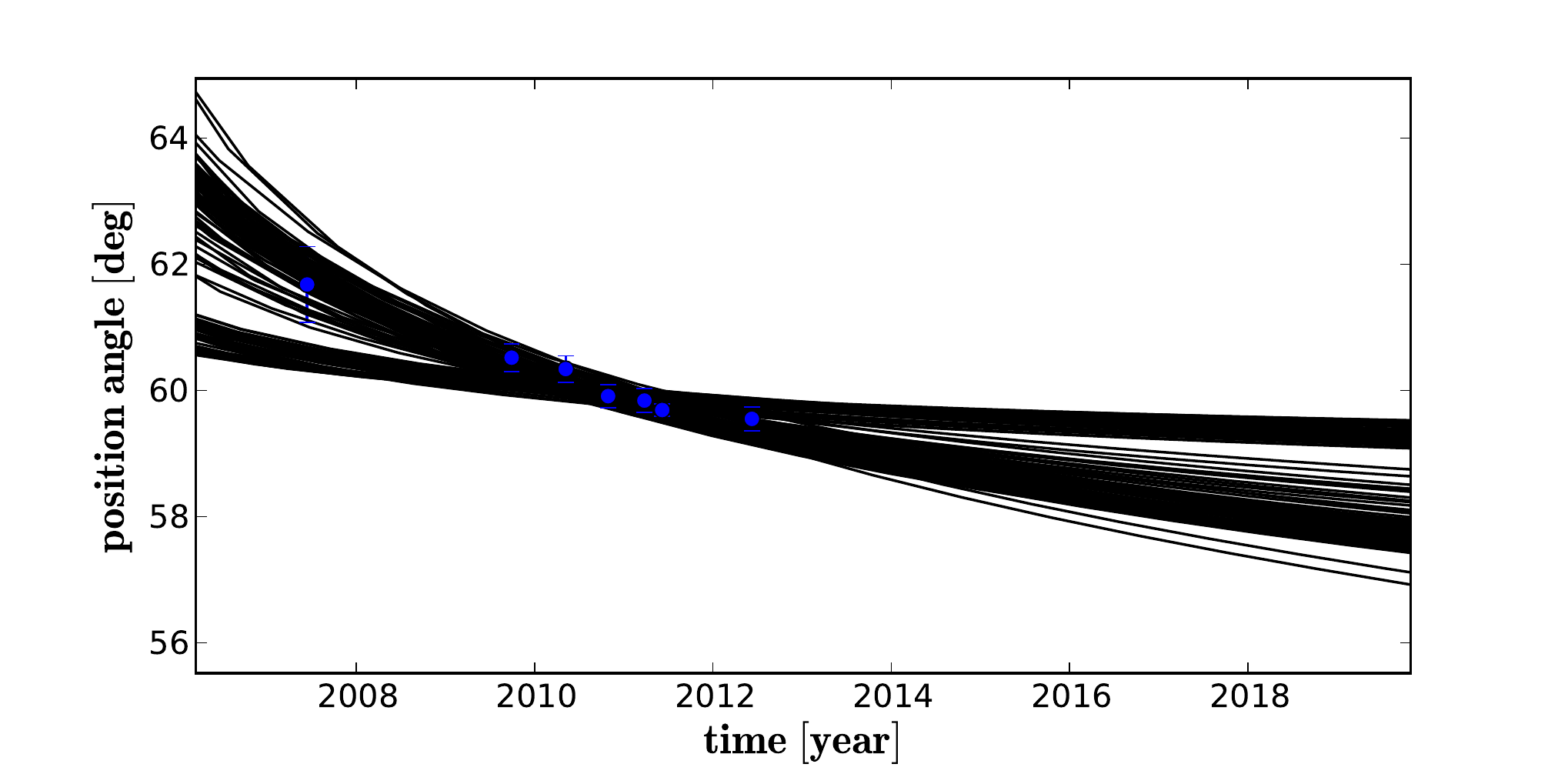}
\label{pztel-worst-best-orbits-pa}
}

\caption[]{Separation and position angle development of the 300 orbits with the largest reduced $\chi^2$ out of the sample of the 1\,\% best fitting orbit solutions of the PZ\,Tel system shown in Fig.~\ref{fig:orbit-corr-pztel}} 
\label{pztel-worst-best-orbits}
\end{figure*}

For the PZ\,Tel system we fixed the total system mass to 1.2\,M$_\odot$ as was already done by us in \cite{2012MNRAS.424.1714M}. This is not only sensible given the mass estimates for the primary star and companion discussed in section \ref{sec: target: pztel}, but also enables us to compare the results with our previous work. In our previous analysis of the system we discussed an infrared excess found in the photometry of the primary star with Spitzer/MIPS (\citealt{2004ApJS..154...25R}) at 70$\mu$m by \cite{2008ApJ...681.1484R}. This prompted us to consider a circumstellar or circumbinary disc in the system. We showed subsequently that we could only recover orbit solutions that were consistent with a circumbinary disc with an inner radius larger than 46\,au. There were no orbit solutions recovered that would allow for the existence of a circumstellar disc around PZ\,Tel\,A. Since then new far infrared photometry of the PZ\,Tel system was published by \cite{2014A&A...565A..68R}. They used Herschel/PACS (\citealt{2008SPIE.7010E...3P}) to study the system at 70$\mu$m, 100$\mu$m and 160$\mu$m. They could detect no significant infrared excess in the PZ\,Tel system and thus exclude the existence of a circumbinary disc. The discrepancy between the findings of Herschel/PACS and Spitzer/MIPS might be due to the larger pixel scale of the latter (3.3\,arcsec/pixel versus 5.3\,arcsec/pixel). It is conceivable that a background source might have contaminated the original Spitzer measurement. Due to these new findings we did not restrict the semi-major axis of the system by the presence of a potential circumbinary disk, but instead the semi-major axis was only restricted to values smaller than 23.3\,arcsec for the stability considerations indicated in section \ref{sec: LSMC}.\\
The results of our LSMC fits are shown in Fig.~\ref{fig:orbit-corr-pztel}. As was the case for the GQ\,Lup system we show the orbit parameters of the best fittings solutions as a function of eccentricity. In the case of the PZ\,Tel system we show all solutions with a reduced $\chi^2$ smaller than 2. This is equivalent to showing the best 0.6\,\% of all orbit solutions. We choose the absolute $\chi^2$ cutoff in this case instead of showing the best 1\,\% of all orbits because the 1\,\% best-fitting orbits would have included solutions with an unreasonably large reduced $\chi^2$ ($\chi^2_{red}\,>\,2$). \\
We recover well-fitting orbits with semi-major axes between 0.379\,arcsec and 23.3\,arcsec, corresponding to orbital periods between 78.7\,yr and 37938.3\,yr. All of our recovered orbit solutions are highly eccentric with eccentricities ranging between 0.622 and 0.99991. In general, we observe in Fig.~\ref{a-vs-e-pztel} that in the case of increasing eccentricities the corresponding semi-major axes are also increasing. However, the distribution of semi-major axes shows a strong peak at 0.443\,arcsec and 91\,\% of all good solutions show semi-major axes smaller than 5\,arcsec. If we only consider eccentricities smaller than 0.9, the semi-major axes can actually be constrained to a much narrower range between 0.41\,arcsec and 5.3\,arcsec.\\
In Fig.~\ref{i-vs-e-pztel} we show possible inclinations as functions of eccentricity. We can see a similar trend as was the case for the semi-major axes. For increasing eccentricities the inclination is also increasing, i.e. for extreme eccentric systems, close to pole-on orbits are possible. In general, we find orbit solutions with inclinations between 91.3$^\circ$ and 168.1$^\circ$ with a strong peak at 93.2$^\circ$. If we again only consider orbits with eccentricities smaller than 0.9, then the range of inclinations decreases to values between 91.3$^\circ$ and 102.2$^\circ$.\\
The other two angular orbit elements are shown in Fig.~\ref{node-vs-e-pztel} and Fig.~\ref{peri-vs-e-pztel}, respectively. If all orbit solutions are considered then the longitude of the ascending node can not be constrained. However, it shows a very strong peak at 57.4$^\circ$, and 92\,\% of all solutions are located between 50$^\circ$ and 70$^\circ$. If we look at the solutions with eccentricities smaller than 0.9 we find a very narrow range for the longitude of the ascending node between 52.7$^\circ$ and 59.3$^\circ$. \\
For the argument of the periastron we find solutions between 122.2$^\circ$ and 306$^\circ$. This gets only slightly narrower if only solutions with \textit{e}$<$0.9 are considered and encompasses then the range between 179.1$^\circ$ and 288.0$^\circ$. A peak of the distribution can be observed at 187.3$^\circ$. The time of the periastron passage is in principle not well constrained and shows values between the years 0 and 5000. This does not depend on the corresponding eccentricity of the orbit solutions. However, the distribution shows a dominant peak for the epoch 2003.5. This is very consistent with the non-detection of the companion in the VLT/NACO observation of mid-2003 by \cite{2005ApJ...625.1004M} as was already mentioned by us in \cite{2012MNRAS.424.1714M}. \\
Overall our new astrometric measurement of June 2012 fits very well with the recovered orbits of our previous study. Thus we come to similar conclusions with only small changes for the possible orbits of the PZ\,Tel system. One of these changes is that we now consider orbits up to semi-major axes of 23.3\,arcsec, i.e. our orbit solutions are no longer truncated by the existence of a possible circumbinary disk. We show the three best fitting orbit solutions in Fig.~\ref{orbits-pztel} and the corresponding orbit elements in Tab.~\ref{tab: orbit-elements-pztel}. All of these orbits have extreme eccentricities close to 1. If the system is not in the process of flying apart, then it would seem that less eccentric orbits are more likely to produce a stable system. \\
As we did in the previous section for the GQ\,Lup system, we show the 300 orbit solutions from the best-fitting sample with the highest reduced $\chi^2$ in Fig.~\ref{pztel-worst-best-orbits}. The orbits as well as the astrometric measurements of the system are plotted in separation and position angle versus time. We can conclude that with our current measurement accuracy, any new measurement taken at present would further improve our ability to narrow down the orbit elements of the system. The PZ\,Tel system should thus be monitored continuously.

\begin{table}
  \caption{Orbit elements and $\chi^2_{red}$ of the best-fitting orbits of the substellar companion of PZ\,Tel around its host star shown in Fig.~\ref{orbits-pztel}}
  \begin{tabular}{@{}lccc@{}}
  \hline   
 	Nr. 							& 1						&  2 						& 3 				\\
 	\hline
	a\,[arcsec] 			& 3.02				&	2.40 					& 1.10			\\
	e									& 0.996				& 0.991					& 0.965			\\
	P\,[yr]						& 1766.6			&	1256.0				&	386.9		\\	
	i\,[$^\circ$]					& 109.6				&	104.0					&	100.6			\\
	$\Omega$\,[$^\circ$]		& 43.1				&	47.6					&	51.1			\\
	$\omega$\,[$^\circ$]		& 154.9				&	160.9					& 169.1			\\
	T$_0$\,[JD]				& 1807003.6		& 1993288.4			& 2451803.7	\\
	$\chi^2_{red}$		&	0.34				&	0.34					&	0.34			\\
 \hline\end{tabular}

\label{tab: orbit-elements-pztel}
\end{table}

\section{Detection limits}
\begin{figure*}

\subfloat[]{
\includegraphics[scale=0.43]{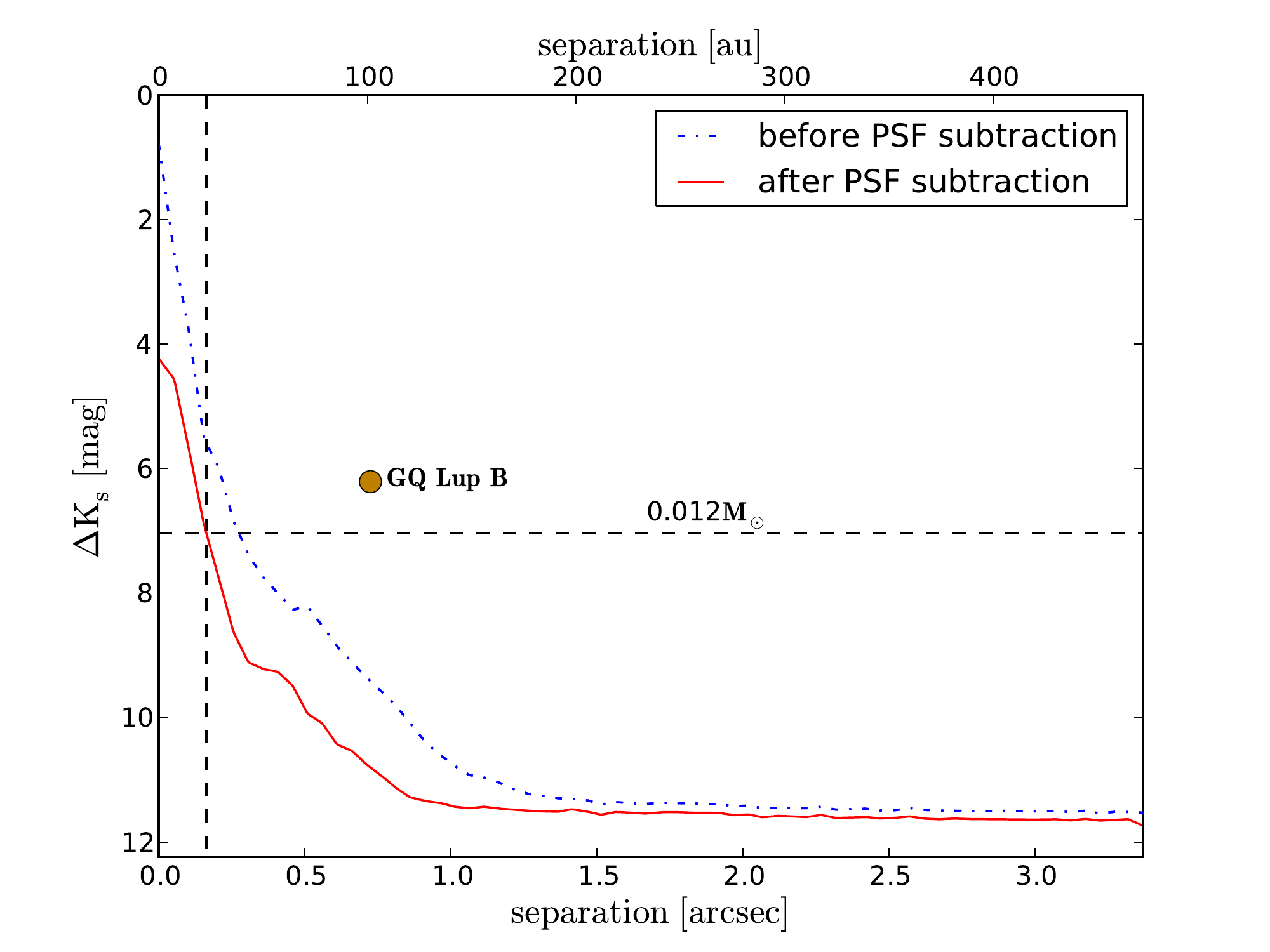}
\label{gqlup-dr-pm}
}
\subfloat[]{
\includegraphics[scale=0.43]{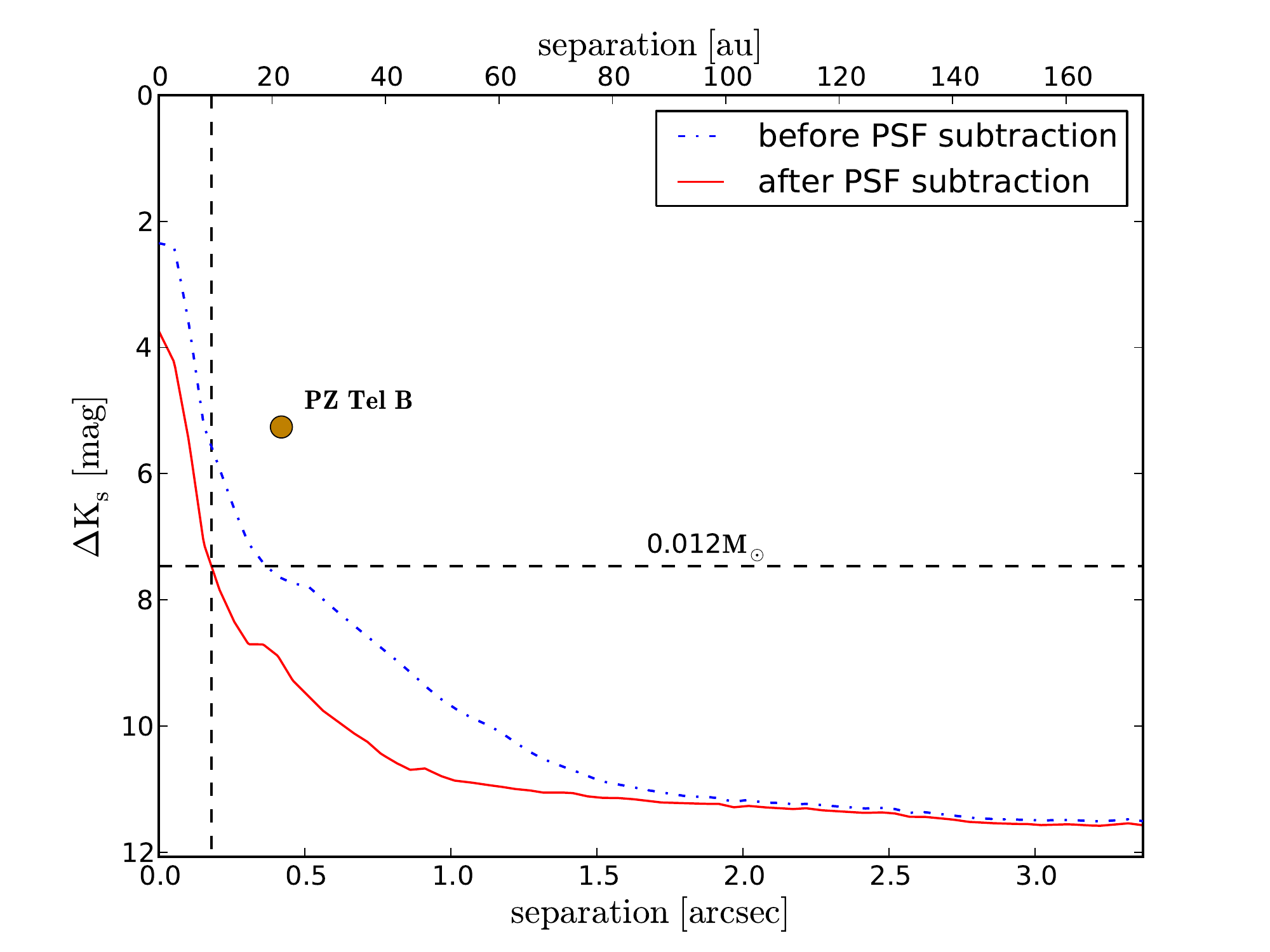}
\label{pztel-dr-pm}
}

\subfloat[]{
\includegraphics[scale=0.43]{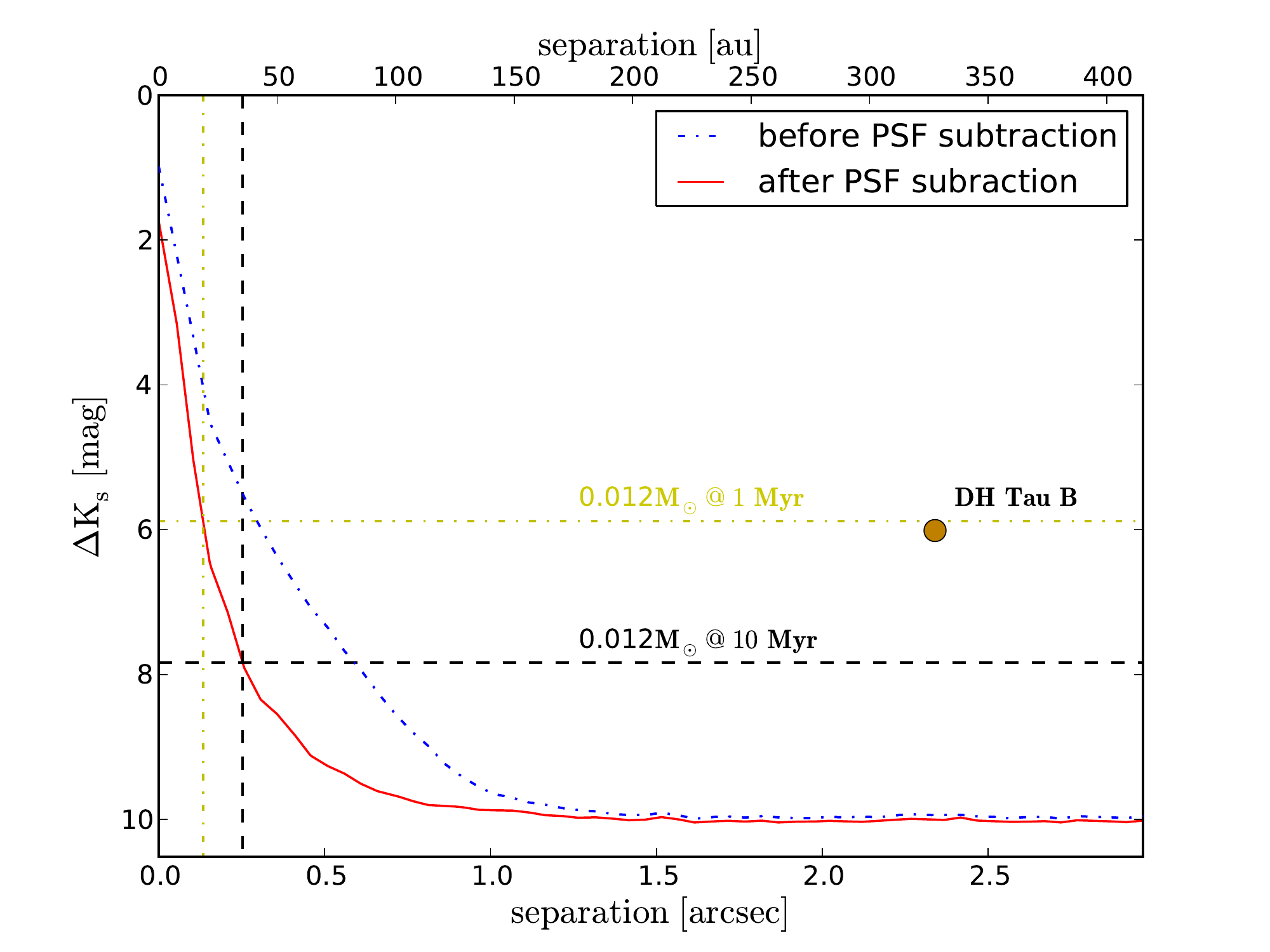}
\label{dhtau-dr-pm}
}
\subfloat[]{
\includegraphics[scale=0.43]{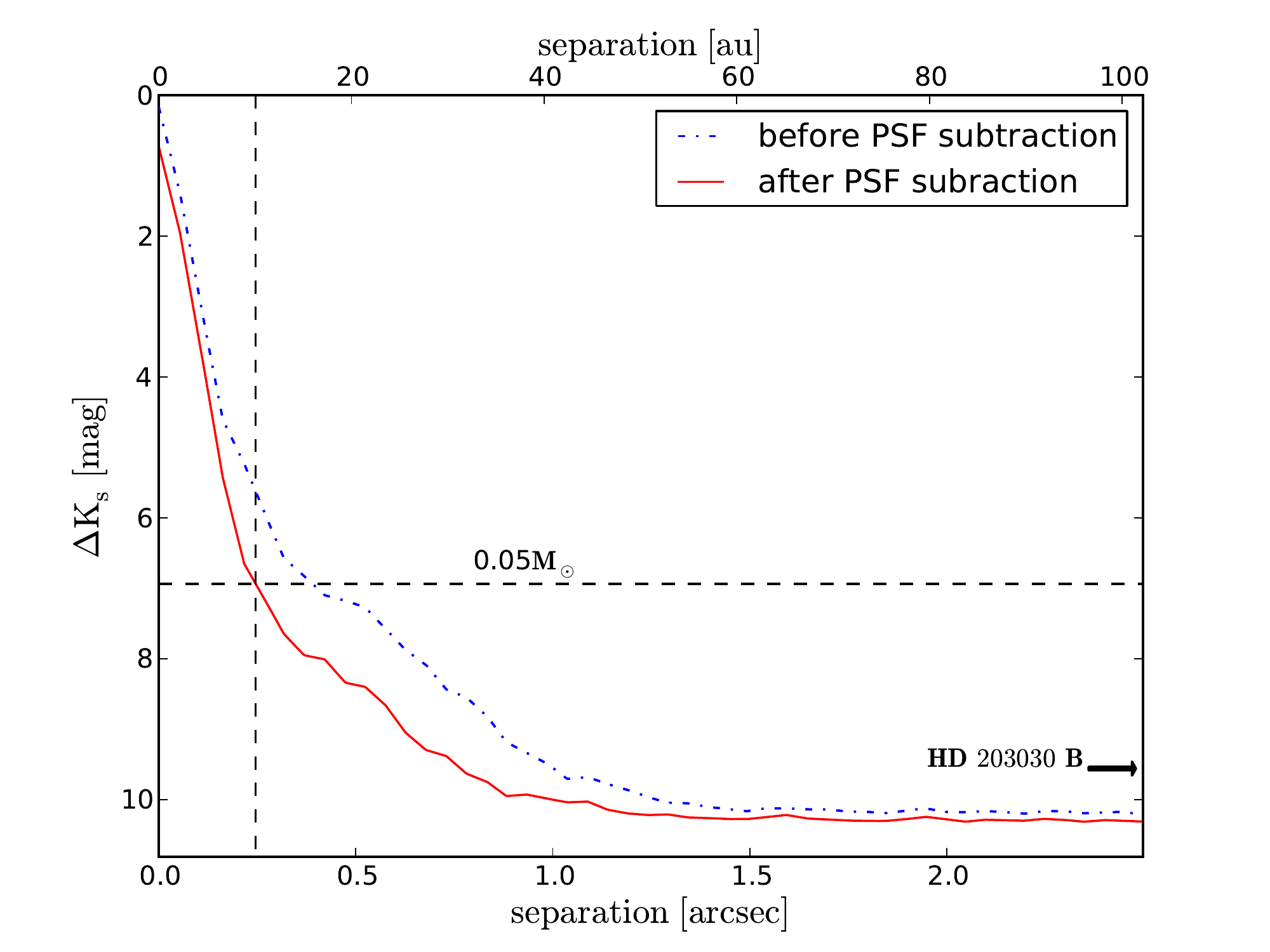}
\label{hd203030-dr-pm}
}

\subfloat[]{
\includegraphics[scale=0.43]{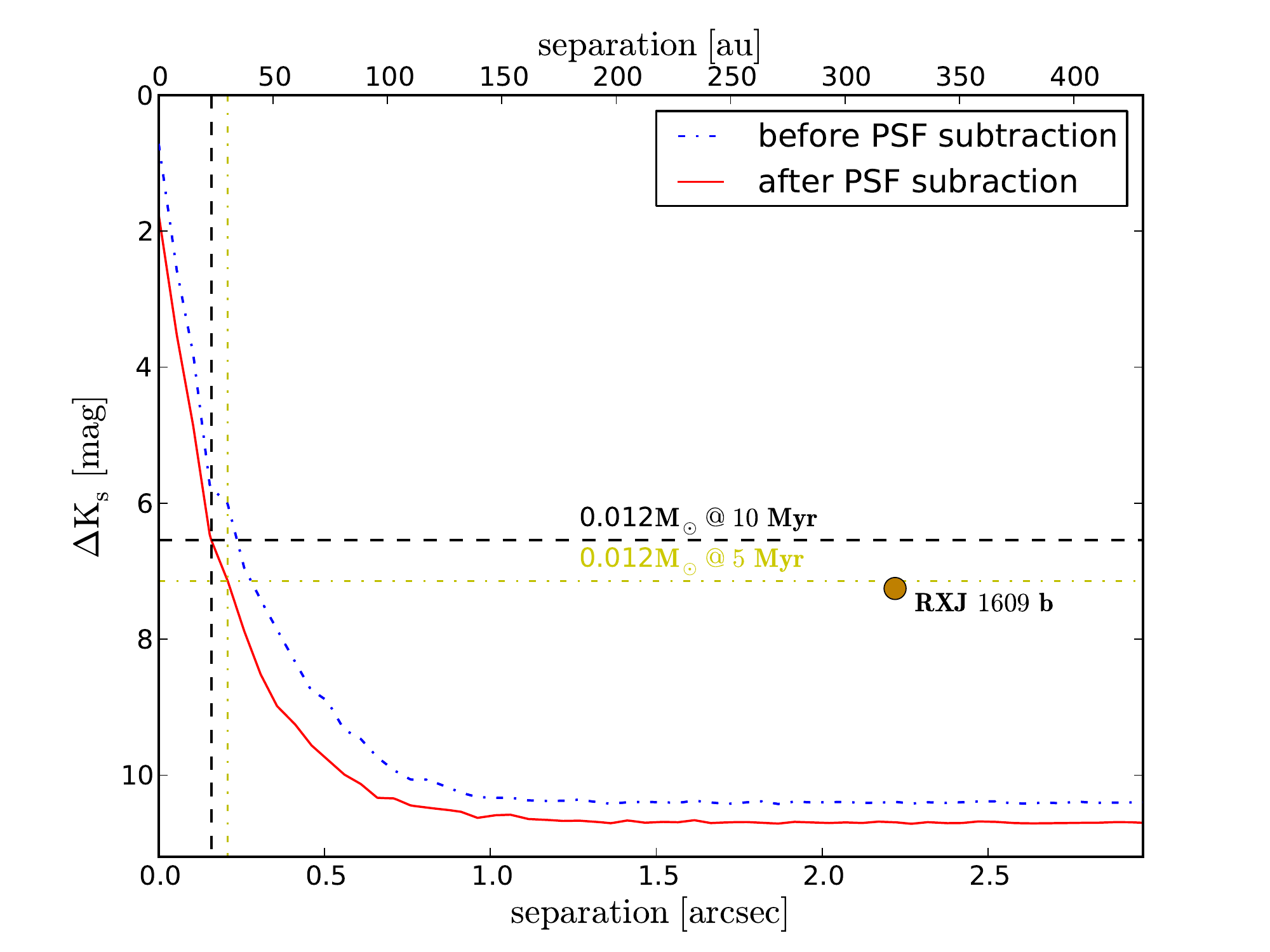}
\label{rxj1609-dr-pm}
}
\subfloat[]{
\includegraphics[scale=0.43]{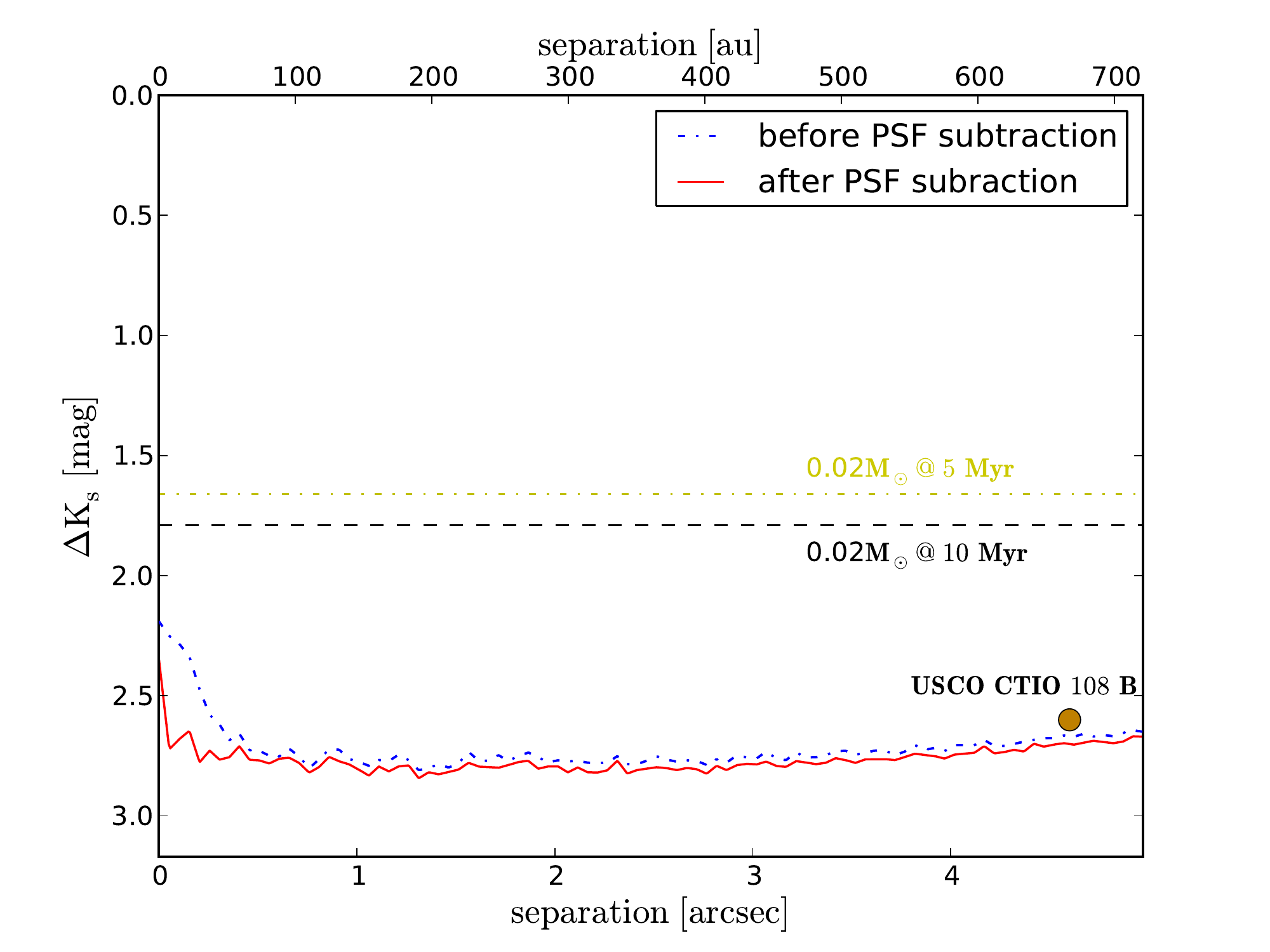}
\label{usco108-dr-pm}
}

\caption[]{Dynamic range plot for our NACO images before and after PSF subtraction. All objects above the solid (red) or dash-dotted (blue) lines are detectable. These lines mark the detection limit for a signal-to-noise of 5. The positions of the companions are indicated by the filled circles. Mass estimates are done utilizing the evolutionary models by \cite{2000ApJ...542..464C}.} 
\label{dr-diagram}
\end{figure*}

For all mass calculations in this section, the DUSTY models by \cite{2000ApJ...542..464C} were used. The DUSTY models include condensed dust particles of various species in the equation of state and the radiative transfer equations. They hence include the dust opacity, which is believed to have a major influence on the infra-red (IR) colors below an effective temperature of 2000\,K (see also \citealt{2000ApJ...542..464C}). It was shown in the reference publication that these models reproduce the infrared colors and flux of late M and L dwarfs accurately, i.e. they are well suited to predict the major physical parameters of all companions discussed in this work. There are meanwhile more accurate models to describe T (methane) dwarfs and giant extrasolar planets (especially hot Jupiters), as discussed in \cite{2003A&A...402..701B}. However, the predicted photometry of the DUSTY models does not vary much ($\sim$\,0.1\,mag) from these models and hence the calculated mass limits should be fairly accurate even towards lower masses.\\
The dynamic range plot for the GQ\,Lup system is shown in Fig.~\ref{gqlup-dr-pm}. In order to detect the lowest possible masses we median combined all our recent observation epochs between 2008 and 2012. Given the 2MASS magnitude of GQ\,Lup\,A of 7.096\,$\pm$\,0.020\,mag and the distance of 140\,$\pm$\,50\,pc we calculated an absolute magnitude for the primary star of 1.365\,$\pm$\,0.776\,mag. The age of the GQ\,Lup system is probably between 1 and 3 \,Myr as mentioned in section \ref{sec: target: gqlup}. We thus utilized models for an age of 1\,Myr. From our calculations we can exclude additional sub-stellar companions with a mass down to 0.0040$\pm$0.0017\,M$_{\odot}$ at distances down to 0.25\,arcsec (35\,au). In the background limited region outside of 1\,arcsec (140\,au) we can exclude objects down to masses of 0.0017\,$\pm$\,0.0003\,M$_{\odot}$.\\
The PZ\,Tel system is most likely older than the GQ\,Lup system as reviewed in section \ref{sec: target: pztel}. We thus utilized model tracks for an age of 10\,Myr. Given the 2MASS magnitude of PZ\,Tel\,A of 6.366\,$\pm$\,0.024\,mag and the distance of PZ\,Tel of 51.5\,$\pm$\,2.5\,pc we calculated an absolute magnitude of 2.807\,$\pm$\,0.107\,mag. Our calculations enable us to exclude additional sub-stellar companions down to masses of 0.0079\,$\pm$\,0.0003\,M$_{\odot}$ at distances down to 0.25\,arcsec (13\,au). In the background limited region outside of 1\,arcsec we can exclude additional objects down to 0.0043\,$\pm$\,0.0001\,M$_{\odot}$. The corresponding dynamic range plot is shown in Fig.~\ref{pztel-dr-pm}.\\
The dynamic range plot for DH\,Tau is shown in Fig.~\ref{dhtau-dr-pm}. The age of the DH\,Tau system is still a matter of discussion. As pointed out in section \ref{sec: target: dhtau}, the age of the primary is believed to be between 0.1 and 4.4\,Myr, while the age of the companion measured by spectroscopy is estimated to be between 3 and 10\,Myr. Subsequently, model tracks for an age of 1\,Myr and 10\,Myr were used to calculate the detection limits. Considering the distance of DH\,Tau of 140\,$\pm$\,10\,pc, and the apparent magnitude in the \textit{K}-band of 8.178\,$\pm$\,0.026\,mag (\citealt{2003yCat.2246....0C}), the absolute magnitude of DH\,Tau in the \textit{K}-band is 2.446\,$\pm$\,0.157\,mag. Utilizing this magnitude and the model plots for 1\,Myr, the minimum detectable mass down to a separation of 0.25\,arcsec (35\,au) is 0.0034\,$\pm$\,0.0003\,M$_{\odot}$. This changes to a larger mass of 0.0116\,$\pm$\,0.0007\,M$_{\odot}$ if the model tracks for older objects with an age of 10\,Myr are used. At a separation of 0.5\,arcsec (70\,au), masses down to 0.0022\,$\pm$\,0.0001\,M$_{\odot}$ and 0.0067\,$\pm$\,0.0003\,M$_{\odot}$ are detectable for young and old objects respectively. In the background-limited region outside of 2\,arcsec (280\,au) and up to 6.5\,arcsec (910\,au), minimum mass objects of 0.00181\,$\pm$\,0.00006\,M$_{\odot}$ and 0.0056\,$\pm$\,0.0002\,M$_{\odot}$ would have been detected. \\
The dynamic range plot for HD\,203030 is shown in Fig.~\ref{hd203030-dr-pm}. The VLT/NACO image of HD\,203030 was taken in the NB\,2.17 filter. The model tracks for the \textit{K}-band were utilized to calculate detection limits. Since the age range of HD\,203030 is between 130\,Myr and 400\,Myr (\citealt{2006ApJ...651.1166M}), an interpolation was made between the model tracks for 100\,Myr and 500\,Myr to calculate the magnitudes for 300\,Myr. Considering the Hipparcos parallax of 24.48\,$\pm$\,1.05\,mas (40.85\,pc), the primary star exhibits an absolute magnitude in the \textit{K}-band of 3.59\,$\pm$\,0.068\,mag (\citealt{2003yCat.2246....0C}). Given the model tracks and the calculated dynamic range, objects with masses down to 0.047\,$\pm$\,0.001\,M$_{\odot}$ would have been detected down to an angular separation of 0.25\,arcsec (10\,au). Less massive objects of 0.032\,$\pm$\,0.001\,M$_{\odot}$ could have been detected outside of 0.5\,arcsec (20\,au). In the background-limited region outside of 2\,arcsec (82\,au) and up to 6.6\,arcsec (270\,au), objects with masses down to 0.0191\,$\pm$\,0.0004\,M$_{\odot}$ would have been detected. Objects outside of 6.6\,arcsec could only be detected to the north and south up to a separation of 13.2\,arcsec (539\,au), and to the east up to a separation of 22\,arcsec (899\,au), due to the placement of HD\,203030 in the field of view of the NACO S\,27 detector.\\
The dynamic range plot for 1RXS\,J160929.1-210524 is shown in Fig.~\ref{rxj1609-dr-pm}. As discussed in section \ref{sec: target: rxj1609}, the age of US is estimated to be between 5\,Myr and 6\,Myr, but was recently re-evaluated and could be up to 13\,Myr (11\,$\pm$\,2\,Myr , \citealt{2012ApJ...746..154P}). Consequently, model tracks for an age of 5\,Myr and 10\,Myr were used to calculate detection limits. Considering the distance of US of 145\,$\pm$\,20\,pc and the apparent magnitude of the primary star in the \textit{K}-band of 8.916\,$\pm$\,0.021\,mag (\citealt{2003yCat.2246....0C}), the absolute magnitude of the primary in the \textit{K}-band is 3.11\,$\pm$\,0.30\,mag. Utilizing this magnitude and the model tracks for a younger age, objects down to 0.0059\,$\pm$\,0.0006\,M$_{\odot}$ would have been detected down to an angular separation of 0.25\,arcsec (36\,au). If US is indeed older, this changes towards slightly higher masses of 0.0086\,$\pm$\,0.0013\,M$_{\odot}$. At an angular separation of 0.5\,arcsec (73\,au), lower mass objects down to 0.0036\,$\pm$\,0.0003\,M$_{\odot}$ and 0.0051\,$\pm$\,0.0003\,M$_{\odot}$ for the two different ages respectively would have been detected. In the background-limited region outside of 2\,arcsec (290\,au) and up to 6.5\,arcsec (943\,au), the minimum detectable mass is 0.0029\,$\pm$\,0.0002\,M$_{\odot}$ for an age of 5\,Myr and 0.0041\,$\pm$\,0.0003\,M$_{\odot}$ for an age of 10\,Myr.\\
The dynamic range plot for UScoCTIO\,108 is shown in Fig.~\ref{usco108-dr-pm}. UScoCTIO\,108 is a member of US like 1RXS\,J160929.1-210524, and hence is located at approximately the same distance of 145\,$\pm$\,20\,pc, and has a similar age. Consequently, detection limits were again computed for ages of 5\,Myr and 10\,Myr. Given the distance of US, and the apparent magnitude of UScoCTIO\,108\,A in the \textit{K}-band of 12.51\,$\pm$\,0.13\,mag (\citealt{2008ApJ...673L.185B}), it exhibits an absolute magnitude in the \textit{K}-band of 6.70\,$\pm$\,0.33\,mag. Since UScoCTIO\,108\,A is very faint there is no significant difference in detection limits between 0.25\,arcsec (36\,au) and 2\,arcsec (290\,au). For an age of 10\,Myr, all objects down to masses of 0.015\,$\pm$\,0.001\,M$_{\odot}$ would have been detected in the field of view of the detector up to an angular separation of 6.5\,arcsec (943\,au). For a younger age of 5\,Myr, this limit is slightly lower with 0.013\,$\pm$\,0.002\,M$_{\odot}$. In general, the detection limits are not as low as in the other VLT/NACO images given the young age of the system. This is because the very faint primary was used for AO corrections, which were hence less optimal than the AO corrections for the other discussed targets with brighter primary stars as reference sources.

\section{Formation scenarios of the sub-stellar companion to GQ\,Lup}

\subsection{In-situ formation via core accretion or gravitational instability}

While it remains uncertain if the companion to GQ\,Lup has a mass below the Deuterium burning limit, it is nonetheless in principle possible that it formed like a planet. Thus we consider the possibility of in-situ formation via either core accretion or gravitational instability of the protoplanetary disk. As discussed in section \ref{sec: target: gqlup} the likely inclination of the host star and thus most probably the inclination of the circumstellar disk is \textit{i}\,=\,27$\pm$\,5$^\circ$. If the companion would have formed in this disk one would expect it to be on a low eccentricity orbit with a similar inclination. However, as discussed in section \ref{sec: orbit: gqlup}, we can exclude all orbits with low eccentricity and inclination since they do not fit with our astrometric measurements. For an inclination of $\sim$27$^\circ$ we find only possible orbit solutions with eccentricities larger than $\sim$\,0.6. It should be mentioned though, that if the disk inclination is higher (as suggested by \citealt{2008A&A...489..349S}) circular orbits would still be possible.\\ 
However, \cite{2010AJ....139..626D} state that given the disk properties that they observe it is unlikely that the circumstellar disk around the host star is massive enough to have formed an object of the companion's estimated mass range at a projected separation of $>$\,100\,au. Thus from the astrometry and the estimated disk properties it seems unlikely that the companion has formed in-situ in a planet-like fashion. \\ 

\subsection{Inward or outward formation and successive scattering}

The next logical step would be to consider that the companion has formed either closer to the star or further away and was then scattered into its current orbit by dynamical interaction with a potential third body which has formed at a similar separation from the host star. Since we find a large number of eccentric orbit solutions, and our best fitting orbits are in fact all eccentric, we can not exclude such a scenario with our astrometric study. However, as already mentioned in the previous section, it is doubtful whether the disk is massive enough for two (sub)-stellar companions to form at even larger projected separations from the host star. However, the disk mass could have been larger in the past. If the disk would have been massive enough, one would expect that gravitational instability would dominate the formation process at such large distances from the central star rather than core accretion. In the recent study by \cite{2013A&A...552A.129V} formation of wide sub-stellar companions via gravitational instability was examined via numerical simulations. They failed to produce any objects in the Jupiter to brown dwarf regime on wide orbits around host stars with $\leq$\,0.7\,M$_\odot$ , i.e. the likely mass of GQ\,Lup\,A. Thus formation of two such objects, on larger projected separations than we observe the companion currently at, seems unlikely.\\
It would still be possible that two sub-stellar companions have formed within few au of the host star. Such a formation process, regardless whether dominated by core accretion or gravitational instability, would leave distinctive signatures such as holes or gaps in the circumstellar disk. However, \cite{2010AJ....139..626D} report that their best fitting disk model does not include gaps or holes and that they cannot find any indications for such structures in their data. Thus the formation of two sub-stellar companions in close proximity to the host star seems also unlikely.   

\subsection{Star like formation from collapse in protostellar cloud}

It is possible that the host star and the companion formed like a binary star system by core collapse in the protostellar cloud. This formation scenario would not require that the plane of the orbit and of the circumstellar disk are aligned. For example, the study by \cite{1994AJ....107..306H} shows that main sequence solar-type binaries with separations larger than 40\,au commonly show spin-orbit misalignment. As shown in section \ref{sec: orbit: gqlup} we find a number of relatively well constrained orbit solutions with low eccentricities, which would be expected if no additional third body disturbs the system. Thus from the astrometry we can not exclude such a formation scenario.\\
However, in a most recent study by \cite{2014ApJ...783L..17Z} it is argued that the accretion rates that are calculated from photometry of the companion in the UVIS with HST/WFC3 are about an order of magnitude higher than what is observed in stars of similar age. They thus argue that they find it unlikely that the companion has formed via collapse of a protostellar core but rather that it formed via gravitational instability in a protoplanetary disk.   


\section{Conclusions}

\begin{figure}

\includegraphics[scale=0.45]{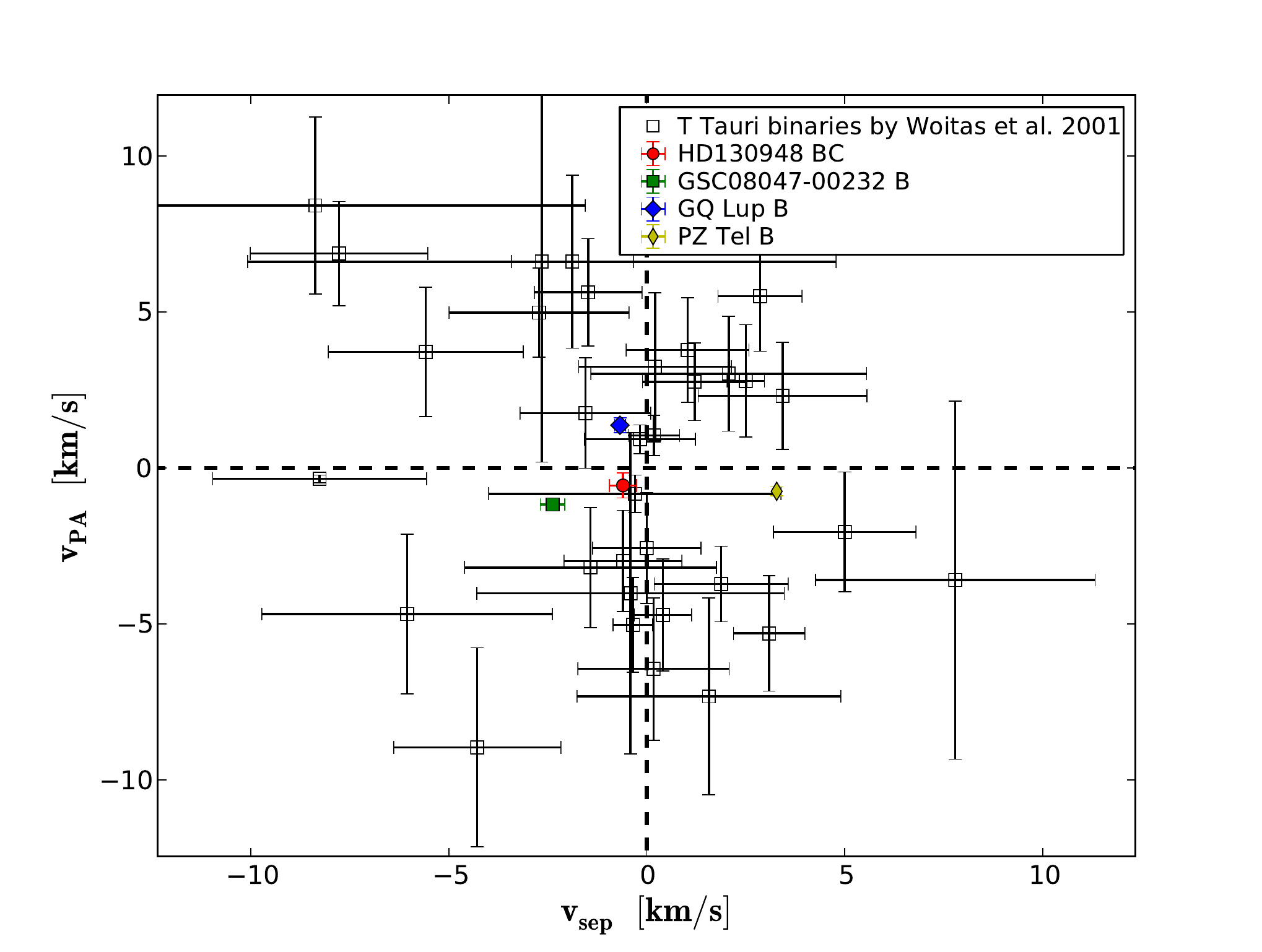}

\caption[]{Projected orbit velocities in separation and position angle of close ($<$\,100\,au projected separation) T Tauri binary stars by \cite{2001A&A...369..249W} as well as the same velocities for directly imaged sub-stellar companions discussed in this study and previous studies by \cite{ginski-hd130} and \cite{2014MNRAS.438.1102G}.} 
\label{woitas}
\end{figure}

In this study we presented new astrometric measurements of 6 directly imaged sub-stellar companions to stars or brown dwarfs. We showed for the first time with a high significance of 5.3\,$\sigma$ that the sub-stellar companion to the young nearby star GQ\,Lup shows differential motion as compared to its host star. This differential motion is consistent with slow orbital motion although no orbit curvature was detected yet. With our statistical LSMC approach we find best fitting orbits with eccentricities between 0.21 and 0.69 with corresponding orbit periods between 786 and 1734 years. While these orbit solutions produce the lowest reduced $\chi^2$ we can not yet exclude less or more eccentric orbits as detailed in section \ref{sec: orbit: gqlup}. Furthermore, we find that our astrometric solutions together with the known spin of the host star and the properties of the circumstellar disk might point towards a star-like formation of the companion. This is, however, in disagreement with the high accretion rates measured for the companion which favor a formation via gravitational instability in the protoplanetary disk. In addition to these considerations, we detected an additional companion candidate 6.9\,arcsec to the east of GQ\,Lup\,A, which is likely a background object given our astrometry of the object.\\
For the PZ\,Tel system we found that our new astrometric measurement agrees very well with the expectations raised in our previous study of the system. We found that the orbit analysis produced also similar results as in previous studies. However, with the new evidence that there is in fact no circumbinary disc present in the PZ\,Tel system, we expanded our orbit analysis to significantly larger semi-major axes. We find best fitting orbits with eccentricities above 0.9 and orbit periods between 387 and 1767 years. In general we can exclude orbit solutions with eccentricities smaller than 0.62. We also showed that the parameter space for possible orbit solutions can be further constrained by measurements taken within the next few years.\\ 
There has been a recent study by \cite{2014MNRAS.437.2686P} indicating that an unseen inner companion could introduce large errors in relative astrometry by periodic displacement of the primary star. This could lead to apparent orbits with high eccentricities when in fact the orbits of the outer companions are close to circular. Our best fit orbit of the GQ\,Lup system exhibits an eccentricity of 0.52 and a semi-major axis of 75.6\,au. Using the method outlined in \cite{2014MNRAS.437.2686P}, we calculated that an inner companion with a mass of 0.061\,M$_{\odot}$ on an orbit with a semi-major axis as small as 10.5\,au could in principle introduce the calculated eccentricity when in fact the orbit of GQ\,Lup\,B is circular or nearly circular. Given our deep imaging data of the system we can exclude masses of 0.060\,$\pm$\,0.024\,M$_{\odot}$ at this projected separation with a detection limit of 5\,$\sigma$. Thus considering the uncertainties of this detection limit we cannot firmly rule out that the observed eccentricity is caused by an inner companion undetected by imaging. However, given that no evidence is found for gaps or holes in the circumstellar disk around GQ\,Lup\,A, it is unlikely that such a companion would exist at such a small separation. Thus we do not believe that our recovered orbit solutions for the GQ\,Lup system are significantly influenced by an unseen inner planetary companion. For the PZ\,Tel system, \cite{2014MNRAS.437.2686P} calculated that an object of 0.124\,M$_{\odot}$ at a semi-major axis of 5.5\,au could cause the high eccentricities reported in our previous study. Given our most recent observations we can exclude objects down to 0.025\,$\pm$\,0.003\,M$_{\odot}$ at this distance, therefore the observed high eccentricity is likely not caused by an additional inner companion. However, it is in principle possible, that an inner companion exists on a highly inclined orbit and is thus not detectable at the time of our observations. In any case, such a companion would also introduce only a very small or no astrometric signal when it is behind or in front of the parent star.\\
In order to determine if our measured orbital velocities are reasonable, we compared them to orbit velocities of close ($<$100\,au projected separation) T Tauri binary systems as determined by \cite{2001A&A...369..249W}. The results are shown in Fig.~\ref{woitas}. In addition to the GQ\,Lup and the PZ\,Tel system, we show orbit velocities of the HD\,130948 system and the GSC08047 system which were determined in previous studies by us (\citealt{ginski-hd130} and \citealt{2014MNRAS.438.1102G} respectively). All our systems show small orbit velocities as compared to the stellar binary systems as would be expected of low mass objects on wide orbits. Thus we are confident in the determined orbital velocities.\\
In addition to the orbital motion which we detected in the GQ\,Lup system and PZ\,Tel system, we could show that the DH\,Tau system and the HD\,203030 system are not showing significant differential motion. The RXJ1609 system also exhibits no apparent signs of significant orbital motion. However, due to the apparent offsets in the different astrometric data sets it is not possible to determine with any certainty if orbital motion is present over the full time period covered. In the case of the UScoCTIO108 system it was also not possible to detect orbital motion due to the large uncertainties of the originial astrometric measurement by \cite{2008ApJ...673L.185B}. In fact it is not yet possible to decide whether the companion and the primary star are co-moving at all. Given our new and much more precise astrometric measurement, such a determination is now possible with one additional well calibrated measurement of similar precision. 

\section*{Acknowledgments}

This work is based partially on the PhD Thesis of Christian Ginski. We want to thank T. L\"ohne for fruitful discussion of the GQ\,Lup disk parameters. We thank the ESO Paranal Team and ESO Users Support group. CG and MM thank DFG for support in project MU 2695/13-1. RN and RE would like to thank DFG for support in the Priority Program SPP 1385 on the "First Ten Million Years of the Solar System" in project NE 515 / 34-1 and 34-2. RN and RE would also like to thank DFG for support in project NE 515 / 36-1. RE would in addition like to thank the Abbe School of photonics for the Ph.D. grant. RN and AB would like to thank DFG for support in NE 515 / 32-1. NV acknowledges the support by project DIUV 38/2011. This research has made use of the SIMBAD database as well as the VizieR catalogue access tool, operated at CDS, Strasbourg, France. This research has made use of NASA's Astrophysics Data System Bibliographic Services. This research has made use of the AstroBetter blog and wiki. CG wishes to express special thanks to John Hunter, author of Matplotlib, which was used for the creation of all diagrams in this work. CG expresses special thanks as well to Donna Keeley for language editing of the manuscript.

\bibliographystyle{mnras}
\bibliography{myBib}

\label{lastpage}

\end{document}